\documentclass[preprint,12pt,authoryear]{elsarticle}
\usepackage[T1]{fontenc}
\usepackage{amsmath}
\usepackage{amssymb}
\usepackage{xcolor,textcomp}
\usepackage{subcaption}

\graphicspath{{figures/}{.}}
\usepackage{hyperref}

\begin{document}

\begin{frontmatter}

\title {Self-assembled versus biological pattern formation in geology}

\author[1,2]{Julyan H. E. Cartwright\corref{cor1}}
\author[3]{Charles S. Cockell}
\author[4]{Julie G. Cosmidis}
\author[5]{Silvia Holler}
\author[1]{F. Javier Huertas}
\author[6]{Sean F. Jordan}
\author[3]{Pamela Knoll}
\author[7]{Electra Kotopoulou}
\author[3]{Corentin C. Loron} 
\author[3]{Sean McMahon}
\author[8]{Anna Neubeck} 
\author[9]{Carlos Pimentel\corref{cor1}}
\author[1]{C. Ignacio Sainz-D\'{\i}az}  
\author[10]{Piotr Szymczak\corref{cor1}}
\affiliation[1]{organization={Instituto Andaluz de Ciencias de la Tierra, CSIC, 18100 Armilla, Granada, Spain}}
\affiliation[2]{organization={Instituto Carlos I de F{\'{\i}}sica Te{\'o}rica y Computacional, Universidad de Granada, 18071 Granada, Spain}}
\affiliation[3]{organization={School of Physics and Astronomy, University of Edinburgh, Edinburgh, UK}}
\affiliation[4]{organization={Department of Earth Sciences, University of Oxford, Oxford OX1 3AN, UK}}
\affiliation[5]{organization={Cellular Computational and Biology Department, CIBIO, Laboratory for Artificial Biology, University of Trento, Via Sommarive 9, Povo, 38123, Italy}}
\affiliation[6]{organization={Life Sciences Institute, School of Chemical Sciences, Dublin City University, Ireland}}
\affiliation[7]{organization={Laboratoire Écologie, Systématique et Évolution, Université Paris-Saclay, France}}
\affiliation[8]{organization={Department of Earth Sciences, Uppsala University, Uppsala, Sweden}}
\affiliation[9]{organization={Departamento de Mineralog\'{\i}a y Petrolog\'{\i}a, Facultad de Ciencias Geol\'ogicas, Universidad Complutense de Madrid, 28040 Madrid, Spain}}
\affiliation[10]{organization={Institute of Theoretical Physics, Faculty of Physics, University of Warsaw, Poland}}
\cortext[cor1]{Corresponding authors: julyan.cartwright@csic.es (Julyan H. E. Cartwright), cpimentelguerra@geo.ucm.es (Carlos
Pimentel), piotrek@fuw.edu.pl (Piotr Szymczak)}

\date{Draft version 1.6 of \today.}

\begin{abstract}
Both abiotic self-organization and biological mechanisms have been put forward as the origin of a number of geological patterns. It is important to comprehend the formation mechanisms of such structures both to understand geological self-organization and in order to differentiate them from biological patterns---fossils and bio-influenced structures---seen in geological systems. Being able to distinguish the traces of biological activity from geological self-organization is fundamental both for understanding  the origin of life on Earth and for the search for life beyond Earth.
\end{abstract}

\end{frontmatter}

\tableofcontents

\section{Introduction}\label{sec:intro}

To what extent are patterns in rocks and minerals self-organized; products of internal dynamics rather than outside agencies?  And can we further discriminate between the products of biological and non-biological outside agencies? 
The geological record contains objects and structures whose morphologies and patterns evoke biology, due to apparent organization, non-randomness, or because they do not obey simple crystallographic rules, etc; thus one of the questions we must concern ourselves with here is what makes us view something as lifelike, or not. 

\subsection{The concept of pattern}
At the heart of the concept of a pattern is some regularity that we can perceive. This may be a periodic regularity, either in space or in time, such as a symmetry---an invariance under a  transformation like translation, reflection, rotation, or scaling--- or some even more general form of regularity, such as coherence, correlation or a well-defined scale. 
Well-correlated structures, even if not symmetric, can still form a pattern; for example, an apparently disordered field of irregularly shaped spots that nonetheless share a characteristic size and characteristic spacing will still be perceived as a pattern. 

A pattern is then that aspect of the morphology of a structure that allows us to recognize it as belonging to some class. That is to say, it is some aspect, some regularity of its form, so that we recognize---for instance---many different forms of the number 2, or the letter "a" as all being 2’s and a’s. 
Consider the recognition of a face that we humans do so well, so that merely drawing two dots and a line beneath them evokes a face for us. We speak of pattern formation in biology, for example, with embryogenesis, vegetation patterns, and ecology, among others.  In computer science, 
pattern recognition refers to assigning a label to an object and thus making a classification. Likewise, in this work we speak of geological patterns and geological pattern formation and contrast these against biological pathways that can lead to the same outcome. Here we consider  formation processes that lead, in a predictable manner, to structures having a certain aspect of form that allows us to classify them into particular categories.

Are these patterns seen in geological systems the result of biological action, or is their apparent organization that of a system self-assembled in a given physicochemical environment?  The definition of what we regard as \emph{the system} is crucial when discussing self-organization and self-assembly. If the whole universe is the system, then everything is self-organized. We are instead interested in cases where the system does not contain many components, yet its behaviour is more complicated than might naively be expected; it is in such situations that we tend to speak of self-organized behaviour. 

\subsection{The term biomorph}

The terms \emph{biomorph} and \emph{biomorphic} are fundamental in this discussion, but they mean different things to different people. It is therefore worth taking a moment to clarify their interdisciplinary origins and  current usage. In 19th-century biology, the term \emph{biomorphic science} is found \citep{mueller1883}. Here, \emph{biomorphic} refers to taking  both structure and function into account. For example,  
``the biomorphic classification of the tissues is based upon the structure, taken in conjunction with the function, the mechanism in connection with its use'' \citep{mcalpineTransverseSectionsPetioles1891}. The noun \emph{biomorph} itself was introduced---perhaps based upon this adjective---in the 1895 book \emph{Evolution in Art: As Illustrated by the Life-histories of Designs} by the biologist, anthropologist, and ethnologist Alfred Cort Haddon, as a term in the study of the history of art:  ``the biomorph is the representation of anything living in contradistinction to the skeuomorph which ... is the representation of anything made, or the physicomorph which is the representation of an object or operation in the physical world'' \citep{haddon1895evolution}. Examples of such biomorphs are found in decorations on ancient pottery from many cultures. In the 20th century, with the rise of abstract art, the term began to be applied to works by artists including Joan Mir\'o, Barbara Hepworth and Henry Moore. Another such 20th century artist is the surrealist painter,  zoologist, and ethologist Desmond Morris. In an article in \emph{New Scientist} magazine \citep{dawkins1986creation},  and later in his book \emph{The Blind Watchmaker}, Richard Dawkins explained how he used one of Morris' biomorph paintings, \emph{The Expectant Valley} (1972),
for the cover of his book \emph{The Selfish Gene}  \citep{dawkins1976selfish}. Dawkins took the name and Morris' concept that biomorphs evolve in the mind, and applied it in a computer algorithm to mimic Darwinian evolution and to produce forms that resembled living organisms: ``On my wanderings through the backwaters of Biomorph Land, I have encountered fairy shrimps, Aztec temples, Gothic church windows, aboriginal drawings of kangaroos, and, on one memorable but unrecapturable occasion, a passable caricature of the Wykeham Professor of Logic'', he wrote  \citep{dawkins1996blind}.

From here, Juan Manuel Garc\'{\i}a Ruiz took the name \citep{garcia1994inorganic} for the abiotic but  lifelike  silica--carbonate structures he had observed in laboratory experiments  \citep{ruiz1981teoria} (Section \ref{sec:biomorphs}). It was later suggested that these structures would likely be misinterpreted as fossils if found in natural mineral settings \citep{garcia2002morphology,garcia2003self}. Thus, the term \emph{biomorph} entered geology, where it now denoted an abiogenic form or object resembling a living or extinct organism.

\subsection{The focus of this review}

There are numerous structures in geological systems whose formation mechanisms are debated between abiotic and biological processes. This question is important in itself for understanding the extent to which biology has shaped, and continues to shape, the geology of our planet. It is also crucial for determining when a particular structure found in a geological context may be taken as evidence for life, both on Earth and in the search for life on other worlds. 

The word \emph{fossil} is associated with palaeontology, designating forms or objects that are either the remnants of organisms themselves, or the consequences of their activity as observed in the geological record (trace fossils). But the term \emph{fossil} has not always been used in this way. The word itself is from the Latin \emph{fossilis}, ``dug (up)'', from \emph{fodere}, ``to dig''. In the past, geologists used \emph{fossil} to denote any object that had a biological aspect to it (e.g., a shape evoking that of a living organism), regardless of whether or not its biogenicity was established. These putatively biogenic objects are often termed pseudofossils today. Thus, this review can be viewed as an enquiry into how to distinguish pseudofossils from actual fossils by comparing patterns produced by physical self-organization with those produced by biological processes.

Pertinent to this question, there are not only features in geology that look biogenic but may in fact be abiogenic, but also the reverse: features that look abiogenic but may be biogenic. In the following, we illustrate this problem with examples of both types.

\section{Geological patterns whose origin may be abiotic or biological; Morphologies common to abiotic and biological patterns}\label{sec:morphologies}\label{sec:geostructures}

Some structures previously viewed as fossils are pseudofossils, now accepted to have formed from purely physical and chemical processes. A famous example is \emph{Eozo\"on canadense}, described in the 19th century as a gigantic foraminiferan, composed of bands of calcite and serpentine.
Eozo\"on is now understood as an abiotic textures in metamorphic rocks \citep{hahn1876xxiv,o1970eozoon,adelman2007eozoon}.
However, there are also numerous instances in which the origin of a given geological pattern is currently debated between biological and abiotic interpretations.

Certain patterns are found both in abiotic settings and in biological systems. 
What characteristic morphologies are seen in geological environments, and what abiotic and biological mechanisms produce these structures?

\subsection{Filaments and tubes}\label{sec:tubes}

\begin{figure}[t!]
 \begin{subfigure}{0.49\textwidth}
    \includegraphics[width=\linewidth]{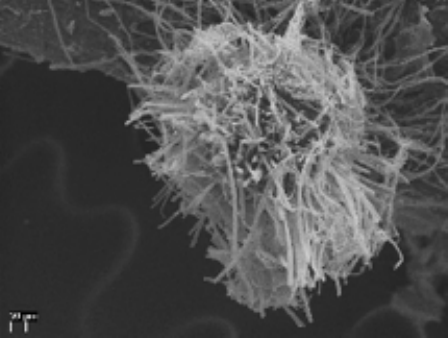}
    \caption{} 
  \end{subfigure}
  \begin{subfigure}{0.49\textwidth}
    \includegraphics[width=\linewidth]{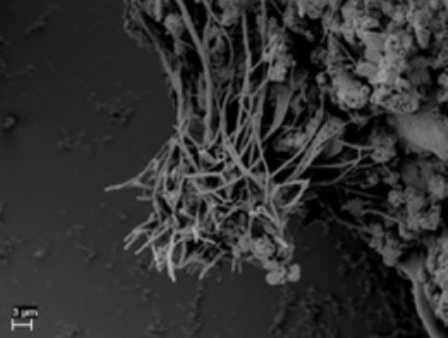}
    \caption{} 
  \end{subfigure}
  \begin{subfigure}{0.34\textwidth}
    \includegraphics[width=\linewidth]{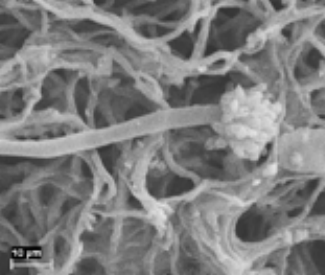}
    \caption{} 
  \end{subfigure}
  \begin{subfigure}{0.32\textwidth}
    \includegraphics[width=\linewidth]{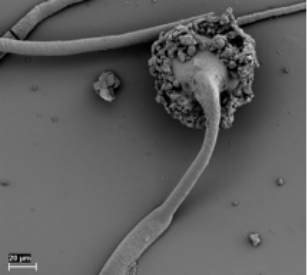}
    \caption{} 
  \end{subfigure}
  \begin{subfigure}{0.32\textwidth}
    \includegraphics[width=\linewidth]{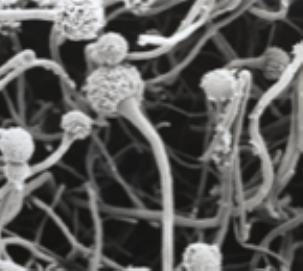}
    \caption{} 
  \end{subfigure}
\caption{
Biological and abiotic tubes.
SEM images of tubular colonies of a) \emph{Aspergillus niger} feeding on straw (image: Anna Neubeck) and b) manganese oxide chemical gardens (image: \citeauthor{Huld2023Chemical}, \citeyear{Huld2023Chemical}). The \emph{A.\ niger} tubes ranged in size between 2-5 $\mu$m in diameter, and the manganese oxides were found in ranges between 1--20 $\mu$m in size.
SEM images of c) \emph{A.\ niger} (image: Anna Neubeck), d) manganese oxide chemical gardens (image: Anna Neubeck), and e) \emph{A.\ niger} closeup (image: D. Gregory \& D. Marshall, CC BY 4.0). 
}
\label{fig:anna}
\end{figure}

We group filaments, that is, threads or fibres with a solid core, together with tubes, i.e., hollow cylindrical bodies, since in a geological context it is often difficult to distinguish one from the other when dealing with smaller scales where a filament and a blocked tube can look identical.

These shapes constitute, alongside spherical and ovoid shapes (Sec.~\ref{sec:vesicles}), one of the oldest morphologies produced by life on Earth. In fact, the oldest putative and disputed body fossils assemblage, the 3.46 Ga Apex chert biota \citep{schopf_microfossils_1993, Schopf_evidence_2007, Wacey_Chert_2016}, is dominated by  structures in the form of carbonaceous, trichome-like filaments interpreted as remains of multi-cellular cyanobacteria and other prokaryotes. Similar trichome-like forms as well as  simpler, carbonaceous tubes commonly interpreted as the tubular sheaths of filamentous cyanobacteria are also part of the slightly younger putative biotas of the 3.4 Ga Strelley Pool Chert in Western Australia \citep{Schopf1987Early, schopf1992oldest} and the 3.3 Ga Kromberg Formation in South Africa \citep{Walsh1985Filamentous, Walsh1992Microfossils, Schopf2002Laser–Raman}. Equivalent organic tubular structures continue to be common among other Proterozoic biotas, such as the 1.88 Ga  Gunflint Iron Formation \citep{Tyler1954Occurrence, Schopf1965Electron}, the Tonian (0.85 Ga) Bitter Springs Formation of central Australia \citep{Schopf1968MicrofloraOT}, or as part of many Ediacaran assemblages \citep{Arvestål2020Organic}. However, also new, more complex and often larger forms appear as early as the Archean, which invited  a Metazoan interpretation of certain forms \citep{Butterfield2015Early, Kazmierczak2016Multiyear}. With the beginning of the Cambrian, even larger organic tubular structures, such as the pterobranchs, entered the fossil record.

Numerous extant and fossil single-celled eukaryotes, including some representatives of the foraminifera, tintinnids, testate amoebae, fly larvae, priapulids, phoronids, and gastropods produce tubular integuments by  \emph{agglutination}, i.e., the binding together of foreign material, typically mineral grains, within an organic or organic--mineral cement.
Distinct from agglutination is controlled biomineralization (Sec.~\ref{sec:biomin}).  
Widespread metazoan biomineralization, arguably the most obvious process for producing fossilizable tubular biological  structures of various sizes, began in the late Ediacaran before it intensified during the early Cambrian, during the interval widely known as the Cambrian explosion. The oldest unequivocal eukaryotic biomineralized hard parts, phosphatic scale-like microfossils, date to the Cryogenian \citep{Cohen2011Phosphate, Wood2018Exploring}, whereas it was in the late Ediacaran that larger biomineralized structures first appeared.  By contrast, microbially induced biomineralization, including calcified cyanobacteria (e.g., Girvanella), iron-oxidizing bacterial filaments (such as Gallionella-like twisted stalks), and stromatolitic carbonates, records much older biologically mediated mineral precipitation extending back through the Proterozoic and into the Archean \citep{astafievaIronBacteriaGallionella2020, zhangGirvanellaFossilsPhanerozoic2024}. Analogous to the earliest organic and agglutinated fossils, tubular structures were among these first biomineralized forms, such as the cosmopolitan taxon Cloudina and the similar Sinotubulites (see the recent review by \citet{Yang2022Taxonomic}). New tubular forms, notably the anabaritids, began to appear in the early Cambrian and rapidly diversified thereafter. 

The prevalence of tubes during this time interval, the late Ediacaran to early Cambrian, inspired \citet{Budd2016Ecological} to call this stage of Metazoan evolution \emph{tube world}. Most of these early tubes were made of calcium carbonate, but tubes of calcium phosphate (e.g., Hyolithellus) appeared alongside the anabaritids in the early Cambrian. The tubes of Cloudina, anabaritids, hyolithelminths, and related forms functioned as complete shells enclosing the entire organism, as is also the case for many later-evolving taxa with tubular shells (e.g., scaphopods, serpulids). At the same time, biomineralized tubular elements became common as parts of larger skeletons in other clades, including arthropod spines, tubular spines or processes in brachiopods and bivalves, elongate tubular extensions of the brachiopod pedicle opening, and the hollow stem of crinoids.

Tubular trace fossils in the form of various types of burrows are ubiquitous  in the Phanerozoic. They started to appear in the late Ediacaran with the advent of Metazoan exploration, i.e., burrowing into the sediment; the oldest simple traces, \emph{Planolites sp.}, are from the late Ediacaran.

Traces in sediments, while often indicative of biological activity, can also be formed through abiotic processes. 
Ambient inclusion trails (AITs) found in cherts and authigenic minerals are one such instance (Sec.~\ref{sec:ait}).
Degassing tubes are another notable example of such non-biological formations. These tubular structures arise from the release of gases from sediments or volcanic activities and can vary significantly in size, ranging from micrometres to several decimetres in diameter. This wide size range complicates the interpretation of sedimentary traces as evidence of past life, as degassing tubes can resemble the burrows and trails left by ancient organisms. Additionally, other geological processes, such as fluid escape structures or the movement of subterranean water, can create similar features in sediments, further challenging the identification of biologically-derived traces. \citep{Stanulla2017Structural} 

Chemobrionic tubular structures (Sec.~\ref{sec:chemgardens}) tend to form in environments where chemical gradients exist, often through the interaction of different solutions or gases.  These structures emerge when mineral precipitation occurs in zones of sharp chemical contrast, such as where alkaline and acidic solutions meet. For example, in hydrothermal vents, chemobrionic formations arise as mineral-rich waters mix with the surrounding seawater, leading to the deposition of complex, sometimes tubular minerals. The resulting structures can resemble biological tubes but are purely inorganic. Their formation is influenced by factors like temperature, fluid dynamics, and the concentration of dissolved minerals.  Laboratory experiments   create chemobrionic tubular structures by mimicking these natural conditions \citep{barge2015chemical}. 

Other examples of  abiotic tubular structures are zeolites, naturally occurring minerals with a porous structure that can form micrometre-sized tubular crystals. These crystals are formed through the crystallization of aluminosilicate compounds in specific geological environments, such as volcanic ash beds or hydrothermal systems \citep{Halliwell2022Hierarchical}. The tubular structures of zeolites are often hexagonal, but there exist  biological structures that are likewise  formed of  polygonal tubes. 
Different types of laboratory-made self-assembled abiotic structures such as silica-carbonate biomorphs (Sec.~\ref{sec:biomorphs}) and carbon-sulfur biomorphs (Sec.~\ref{sec:carbon-sulfur}) can adopt filamentous morphologies.
In many instances, similar tubular morphologies can be produced through both abiotic and biological mechanisms, requiring  careful diagnosis for their differentiation.
An example is iron--mineral tubules  \citep{barge2016self,podbielski2025troubles}.

\subsubsection{Igneous glasses and their weathering products}\label{sec:glasses}

\begin{figure}[tbp]
\begin{center}
\includegraphics[width=\columnwidth]{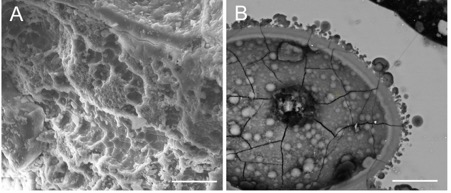}
\end{center}
\caption{Etching features within basaltic glass
A) Etching features observed by secondary Scanning Electron Microscopy (SEM) in basalt glasses observed by \citet{cockell2009alteration,cockell2009bacteria} (scale bar 15 $\mu$m), B) Back-scattered SEM images of pitted features and tubule-like etchings around the rim of a vesicle in weathered basalt glass sliced and polished through a vesicle (scale bar 15 $\mu$m). The material within the vesicle is palagonite.
}\label{fig:basaltic1}
\end{figure}

\begin{figure}[tbp]
\begin{center}
\includegraphics[width=\columnwidth]{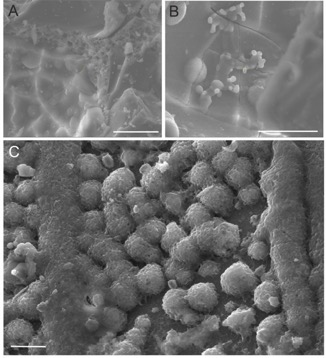}
\end{center}
\caption{
Biomorphic features in basalt glass and its palagonite weathering products observed in secondary SEM. A) Hemispherical structures on palagonite rinds in terrestrial weathered basalt glass, Iceland (from \citet{cockell2009bacteria}) (scale bar 20 $\mu$m), B) Manganese-rich biomorphs in the same weathered glass as (a) (scale bar 20 $\mu$m), C) Biofilm-like spherical assemblages in weathered basaltic glass from the mid-Atlantic Rift (from \citet{cockell2010microbial})  (scale bar 5 $\mu$m).
}\label{fig:basaltic2}
\end{figure}

Basalt glasses and other volcanic rocks, and their secondary weathering products, are the substrates in which some signatures of early life on Earth are found and in which we would search for life on Mars \citep{furnes2004early,wacey20063,fisk2006iron,mcloughlin2007biogenicity}. Yet within them, we find a variety of patterns and self-organizing geological structures which have equivocal connections with life (e.g., \citet{thorseth1992importance,thorseth1995microbes,thorseth2003microbial,fisk1998alteration,torsvik1998evidence,furnes2005preservation,storrie2004elemental,staudigel1998biologically,banerjee2006preservation}).

Prominent among these structures are pitted and tubular structures found within basaltic glasses (Fig.~\ref{fig:basaltic1}a,b). The former are overlapping irregular indentations in glass comprising pits whose diameters are often similar to microorganisms (micron-sized), raising the question of their association with microorganisms \citep{thorseth1992importance}. Tubules are cylinders, often several microns long, formed perpendicular to fractures or vesicle surfaces. Due to the presence of microorganisms within these structures, it has been suggested that they are produced by microorganisms, but it is also possible that these structures, abiotically produced, preferentially trap biological material. In particular, the discovery of these features in ancient rock samples has made their provenance---that is to say, whether they proceed from biological or abiotic processes---important in assessing evidence of early life. 

Both pitted textures and tubules have plausible non-biological explanations. Pitted textures may represent non-homogeneous weathering of glasses, caused by irregularities in the glass or the fluids on their surfaces. Tubular structures, although elongate and regular, could represent leaching and weathering along stress fractures around the rims of vesicles.
In addition to features that involve the etching of glasses into organized structures, glasses also host self-organized structures on their surfaces in positive relief. For example, palagonite rinds, the weathering products of basalt glass, frequently exhibit swellings on their surface that resemble coccoid microorganisms and are one to several microns in size (Fig.~\ref{fig:basaltic2}a). The lack of separation between the hemispherical forms and the underlying palagonite suggest that they are part of the rind and not microorganisms growing on their surfaces. Other features show more irregular morphologies. \citet{cockell2009bacteria} describe manganese-rich deposits at the micron scale on the surface of palagonites (Fig.~\ref{fig:basaltic2}b). These features have tentacle-like morphologies that exhibit separation reminiscent of cell division. The possibility that these are features produced by manganese-oxidizing bacteria cannot be ruled out, but their irregularity makes their biological provenance questionable.
In a study of mid-Atlantic Ridge basaltic glasses that are $<200\,000$ years old, \citet{cockell2010microbial}  observed sheets of coccoid structures strongly reminiscent of biofilms (Fig.~\ref{fig:basaltic2}c). However, these sheets of structures, across a lateral transect, grade into subdued form and then into a continuous palagonitic weathering rind, suggesting they are not biological. The spheres are $\sim4$~$\mu$m in diameter, larger than many prokaryotes, and nanoSIMS does not reveal carbon enrichment. Determining the provenance of such structures is complicated by the unequivocal evidence of diatoms and other biological material deposited on the glass from seawater. 

These findings show that basalt glasses can give rise to diverse structures whose organizational characteristics confound the search for life, since they have morphologies and sizes similar to microorganisms. Broadly, they fit into two types of structures: pitting and etched structures within the glass and three-dimensional structures deposited on the surface of glass, particularly associated with secondary weathering products such as palagonite. These structures show that at micron scales, interactions between glasses, fluids and local gradients of ions leached from the glass and brought in by fluids can lead to diverse geological structures with a complexity that brings them within the vista of structures we would seek as evidence of life. Clearly there is more work to be done in understanding the emergence and self-organisation of the chemical products of basalt glass--water interactions and their relationship to biota.

\subsubsection{Clay structures}\label{sec:clay}

Clay is a naturally occurring material primarily composed of fine-grained minerals that are usually plastic when mixed with the right amount of water and harden when dried or fired. Clay minerals are hydrated phyllosilicates characterized by particles roughly a few micrometres in size \citep{Bergaya2013General}. These minerals are found in soils, continental and marine sediments, and in hydrothermal alteration zones. In these environments, clay minerals often coexist with microorganisms and organic matter, which are frequently involved in their formation. Therefore, the physico-chemical conditions—such as temperature, pressure, and fluid composition (including electrolytes and ligands)—that may be permanent or seasonal, lead to formation processes governed by kinetics rather than thermodynamic equilibrium. This results in euhedral crystals of clay minerals being less common than particles with low crystallinity and small size. The involvement of microorganisms in the formation of clay minerals is supported by evidence from various processes within Earth’s crust (see the review of \citet{Cuadros2017Clay}). The interaction between clay minerals and microorganisms extends beyond simply controlling crystal shape.  There are arguments both supporting and opposing the idea that some textures and particle arrangements—such as tubules, fibers, spherules, honeycomb textures, or booklets—may have a biotic origin. 

The most common morphology of clay minerals is platy. However, the typical morphology of many phyllosilicates and related minerals is tubular. The planar structures of tetrahedral and octahedral sheets differ in lattice constants, which creates structural stresses when they are assembled in layers. To reduce these stresses, the tetrahedra in either the top or bottom sheet may be slightly rotated to improve coupling between sheets. Alternatively,  stress may be alleviated by the layers curving, as seen in halloysite, imogolite, and chrysotile, or through isomorphic substitutions. However, no biotic origin of these morphologies has been reported yet.

\begin{figure}[tb]
\begin{center}
\includegraphics[width=\columnwidth]{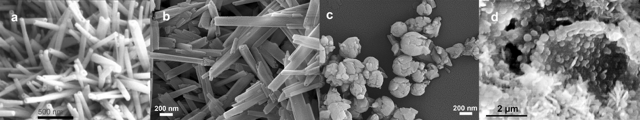}
\end{center}
\caption{SEM images of a) hollow tubular halloysite (Turplu Mine, Balikesir, Turkey); b) prismatic halloysite (Hospital Quarry, Elgin, Scotland); c) spherical halloysite (Pama Mine, Province of Rio Negro, Argentina); d) spherical kaolinite on gel grains, formed in hydrothermal experiments \citep{HUERTAS1993Hydrothermal}. Images a, b, c reproduced from the ‘Images of Clay Archive’ of the Mineralogical Society of Great Britain \& Ireland and The Clay Minerals Society (https://www.minersoc.org/images-of-clay.html)}
\label{fig:clay1}
\end{figure}

Halloysite is a mineral in the kaolin group, notable for its high water content, found in various geological settings. It forms through weathering and hydrothermal alteration of volcanic and crystalline rocks, as well as in laterite profiles and soils derived from volcanic materials \citep{Joussein2005Halloysite}. Halloysite particles can appear as tubes, cylinders, or spirals (Fig.~\ref{fig:clay1}). Its structure consists of a rolled kaolinite-like 1:1 aluminosilicate layer containing water molecules and exchangeable cations \citep{Bailey1990Halloysite}. These layers curve to compensate for the mismatch between the tetrahedral and octahedral sheets, forming tubes \citep{guggenheim_phyllosilicates_2015, Gray2023Morphological}. Further irreversible dehydration may convert tubular particles into prismatic forms \citep{Kogure2013Structure}. Spheroids of halloysite are another common form, frequently found in weathered volcanic pumices and ashes. They develop under high saturation conditions that promote precipitation in confined spaces \citep{Cravero2016origin}. How the halloysite structure adapts to this spheroidal shape remains an open question. During dehydration, spheres may develop apparent polygonal faces \citep{Bailey1990Halloysite}. Halloysite spheroids may be related to the spherical kaolinite identified by \citet{tomura_spherical_1983, tomura_growth_1985} with a radial structure. The rearrangement of the gel or glass under hydrothermal conditions in highly supersaturated solutions,  reducing surface energy, creates metastable spherical particles that subsequently evolve into platy crystals \citep{fiore_morphology_1995, Huertas2004In}. Conversely, \citet{tazaki_microbial_2005} proposed a biotic origin based on  experiments.  Hollow spherical halloysite may form on the surface of bacterial cells at room temperature. Degassing of H\textsubscript{2}S may produce “clay bubbles” and transitional “spherical bio-halloysite” formation.

\begin{figure}[tb]
\begin{center}
\includegraphics[width=\columnwidth]{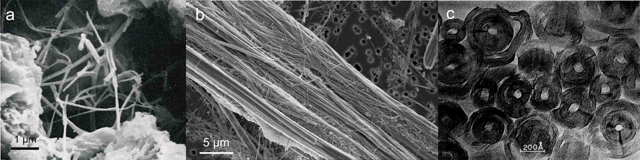}
\end{center}
\caption{SEM images of fibres of a) imogolite (Kyushu, Japan; \citet{Eswaran_1972}) and b) chrysotile (Macizo de Ojén, Málaga, Spain); TEM image of the cross-section of chrysotile tubules (Tasmania; \citet{Yada1971Study}). }
\label{fig:clay2}
\end{figure}

The morphology of imogolite features microscopic, thread-like structures that are several micrometres long, formed by bundles of single-wall aluminosilicate nanotubes (Fig.~\ref{fig:clay2}a). It was first identified by \citet{Yoshinaga1962Imogolite} in volcanic soils in Japan. It is a typical low-crystallinity product of volcanic ash weathering. Its single-layer structure consists of a gibbsite sheet that curls to accommodate isolated silica tetrahedra on the inner surface \citep{CRADWICK1972Imogolite, D2023atomic}. 

Chrysotile, a polymorph of the serpentine mineral group, has a rolled structure that forms fibrous morphologies (Fig.~\ref{fig:clay2}b). The brucite sheet slightly exceeds the lattice dimensions of the silica sheet, causing the layer to curl into scrolls or tubular shapes \citep{guggenheim_phyllosilicates_2015}. Electron microscopy shows curved layer walls forming cylinders, spirals, and cone-in-cone morphologies \citep{Yada1967Study}. The individual tubes or fibrils bundle together to create fibres (Fig.~\ref{fig:clay2}c).

\begin{figure}[tb]
\begin{center}
\includegraphics[width=\columnwidth]{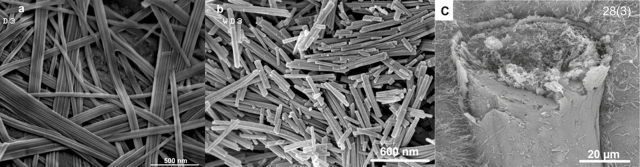}
\end{center}
\caption{SEM images of a) sepiolite (San Bernardino Co., CA-USA) and b) palygorskite (Jizda, Kazakhstan). c) Biomorph of sepiolite-dolomite with external dolomite crust (Madrid Basin, Spain; \citet{Leguey2010role}). Images a, b reproduced from the ‘Images of Clay Archive’ of the Mineralogical Society of Great Britain \& Ireland and The Clay Minerals Society (https://www.minersoc.org/images-of-clay.html).}
\label{fig:clay3}
\end{figure}

Sepiolite and palygorskite are fibrous phyllosilicates characterized by a unique structural arrangement. Their frameworks consist of continuous basal oxygen planes, but with periodic inversions of the tetrahedral chains---every two units in palygorskite and every three in sepiolite---along with discontinuous octahedral sheets \citep{BaileyStructures1980}. This configuration gives rise to pyroxene-like ribbons or polysomes, whose alignment underpins their fibrous morphology (Fig.~\ref{fig:clay3}a,b). The inversion of the silica tetrahedra serves to accommodate the structural mismatch between the tetrahedral and octahedral sheets. According to \citet{Guggenheim2011Structures}, the octahedral strips in these minerals can incorporate significant amounts of water, both as hydroxyl groups, OH\textsuperscript{-}, and molecular water, which aligns with their frequent formation in lacustrine environments. These settings, often subject to evaporation-driven alkaline conditions, promote the precipitation of such minerals \citep{Galan2011Palygorskite}.

Furthermore, \citet{Leguey2010role} proposed that some high-grade sepiolite deposits in the Madrid Basin (Spain) may have originated through biomineralization processes. In these cases, dolomite biomorphs acted as precursors to sepiolite formation (Fig.~\ref{fig:clay3}c). Additional biosignatures, such as nutrient depletion and reducing conditions at mineral--organic interfaces, suggest the presence of clay-organic biofilms that supported microbial activity during mineral genesis.

\begin{figure}[tb]
\begin{center}
\includegraphics[width=\columnwidth]{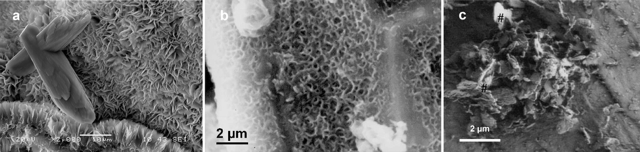}
\end{center}
\caption{SEM images of honeycomb smectite particles arrangements. a) Victoria Land Basin, Antarctica (reproduced from the ‘Images of Clay Archive’ of the Mineralogical Society of Great Britain \& Ireland and The Clay Minerals Society (https://www.minersoc.org/images-of-clay.html)); b) hydrothermal alteration of obsidian \citep{Fiore2001Smectite}; c) “clay hutches” \citep{Lunsdorf2000Clay}.}
\label{fig:clay4}
\end{figure}

Smectites are common phyllosilicate minerals found in meteoric and hydrothermal alteration environments, often associated with the transformation of volcanic glass and other aluminosilicates. Under scanning electron microscopy (SEM), a common feature is the \emph{honeycomb} texture, characterized by a porous, alveolar arrangement of clay lamellae with edge-to-edge or edge-to-face contacts (Fig.~\ref{fig:clay4}a). Numerous studies have shown that honeycomb textures can form through purely physico-chemical processes, without biological involvement. The main mechanisms include 1) 
Incongruent dissolution of volcanic glass: this process releases soluble species (Si, Al, Na, Ca, Mg), causing supersaturation that promotes smectite neoformation \citep{Jongmans1998Isotropic, Celik1999Clay};
2)  Precipitation as lamellar gels: smectites often form as poorly ordered lamellar aggregates, which tend to create open, porous structures in aqueous environments \citep{Lagaly2006Chapter}; 
3) Structural collapse during desiccation: natural drying or drying during SEM analysis can cause partial collapse of the lamellae, resulting in cavities and honeycomb-like morphologies \citep{Deirieh2018Particle}; 
4) Growth in spatially restricted environments: smectite precipitation within existing pores or cavities in the substrate can produce reticulated or alveolar structures \citep{Christidis_Scott_Marcopoulos_1995, Arslan2010Mineralogy}.

These processes have been successfully replicated in laboratory settings under strictly abiotic conditions, confirming their non-biological origin. The altered materials included volcanic glass (Fig.~\ref{fig:clay4}b) \citep{Fiore2001Smectite}, amorphous Al-Mg silicates \citep{Huertas2000Experimental}, and feldspar \citep{Haile2015Experimental}. Numerous other examples further support this conclusion. In contrast, \citet{Lunsdorf2000Clay} added soil to sterilized water and a sterilized glass slide. After 14 days, a biofilm developed, notably containing some house-of-card structures (honeycomb-like structures that these authors called \emph{clay hutches}, Fig.~\ref{fig:clay4}c). According to their interpretation, bacteria generated these structures after being displaced from their original habitat and dispersed in water. Although certain microorganisms can enhance volcanic glass alteration and clay mineral formation—for example, through excretion of organic acids or biofilm formation—their activity is not essential for developing honeycomb textures. In natural settings, the presence of microbial activity and such morphologies may simply indicate overlapping favourable conditions rather than a direct cause-and-effect relationship. 

\begin{figure}[tb]
\begin{center}
\includegraphics[width=\columnwidth]{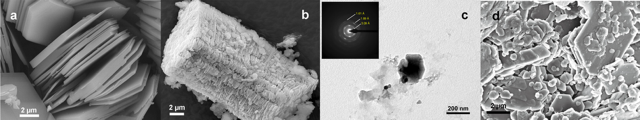}
\end{center}
\caption{Euhedral crystals of kaolinite. SEM image of a) a kaolinite booklet (close view) (UK North Sea) and b) a large kaolinite stack (Georgia, USA). Kaolinite crystals grown at room temperature in aqueous Si-Al solutions, containing c) microorganisms from a peat-moss soil (TEM image and SEAD pattern, from \citet{Fiore2011Bacteria}); d) \textit{Paecylomyces inflatus} fungi (SEM image, from \citet{Pasquale2024Kaolinite}). Images a and b reproduced from the ‘Images of Clay Archive’ of the Mineralogical Society of Great Britain \& Ireland and The Clay Minerals Society (https://www.minersoc.org/images-of-clay.html).}
\label{fig:clay5}
\end{figure}

Kaolinite is one of the simplest clay minerals. A distinctive texture of highly crystalline kaolinite is the \emph{booklet}—stacked, platy crystals arranged like pages of a small open book (Fig.~\ref{fig:clay5}a,b). These structures are typically micrometre-sized, consisting of thin lamellae or plates that can slightly diverge from a common centre, giving the appearance of a booklet. They form through purely inorganic geochemical processes, such as the hydrolysis of feldspar in acidic conditions, the weathering of aluminum-rich rocks, or the hydrothermal alteration of volcanic glass or feldspar. While microbial activity can influence mineral weathering—for example, by producing organic acids—kaolinite itself does not require biological processes for formation. Therefore, its formation is considered abiotic, even if biological factors can accelerate or influence the conditions in which it develops.

Laboratory precipitation of kaolinite was first observed by \citet{Linares1971Kaolinite} from circumneutral Al and Si solutions, with the aqueous Al stabilized by fulvic acids extracted from peat soil. Forty years later, \citet{Fiore2011Bacteria} successfully reproduced this experiment in solutions by inoculating them with  heterotrophic soil bacteria from the same peat soil used by \citet{Linares1971Kaolinite}, resulting in very few euhedral kaolinite crystals (Fig.~\ref{fig:clay5}c). Furthermore, when Si-Al solutions were inoculated with fungi (\textit{Paecylomyces inflatus}), widely grown idiomorphic kaolinite crystals developed (Fig.~\ref{fig:clay5}d) \citep{Pasquale2024Kaolinite}. In both studies, the microorganisms catalysed the transformation of an aluminosilicate gel precipitated from solution into kaolinite. These experiments would be barren without the presence of microorganisms. However, although microorganisms seem key, it cannot be definitively stated that these minerals are of biotic origin. In many natural environments, abiotic and biotic processes interact; a microbial mat may initiate a structure that is later modified by inorganic precipitation. Some morphologies, such as spheres or tubes, can form in both biological and inorganic contexts, making it difficult to determine their origin without detailed analysis.

\subsubsection{Ambient inclusion trails and tubular microcavities}\label{sec:ait}

\begin{figure}[tbp]
\begin{center}
\includegraphics[width=0.9\columnwidth]{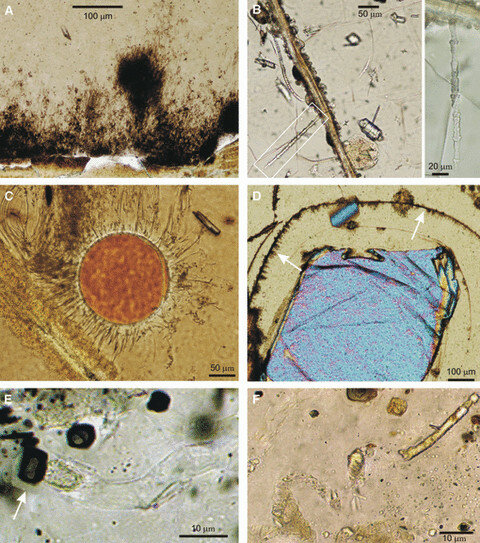}
\end{center}
\caption{ 
Optical microphotographs of tubular microcavities (TMCs) and ambient inclusion trails (AITs). Two different TMCs are depicted, \textit{Tubulohyalichnus simplus} (A, C, and D, arrowed) and \textit{Tubulohyalichnus annularis} (B). E and F are two different ambient inclusion trails.
From \citet{MCLOUGHLIN2010Mechanisms}.}
\label{fig:ait}
\end{figure}

Ambient inclusion trails (AITs; Fig.~\ref{fig:ait}) found in cherts and authigenic minerals can form tubular tunnels in minerals and rocks due to a directional fluid flow and/or local dissolution of minerals within a rock. AITs differ in their presentation of longitudinal striae, consistent diameter, and polygonal cross-section. The origins of AITs are uncertain but are speculated to arise from the migration of crystalline or organic inclusions within sealed substrates. \citet{MCLOUGHLIN2010Mechanisms} argue that AITs in, for example, pillow lavas are unlikely, because, they argue,  localized chemical solution agents, and elevated fluid pressures, necessary to drive this process, are absent. However, they contrast ambient inclusion trails  found in cherts and authigenic minerals with biological tubular microcavities (TMC).
Such biological tubular microcavities, it is argued, would be distinguishable from those owing to abiotic 
processes based upon shape, distribution, and a lack of intersections \citep{finlay2020reviews}.

\subsection{Branching patterns}\label{sec:branching}

Branching patterns are among the most pervasive  growth patterns in nature. Examples from biology---trees, lungs, corals, etc---and from abiotic systems---like lightning, fluid flows, etc---highlight their remarkable breadth across both in scale and medium. Despite appearing in such diverse contexts, the underlying dynamics of branching often exhibits surprising similarities, suggesting the presence of general principles governing its formation. Historically, branching structures have been treated separately within disciplines such as botany, geology, and physics. This limited the recognition of branching as a unifying morphogenetic process. However,  interdisciplinary approaches have begun to reveal shared mechanisms across systems that were once considered unrelated \citep{Fleury-2001}.
Modern theoretical and computational models suggest that many branching structures, whether biological or abiotic, may belong to common universality classes defined by shared mathematical rules and physical constraints, such as diffusion, competition, growth inhibition, and mechanical instability \citep{Fleury-2001, kenkel1996, Hou2023, Saito1989, Heim2017}.
In biology, branching is often highly regulated and hierarchical, serving essential roles in resource transport (e.g., vascular systems in plants and animals), growth and reproduction (e.g., fungi, corals), or environmental sensing and exploration (e.g., root systems, neuronal networks). Abiotic branching, by contrast, typically arises from physical principles such as fluid dynamics, electrical discharge, or chemical diffusion, leading to morphologies that are often more stochastic or fractal in character \citep{Liu2023Dendritic}.

\begin{figure}[tbp]
 \begin{subfigure}{0.474\textwidth}
    \includegraphics[width=\linewidth]{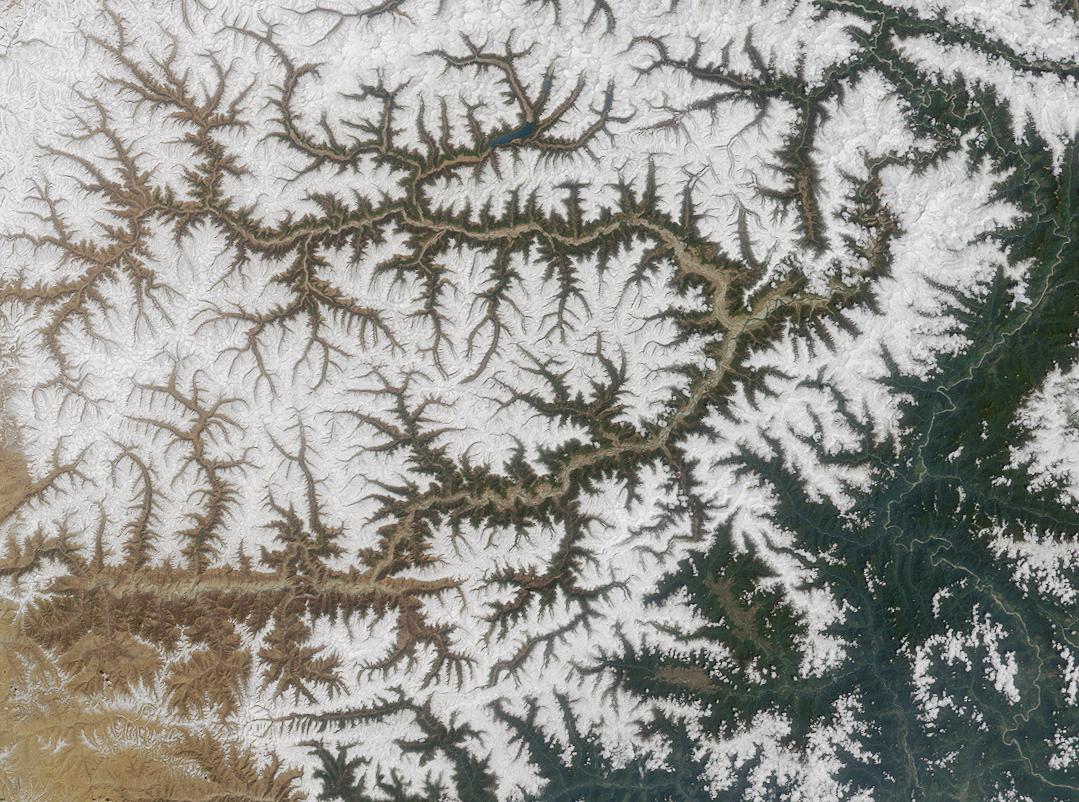}
    \caption{} 
  \end{subfigure}
   \begin{subfigure}{0.526\textwidth}
    \includegraphics[width=\linewidth]{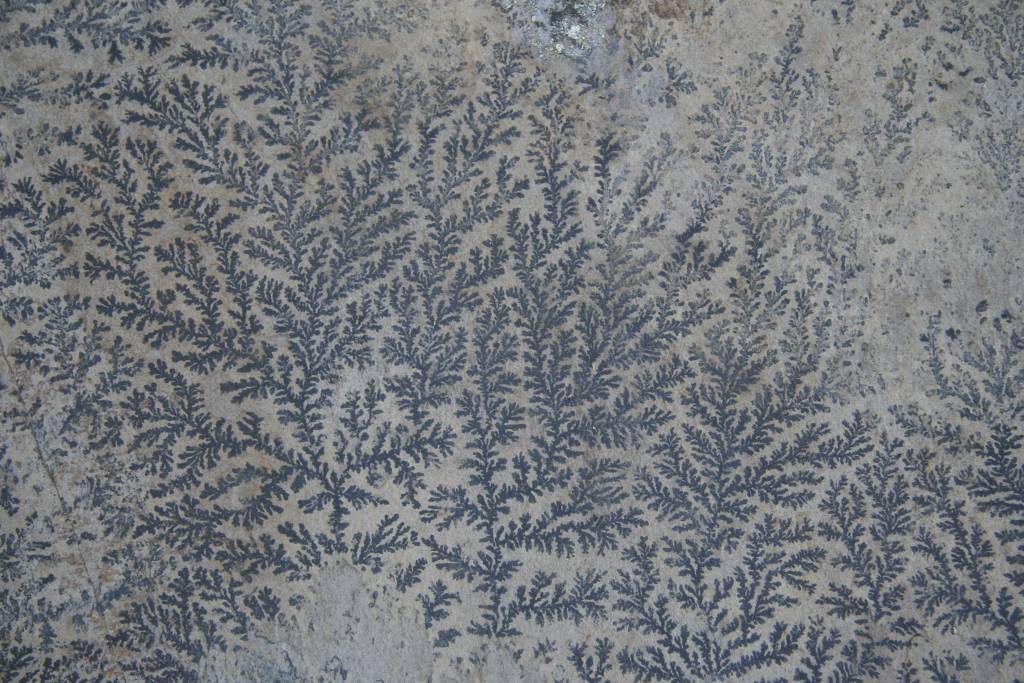}
    \caption{} 
  \end{subfigure}
\caption{Geological branching patterns. 
(a) Dendritic drainage at a km scale: the Yarlung Tsangpo River, Tibet, seen from space: snow cover has melted in the valley system (NASA, public domain). (b) Geological dendrites at a cm scale, Burra Creek, South Australia (David Clarke, CC BY-NC-ND 2.0).
}
\label{fig:branching}
\end{figure}

Across different spatial scales, branching structures appear in diverse and sometimes convergent forms. At a large scale, geological branching patterns are seen in geomorphology in river networks (Fig.~\ref{fig:branching}a). On the metric-decimetric scale, a very striking example of branching geological structures are the fulgurites, which are rocks formed by the action of lightning \citep{cartwright_self-organized}.

Branching networks of tubes are also found both in biology and in abiotic systems. We have noted in Sec.~\ref{sec:tubes} the ubiquity of tubes, arising from both of these sources, in geological contexts, and in many cases tube branching is seen. Microbial activity can produce tubes through filamentous growth, sheaths, or tunnelling behavior in microbialites. Abiotic processes such as gas escape, hydrothermal venting, or permineralization can likewise create tubular features. In many cases, these tubes exhibit branching, contributing additional morphological complexity.

Other ubiquitous geological branching patterns are those called dendrites; `tree-like' (Fig.~\ref{fig:branching}b). Dendrites may arise through a variety of mechanisms. Crystallographic dendrites, like those found in snowflakes or metallic alloys, result from rapid and non-equilibrium crystal growth \citep{nakamura2017eastern}. Mineral dendrites, particularly those composed of manganese or iron oxides, develop via diffusion-limited aggregation on rock surfaces or within fractures \citep{garcia1994inorganic, Neubeck2025Mixed,Hou2025mineral}. Fluid-flow dendrites can form through unstable reaction-diffusion processes in porous or sedimentary environments. Biogenic dendrites may be shaped by microbial mediation, with extracellular polymeric substances often playing a role in guiding mineral precipitation into filamentous or reticulate arrangements \citep{TURNER2005}.
It is not clear whether all geological dendrites are completely abiotic, or some have a biological influence.

\subsubsection{Dendrites and Frutexites} \label{sec:dendrites}

 \begin{figure}[tbp]
\begin{center}
\includegraphics[width=0.9\columnwidth]{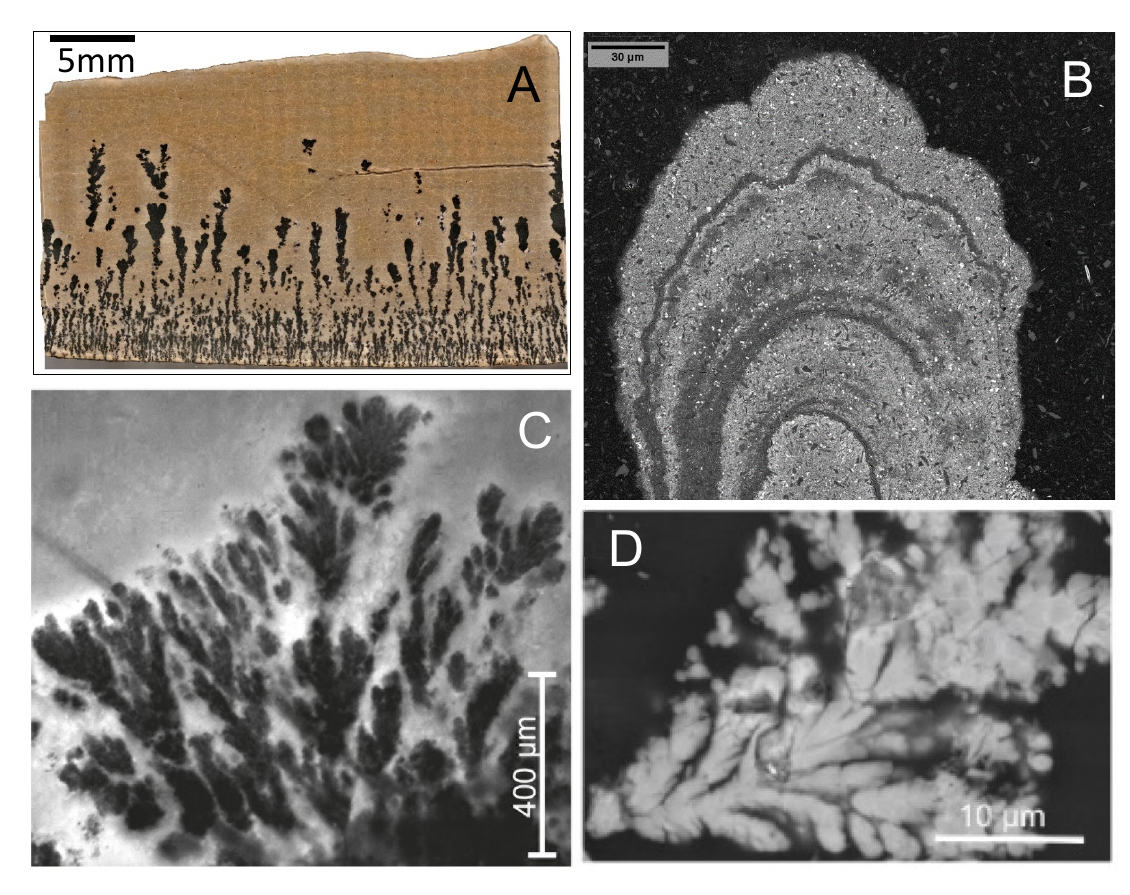}
\end{center}
\caption{(A) Cross-section through three dimensional manganese oxide dendrites in  zeolites and (B) scanning electron microscope–backscattered image showing structure of one of such dendrites with internal banding \citep{Hou2023}, (C) Frutexite-like structure from the tunnel of Äspö Hard Rock Laboratory, Sweden \citep{Heim2017}, (D) dendrite structures from a Mn rock wall deposit in the Ytterby mine, Sweden \citep{Sjoberg2021}.}
\label{fig:dendrite}
\end{figure}

Mineral dendrites are typically blackish-to-reddish branched patterns commonly found as two-dimensional structures on rock surfaces such as beddings, joints and fracture surfaces \citep{Potter1979b, Chopard1991, Garcia1994} and 
 more rarely as three-dimensional structures within a host rock \citep
 {Straaten1978,Potter1979b,Hou2023}.  They consist of amorphous or low-crystalline metal oxide minerals (e.g., manganese or iron oxides). The organic-like form of dendrites  led some naturalists to the conviction that they are fossilized plants~\citep{Dezallier1755,Knorr1755}, 
  however, already by mid-18th century, they were increasingly recognized as patterns emerging as a result of physical processes. For example, \citet{daCosta1757} mentions that similar patterns can be created by inserting oil in between two sandstone plates and then pulling apart the plates, making  this the first known experiment in viscous fingering \citep{cartwright_self-organized}. 

  Classical models explain dendrites as purely abiotic precipitates: manganese-rich water seeping through cracks oxidizes upon contact with air/oxygenated water, causing Mn-oxide minerals to crystallize in a fractal pattern~\citep{Chopard1991, Garcia1994, Hou2025mineral}. However, microbes are known to strongly influence Mn and Fe oxidation in nature, prompting research into possible biogenic pathways for dendrite formation.

Manganese(II) oxidation by dissolved $\text{O}_2$ is kinetically slow at neutral pH, but certain bacteria and fungi can catalyse this reaction, making it orders of magnitude faster~\citep{Emerson1982,Learman2011}. Likewise, microaerophilic iron-oxidizing bacteria rapidly oxidize Fe(II) to Fe(III) minerals under conditions where purely chemical oxidation is limited~\citep{kappler2005,Maisch2019}. This raises the hypothesis that some dendritic Mn/Fe oxide deposits may be microbially mediated mineralization, rather than solely inorganic precipitation. Researchers have begun to investigate whether the presence and metabolism of Mn-oxidizing or Fe-oxidizing microbes during dendrite growth leave identifiable signals in the morphology, chemistry, or isotopic composition of the mineral patterns.

Field evidence of microbial involvement comes from environments where Mn-oxides are actively forming. \citet{Shiraishi2019} studied a Japanese hot spring depositing Mn-oxides and quantified the relative contributions of biotic and abiotic Mn(II) oxidation. They found Mn-oxidizing bacteria at all sites and determined that microbially mediated oxidation dominated over purely chemical processes, by factors of 2 to 24. While Shiraishi’s work focused on stromatolitic crusts, not the classic  dendrite patterns, it demonstrates that microbial Mn oxidation can far outpace abiotic routes in natural systems, implying that ancient dendritic Mn deposits could plausibly be of biogenic origin if formed under similar conditions.

Direct morphological evidence linking microbes to Mn dendrites comes from \citet{Sjoberg2021}, who examined a subsurface Mn deposit in the Ytterby mine (Sweden) that contains dendritic/botryoidal Mn-Fe oxides. Using electron microscopy, they observed organic microbial traces,  including filamentous and cell-like shapes encrusted in Mn-oxide, intimately associated with the earliest mineral aggregates. The Mn precipitates appear to nucleate as globular or \emph{wad-like} nanoparticles entangled with organic filaments, which then coalesce and accrete into larger dendritic or botryoidal forms. The presence of a “chaotic mix of organic traces and early Mn-oxide particles” that become overlain by laminated Mn-oxide layers suggests an iterative growth process: microbes initiate Mn-oxide deposition (forming nanostructures on their cells/EPS), then abiotic precipitation adds layered overgrowths. This iterative biotic--abiotic growth could explain the internal banding or lamination in some dendrites. Sjöberg’s mineral analyses showed a mix of Mn(III/IV) oxides with residual Mn(II) and high Ca content, consistent with biogenic birnessite containing cation substitution. Their study proposes that microbial activity is a \emph{seed} for the non-classical crystallization of Mn dendrites, giving a template that guides subsequent chemical accretion.

A related class of structures, known as Frutexites, shares many morphological and compositional features with dendrites but is usually linked with biotic processes. Frutexites are minute, bush-like microstromatolitic structures, typically rich in iron and/or manganese oxides with minor carbonate (calcite). They were first described by \citet{Maslov1960} as problematic microfossils; meaning that their biological origin was unclear. 
Frutexites occur as sub-millimetre to millimetre-scale dendritic \emph{shrubs} composed of divergent branches \citep{RodriguezMartinez2011}. These microcolumns often exhibit fine laminations and a directional growth pattern, generally growing outward or upward from a surface \citep{Neubeck2021}. In cross-section, Frutexites show finely layered alternating bands of iron/manganese oxide and carbonate minerals, giving them a banded or laminated internal texture.

Frutexites have a deep time range, being reported in geological records from the Proterozoic through the Phanerozoic \citep{RodriguezMartinez2011}. They are especially common in paleoenvironmental settings that are dark, low-energy, and low in oxygen, often where normal organisms struggle to live~\citep{Jakubowicz2014}. They are frequently found encrusting or embedded in marine carbonate rocks such as stromatolites, reefs, and hardgrounds~\citep{Walter1979,Boehm1993,Guido2016}. They occur in cavities of stromatolites, within microbial limestones, in marine hardground surfaces, and in early diagenetic voids like cavities, cracks, veins, and Neptunian dikes (fissures filled with sediment)~\citep{Gischler1996,Reolid2010,Cavalazzi2007Chemosynthetic,Mamet2006}. Major fossil occurrences are often associated with rock cavities or fractures in carbonate platforms. Notably, they tend to grow attached to surfaces, either vertically upward or radially outward from the walls of a cavity. Frutexites appear in ancient deep dysaerobic basins and sheltered settings with very low sedimentation rates. They have been documented on the deep seafloor \citep{cho2018} and in cryptic niches of shallow reefs, such as inside small cavities or synsedimentary cracks where light is minimal \citep{Jakubowicz2014,Myrow1991,Walter1979}. Across these occurrences, common themes are low light, low oxygen, and oligotrophic (nutrient-poor) conditions, indicating Frutexites thrived where photosynthetic organisms could not. They are rarely found alongside abundant benthic fauna; when present with  organisms, they often encrust the skeletal remains after the organism’s death, implying Frutexites colonized surfaces in hostile environments or during early burial~\citep{Jakubowicz2014,Reolid2010}.
Despite their abundance in the fossil record, modern analogues of marine Frutexites are scarce, which has long puzzled researchers~\citep{Jakubowicz2014}. However, similar structures have been observed in a variety of modern extreme or non-marine environments~\citep{Kazmierczak2006,Heim2017,RodriguezMartinez2011}. 

From their first description, Frutexites have attracted debate over whether they are biogenic or abiotic in origin. The resemblance to inorganic dendrites led some early workers to suggest an abiotic origin. For example,~\citet{Wendt1970} argued that Frutexites could be nothing more than chemical precipitation features, mineral \emph{pseudofossils} that only mimic biological shapes. This skepticism was fueled by the fact that Frutexites often lack obvious fossils of microorganisms within them, making it hard to prove a biological cause.  Thus, in the past, Frutexites were sometimes dismissed as diagenetic mineral dendrites rather than true microfossils.

However, mounting evidence from paleobiology, sedimentology, and geomicrobiology now supports a biogenic or biologically induced origin for Frutexites in most cases. First, some Frutexites-bearing rocks contain actual microfossils or microbial structures in close proximity. In a Proterozoic example, \citet{Horodyski1975} reported microscopic fossil cells and filaments embedded around Frutexites-like stromatolites, suggesting the shrubs grew intertwined with microbial mats. Likewise,  modern travertine shrubs studied by \citet{Chafetz1998} were found to entomb bacterial filaments and nanoscopic cell-like objects, directly tying the shrub fabric to microbial presence. These studies interpreted Frutexites as biologically induced mineral precipitates, not abiotic minerals, because the microorganisms were physically preserved within or adjacent to the structures. In younger examples, SEM imaging has occasionally revealed filamentous microstructures or biofilm textures templating the iron oxides of Frutexites.

A further important step forward was the discovery of modern Frutexites-like systems. In caves, hot springs, and the Äspö subsurface tunnel (Sweden), living microbial communities are actively precipitating Fe/Mn oxides into dendritic, shrub-like forms
\citep{Guido2016,Heim2017}. For instance, in  Sicilian submarine caves (Italy), researchers observed mesophilic Fe–Mn oxidizing bacteria flourishing in microcavities and inducing Frutexites precipitation \citep{Guido2016}. In the Äspö tunnel, the iron biofilm’s structures “strongly resemble the dendritic pattern of fossil Frutexites” and are clearly the product of microbial metabolism in the biofilm \citep{Heim2017}. These living examples provide a direct analogue: they demonstrate that microbes can produce identical structures, and thus ancient Frutexites are likely fossilized microbialites. The modern analogues also help explain Frutexites preferred settings; for example, iron-oxidizing bacteria thrive in low-oxygen, Fe-rich waters, exactly the conditions inferred for many fossil Frutexites occurrences. 

Sedimentologists have noted that Frutexites often occur in settings that practically require a biological trigger. Frutexites \emph{shrubs} are commonly found on surfaces that were organic-rich or in the process of organic decay, such as surfaces of coral skeletons undergoing microbial decomposition \citep{Jakubowicz2014}. A detailed study of Middle Devonian Frutexites encrusting rugose corals showed that the shrubs formed during early burial of the coral, in concert with microbial processes breaking down organic matter. Frutexites there exhibit alternating hematite and calcite laminae, which is hard to explain by simple inorganic precipitation. The authors proposed a model of fluctuating redox conditions driven by microbial activity: during certain phases, heterotrophic microbes (sulfate-reducing bacteria) consumed organic matter, raising alkalinity and causing micritic calcite to precipitate; in alternating phases, chemoautotrophic microbes (iron-oxidizing/nitrate-reducing bacteria) flourished, oxidizing $\text{Fe}^{2+}$ to $\text{Fe}^{3+}$ and precipitating iron oxide layers. In essence, the Frutexites laminations record microbially mediated chemical oscillations; a biological \emph{fingerprint} not expected in a purely abiotic scenario. Both the calcitic and ferruginous layers are primary (original) and tied to microbial biofilms, indicating the entire structure grew as a result of microbial mineralization under changing environmental conditions.

Finally, geochemical analyses have provided more direct evidence of biotic origin of Frutexites. A recent study examined Frutexites filling veins in subseafloor ultramafic rocks and found clear geochemical biosignatures. The Frutexites material contained detectable macromolecular organic carbon and had a significantly depleted ratio $^{13}$C/$^{12}$C $ \approx 35$ \textperthousand 
~relative to the surrounding carbonate \citep{Neubeck2021}. Such a strong light carbon isotopic signature is characteristic of biologically synthesized organic matter (e.g., microbial biomass), rather than inorganic carbonate, which is usually much less $^{13}$C-depleted. This isotopic evidence is consistent with a biological origin for  Frutexites, implying that microbes were involved in the formation and left behind their carbon. In modern Frutexites-like biofilms, organic biomarkers have also been detected: for example, the Äspö biofilm contains sterols of algal/fungal origin within the iron-oxide matrix \citep{Heim2017} matrix, indicating incorporation of organic compounds from living organisms. Additionally, it was noted that biogenic Fe/Mn oxides can have distinct mineralogical or chemical traits, such as specific crystal forms or impurities, that differentiate them from purely inorganic oxides
\citep{Huld2023Chemical}, which can be used to infer biological involvement in ancient Frutexites when present.
Given this breadth of evidence, the current consensus in geomicrobiology and sedimentary geology is that Frutexites are most likely microbially induced sedimentary structures  in which microbial biofilms catalysed mineral precipitation \citep{Myrow1991,Boehm1993,Reolid2010}.

However, the specific types of microbes involved remain a topic of ongoing discussion~\citep{Jakubowicz2014}.  Early workers speculated Frutexites-forming organisms could be cyanobacteria or other photosynthesizers \citep{Maslov1960,Walter1979}. But the prevalent occurrence in aphotic environments led later authors to favour chemolithotrophic bacteria, especially iron-oxidizers or manganese-oxidizers, as the main architects \citep{Chafetz1998}.  Modern analogues have confirmed that iron-oxidizing bacteria, like Gallionella, Leptothrix, can indeed build such structures in the dark. Still, Frutexites likely result from a consortium of microbes: iron oxidizers, alongside heterotrophic bacteria that modify local chemistry, and perhaps even archaea or fungi in the mix~\citep{Heim2017}. The precise interplay is complex; e.g., nitrate-reducing bacteria might enable iron oxidation under anoxic conditions, and sulfate-reducers create micro-environments that promote carbonate layers. Research is ongoing to unravel these interactions, but it does seem that life processes are integral to the formation of Frutexites.
That said, some caution remains. Because Frutexites lack unequivocal cellular fossils and can form inorganically under certain laboratory or field conditions, each occurrence is scrutinized. There may be instances where abiotic processes produced Frutexites-like forms; for example, rapid chemical precipitation in the absence of microbes can sometimes create layered Fe--Mn oxide crusts that mimic biogenic textures. 

Frutexites morphologically resemble mineral dendrites, both can display branching, plant-like silhouettes (Fig.~\ref{fig:dendrite}); their tree-like form earned Frutexites the nickname \emph{iron shrubs} \citep{RodriguezMartinez2011,Jakubowicz2014}. 
Differences between these two structures, as highlighted in the literature, primarily concern the biological origin of Frutexites versus the chemical origin of dendrites, as well as the fact that Frutexites are always three-dimensional, columnar structures with internal layering. Attention is also drawn to the internal structure of Frutexites, which may include clotted or peloidal textures and microspherules consistent with microbial precipitation~\citep{Neubeck2021}. Thus, the situation is not  clear-cut. On one hand, as noted earlier, it remains unclear whether biological factors might also play a role in the formation of typical mineral dendrites. Moreover, many mineral dendrites are three-dimensional~\citep{Potter1979b}  and may exhibit internal lamination~\citep{Hou2023}. On the other hand, although microbial formation of Frutexites is likely, a clear connection to a microbial precursor is in many cases lacking~\citep{Heim2017}. 
Microstromatolitic precipitates in hydrothermal environments have been observed to span a morphological and structural spectrum, from clearly bacterially mediated shrub-like forms to patterns indistinguishable from purely abiotic dendrites \citep{Chafetz1999}. Thus it might be most accurate to view a continuum transition between abiotic dendrites and biotic Frutexites.

\subsubsection{The Francevillian ‘biota’} \label{sec:frencevillian}

\begin{figure}[tbp]
\begin{center}
\includegraphics[width=0.9\textwidth]{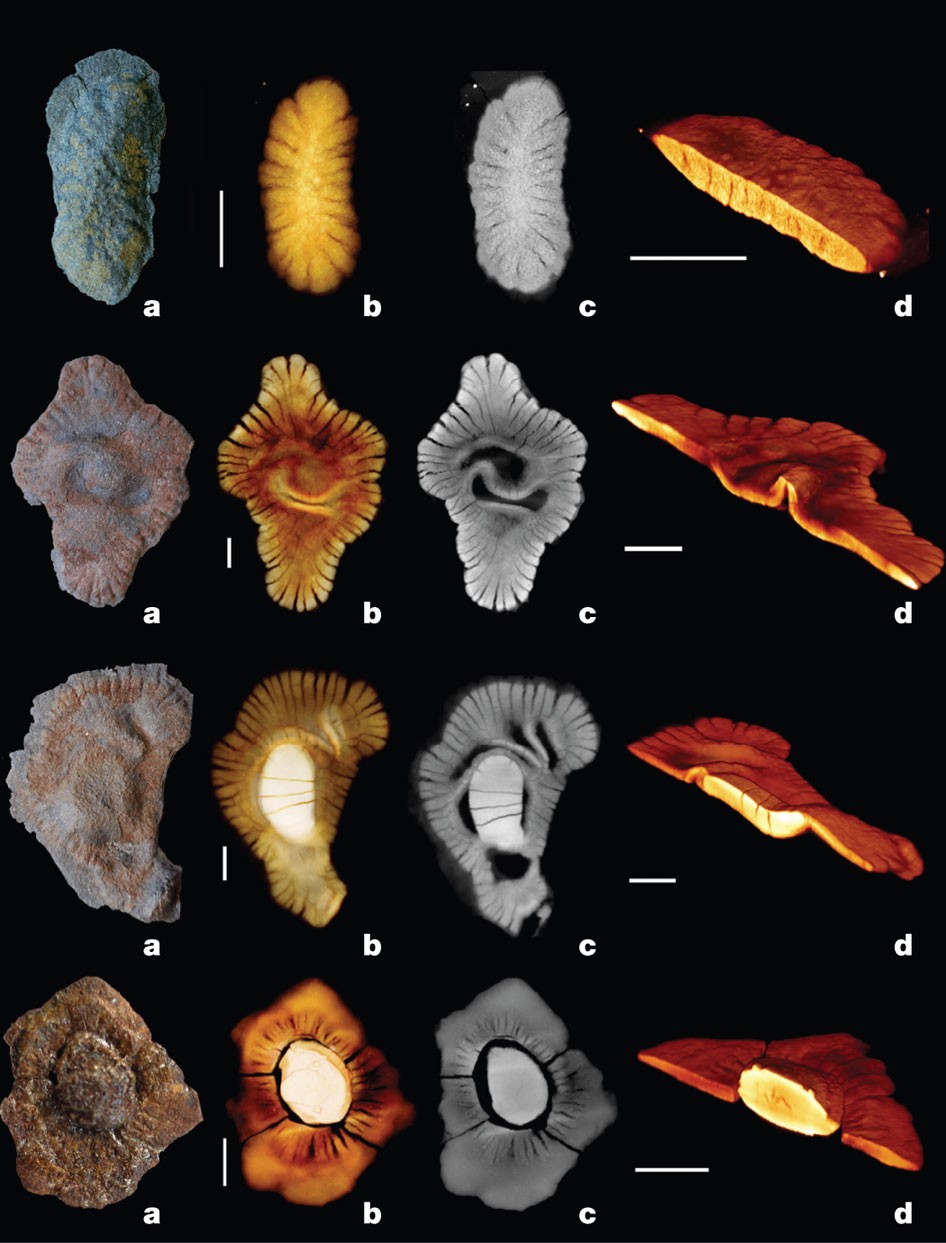}
\end{center}
\caption{Samples from the Francevillian `biota': optical image (a), micro-CT-based reconstruction (b, c) and virtual sections. (image: \citet{albani2010large})
}
\label{fig:francevillian}
\end{figure}

Centimetre-scale  structures have been reported from the ca.\ 2.1 Ga Francevillian B Formation black shale in Gabon (Fig.~\ref{fig:francevillian}). These structures have been interpreted, based on their large size, complex morphologies, isotopic composition, and geochemistry, as fossilized soft-bodied multicellular eukaryotes preserved by pyritization \citep{albani2010large,el20142}. However, the structures lack any diagnostic features of eukaryotes \citep{miao2019new,porter2023frameworks} and are largely out of sync with molecular clock estimates, and fossil evidence, for the emergence of eukaryotes  \citep{porter2020insights}. Their biogenic nature itself remains controversial (e.g., \citet{porter2023frameworks,knoll2011multiple,anderson2016macroscopic,javaux2018paleoproterozoic}). Morphologically, Francevillian structures  resemble  pyrite concretions (e.g., pyrite flowers; see figure 11 in \citet{seilacher2001concretion}) and other authigenic concretions \citep{anderson2016macroscopic}, and similar processes can be suggested to explain their formation in the Francevillian assemblages. The lobate edges of the structures are also reminiscent of viscous fingering patterns (Saffman--Taylor instability; \citet{garcia1993natural,cartwright_self-organized}).

\subsubsection{Brooksella}\label{sec:brooksella}

\begin{figure}[tbp]
\begin{center}\includegraphics[width=0.9\textwidth]{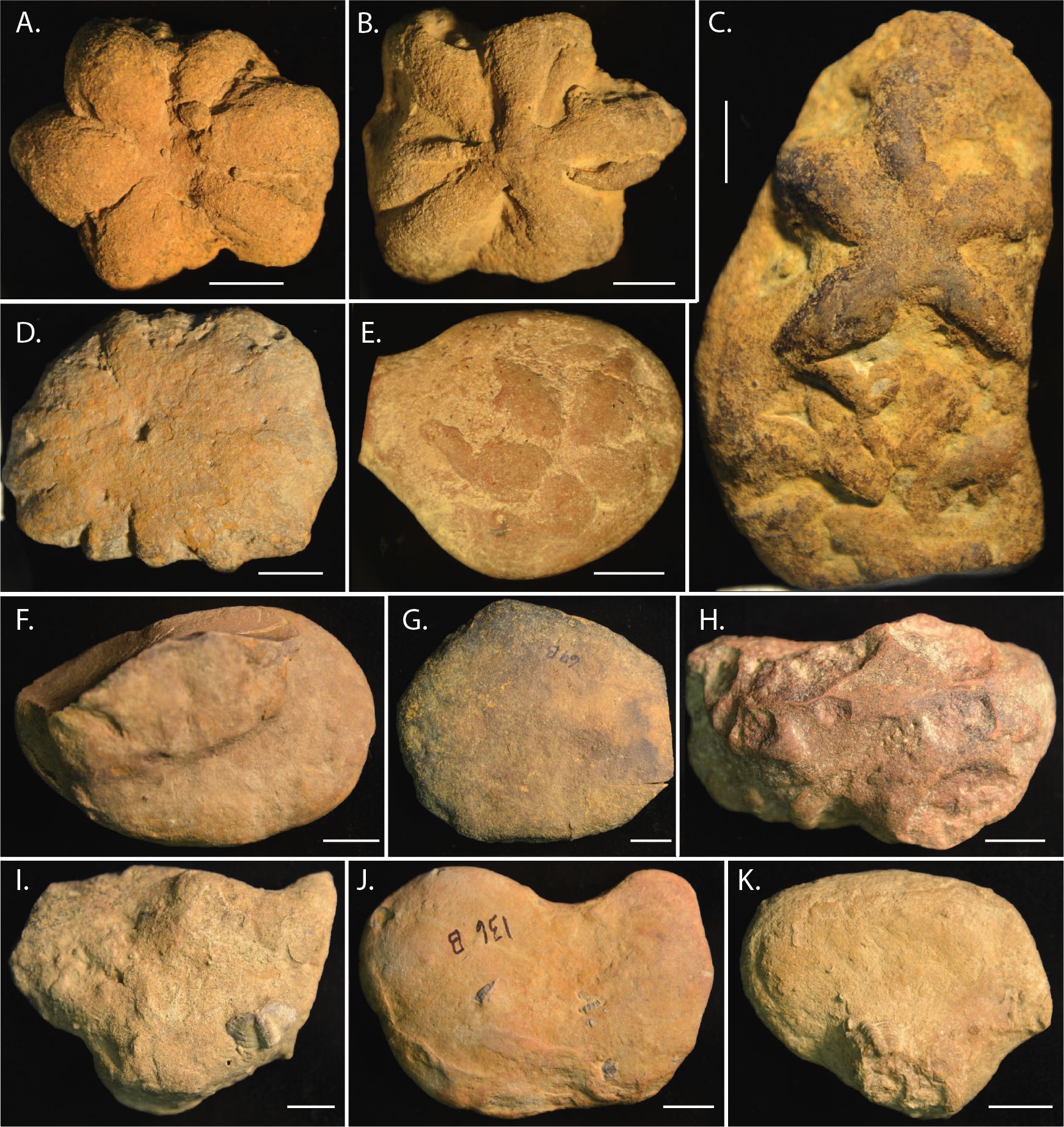}
\end{center}
\caption{Morphological diversity in Brooksella alternata and concretions from Weiss Lake locality.
Brooksella shapes are variable: typical Brooksella have approximately six lobes (A, B); twinned Brooksella can also occur (C); others can have multiple indistinct lobes (D) or lobes that are completely embedded in a concretion (E). Concretions (F–K) also vary in shape, but are mostly round to oblong and many have fossils fragments or whole trilobites embedded in them. Scale bars are one cm.
\citep{nolan2023middle}.
}
\label{fig:brooksella}
\end{figure}

Brooksella, Fig.~\ref{fig:brooksella}, have been attributed variously to 
algae, feeding traces, gas bubbles, and hexactinellid sponges. 
Without internal structure, with strange morphologies and consisting mainly of silica and carbonate, they have long been debated between the biological and the non-biological \citep{nolan2023middle}. 
A diffractogram, \citep{nolan2023middle}, indicates that they are quartz, not opal, and the carbonate is calcite, not dolomite. 
We suspect that they may be non-biological, the result of 
 viscous fingering and osmotic effects \citep{cartwright_self-organized}.

\subsection{Globular forms: spheres, spheroids, spherical shells, and vesicles}\label{sec:vesicles}

Spheres are intrinsically simple shapes that result from any process of radial growth or uniform expansion/contraction focused on a single point. The varieties of spherical objects in geology (beginning with planets themselves) are  too diverse to enumerate. However, spherical shells or vesicles form only under certain conditions. 

\begin{figure}[tb]
\begin{center}
\begin{subfigure}{0.474\textwidth}
    \includegraphics[width=\linewidth]{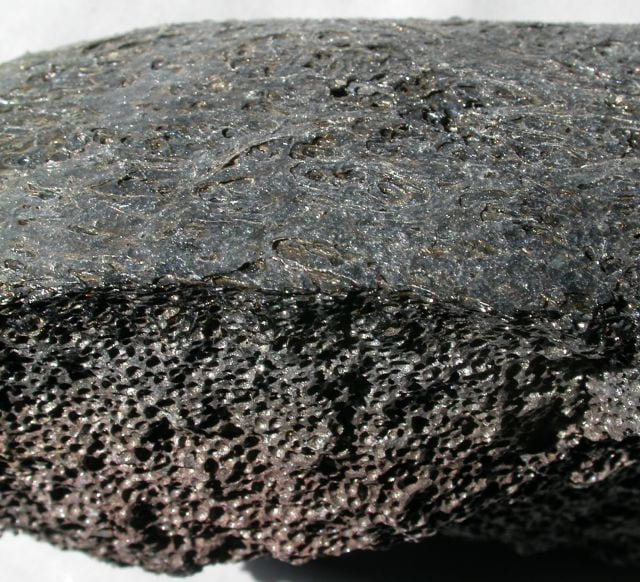}
    \caption{} 
  \end{subfigure}
   \begin{subfigure}{0.387\textwidth}
    \includegraphics[width=\linewidth]{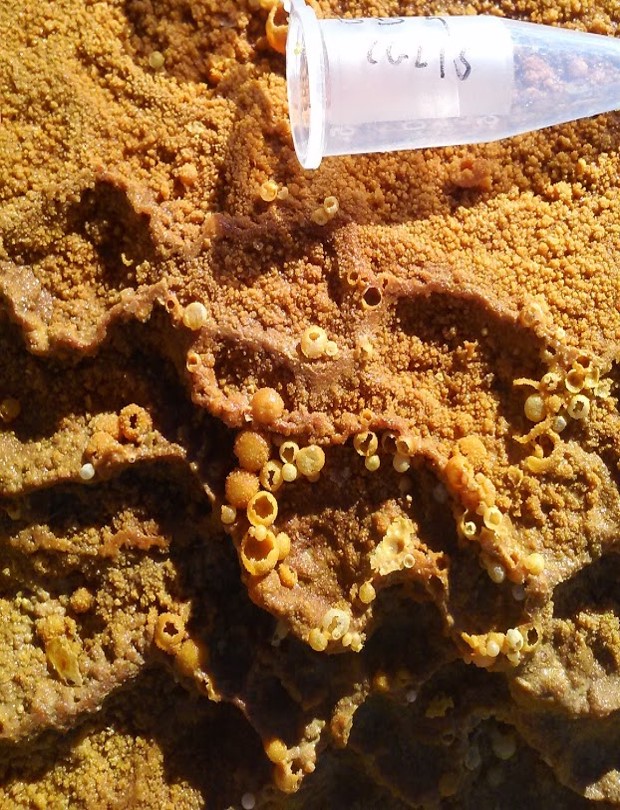}
    \caption{} 
  \end{subfigure}
\end{center}
\caption{Geological vesicles: (a) vesicular basalt (image courtesy of Randy Korotev), and (b) Mineralized gas bubbles at the surface of a travertine at Crystal Geyser, Utah, USA (image: Julie Cosmidis).}
\label{fig:vesicles}
\end{figure}

In geology, vesicles, Fig.~\ref{fig:vesicles}, are small roughly spherical compartments. However, some rocks may also have compartments that are not spherical, in which case they are called vugs. Most common are those found within many types of volcanic rocks, formed by the entrapment of gas bubbles in magma as it cools and solidifies. Lapilli in ash are also spheroidal and sometimes hollow. Vesicular basalt and pumice are particularly common examples having the aspect of a holey cheese (Fig.~\ref{fig:vesicles}a). Geodes can form  as mineral-rich fluids infiltrate these cavities, precipitate, and crystallize. And vesicles are found within hydrothermal vent chimneys where mineral membranes enclose small pores. Mineralized gas bubbles can also form empty spherical vesicles in low-temperature sedimentary environments where rapid mineral precipitation occurs (Fig.~\ref{fig:vesicles}b), for instance in travertine systems \citep{Cosmidis2021Carbonate}. These objects can have a biological origin when the gas is a by-product of microbial metabolism (e.g., oxygen bubbles in stromatolites, \citet{Bosak2009Morphological}).

Vesicles in biology are, likewise, spherical or spheroidal compartments formed by amphiphilic molecules self-arranging and minimizing their energy in aqueous solution. The amphiphilic molecules forming biological vesicles can be of different nature: from simpler fatty acids to more complex phospholipids, such as the ones found in all cell membranes. The existence of a membrane in vesicles is closely related to the need for a boundary to enclose specific material in a compartment. 
This property directly connects  biological vesicles to cells and underlines their importance in an origin-of-life scenario. These organic vesicles can be found in geological settings such as microbial mats and biofilms, organic-rich shales, bioherms and biostromes. 

As both biological and geological vesicles exist due to the physics of bubbles, they have the shared characteristic of a spherical or spheroidal (close to spherical) shape owing to energy minimization characteristic of a membrane.
This is an instance of  uniform contraction to a point, which is what gravity is doing to produce planets, and is what surface tension does to bubbles to produce both biological and geological vesicles.

\subsubsection{Deep-ocean polymetallic nodules}\label{sec:nodules}

\begin{figure}
    \centering
    \includegraphics[width=\linewidth]{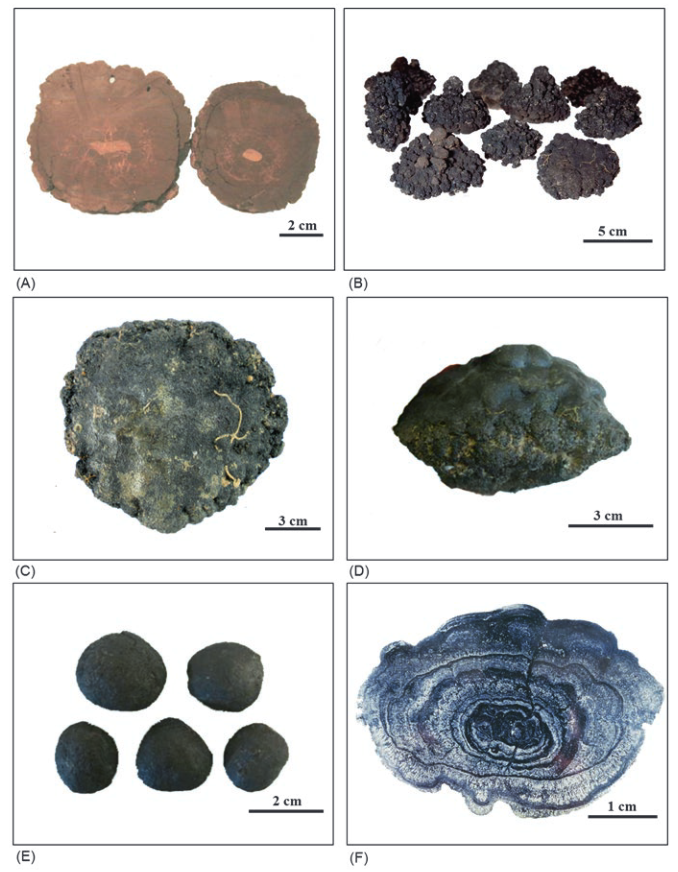}
    \caption{ Deep-ocean polymetallic nodules.
Examples of different sizes and shapes (from \citet{kuhn_composition_2017}).
}
    \label{fig:nodules1}
\end{figure}

Deep oceanic polymetallic nodules are concretions that form on or just below the sediment surface in abyssal plains of the deep oceans \citep{hein_manganese_2016, kuhn_composition_2017, hein_deep-ocean_2020}. They have sizes which are generally between 1--12~cm (Fig.~\ref{fig:nodules1}), although they can range from milli-/micrometric (micronodules) and up to 20~cm, and they can adopt different shapes (e.g., spheroidal, discoidal or botryoidal). The composition is mainly iron and manganese, and they are therefore commonly referred to in the literature as manganese nodules, ferromanganese nodules or Fe--Mn nodules. The formation of these nodules is produced by the slow precipitation of iron oxyhydroxides and manganese oxides around a nucleus, which can be, e.g., shark teeth, cetacean ear bones, rock fragments or fragments of other nodules \citep{glasby_manganese_2006, hein_1311_2014}.  Nodules are typically found in the sediment-covered abyssal plains between 3\,500 and 6\,500~m depth, in regions where the sedimentation rate is very low, less than 1~cm/ky \citep{glasby_manganese_2006, hein_1311_2014, hein_deep-ocean_2020}. In certain regions of the Pacific and Indian Oceans,  nodules can cover more than 50\% of the ocean bed. They have also been described in the Atlantic and polar oceans, although their distribution there is less well known.

\begin{figure}
    \centering
    \includegraphics[width=0.9\linewidth]{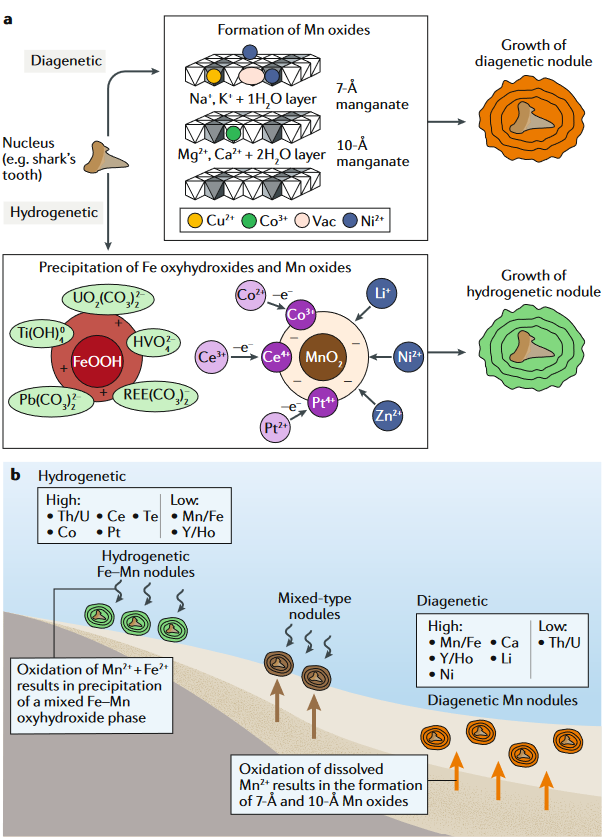}
    \caption{ Deep-ocean polymetallic nodules.
 Possible abiotic formation mechanisms,  a, and environments, b, for Fe-Mn nodules (from \citet{hein_deep-ocean_2020}).
 }
    \label{fig:nodules2}
\end{figure}

Polymetallic nodules were first encountered on the voyage of HMS Challenger between 1872--1876, during which the deep ocean was first studied \citep{challenger_expedition_1872-1876_report_1885}. Since then the nodules have been studied from different points of view: geological, geochemical, environmental and oceanographic. Polymetallic nodules also have been of interest from an economic perspective, with a view to exploit them as ore for different elements considered critical (e.g., Ni, Mn, Co, Li, and rare earth elements; \citet{hein_deep-ocean_2020}). The sequestration of these critical metals during the growth of  polymetallic nodules is due to their high porosity (mean 60\%), their high specific surface area (approx.\ 150~m\textsuperscript{2}/g), the layered and tunnelled structures of the manganese phases and the different surface charges of the precipitated phases, negative for the Mn phases and positive for the Fe phases; \citet{hein_deep-ocean_2020, hein_1311_2014}. 

Recently, it has been proposed that these polymetallic nodules can produce so-called \emph{dark oxygen} on the abyssal seafloor, i.e., where there is only oxygen consumption by abyssal organisms and oxidation of reduced inorganic compounds produced by anaerobic decomposition \citep{sweetman_evidence_2024}. Such oxygen production  is proposed to be due to  high voltage potentials at the nodule surfaces, up to 0.95~V, which could lead to electrolysis of  water, increasing the seawater oxygen content \citep{sweetman_evidence_2024}. However, these results are currently under discussion \citep{downes_contributions_2024}. Nevertheless, although these nodules have been investigated in detail, their origin is still debated, and it is not clear whether it is abiotic or biotic \citep{hein_1311_2014, kuhn_composition_2017}.

The abiotic origin theory provides three possible scenarios for the nodules \citep{bau_discriminating_2014, hein_1311_2014, hein_manganese_2016, kuhn_composition_2017, hein_deep-ocean_2020}: hydrogenetic, diagenetic and a mixed model (Fig.~\ref{fig:nodules2}). In the case of a hydrogenetic model (Fig.~\ref{fig:nodules2}), formation would occur by oxidation of Mn\textsuperscript{2+} and Fe\textsuperscript{2+}, which would form colloids of Mn\textsuperscript{4+} and Fe\textsuperscript{3+} oxides precipitating around a nucleus. This growth process is very slow; a few millimetres every million years) and the sedimentation rate must be almost negligible. In the diagenetic model (Fig.~\ref{fig:nodules2}), formation occurs around a nucleus in the pore space of deep ocean sediments. In these sediments, oxidation of organic matter dissolves Mn oxides and reduces them to Mn\textsuperscript{2+}, along with other associated metals. These metals diffuse upwards through the sediment until they reach oceanic water with a higher oxygen concentration, causing the Mn to re-oxidize and precipitate in the form of various oxides. In this model,  growth is faster, reaching several tens of millimetres every million years. In addition, the sedimentation rate may also be higher than in the hydrogenetic model, although it is always less than 1 cm/ky.  It is important to note that these nodules are found on the sedimentary surface, although their growth is lower than sedimentation rates. Several theories have been proposed to explain this paradox, e.g., that they are raised due to biological or seismic activity, but this aspect remains unresolved \citep{glasby_manganese_2006, hein_deep-ocean_2020}.

\begin{figure}
    \centering
    \includegraphics[width=\linewidth]{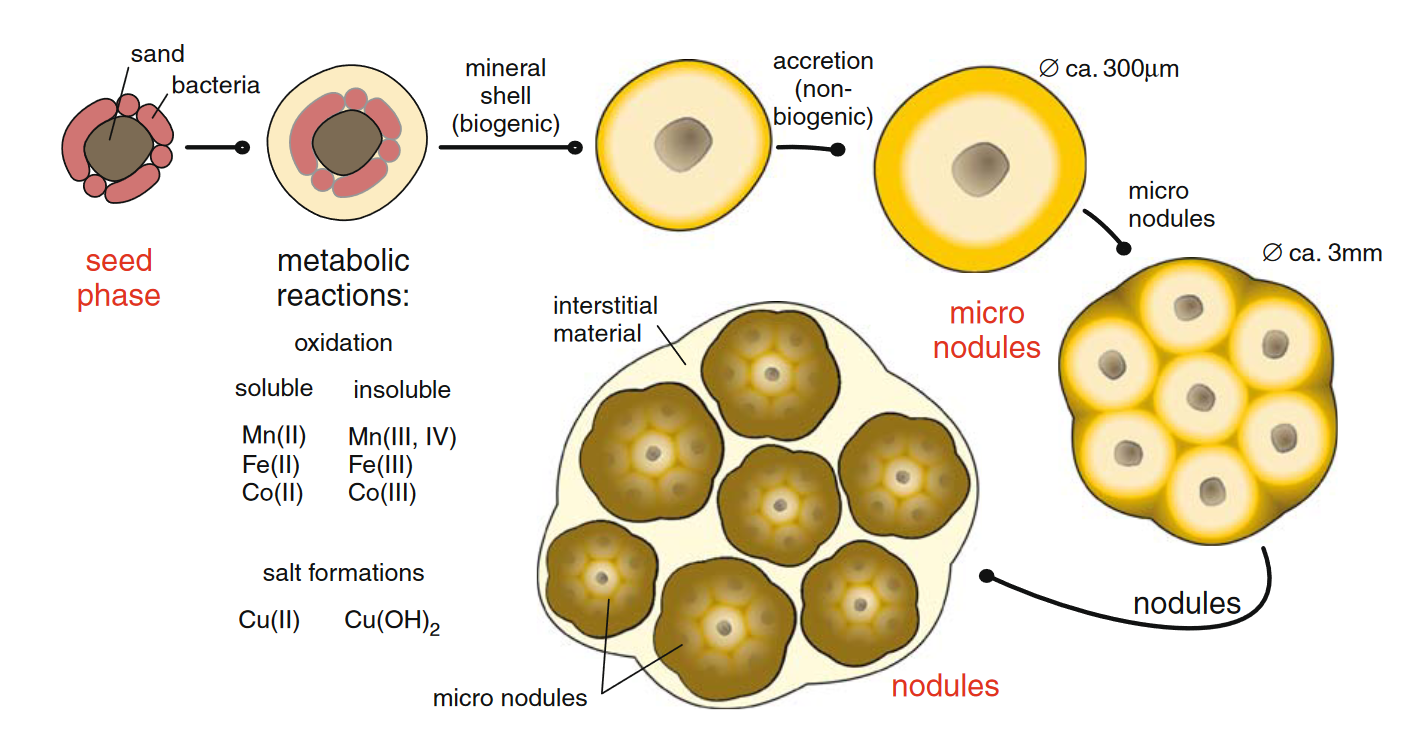}
    \caption{ Deep-ocean polymetallic nodules.
Formation of hydrogenetic nodules mediated by microorganisms (from \citet{wang_biogenic_2009}).}
    \label{fig:nodules3}
\end{figure}

Researchers supporting a biological origin of these nodules base their hypothesis on  communities of organisms found within the nodules; i.e., bacteria and fungi \citep{kuhn_composition_2017, zheng_nitrate-driven_2025}. In these nodules, communities of bacteria that would reduce Mn\textsuperscript{4+} and bacteria that would oxidize Mn\textsuperscript{2+} have been described \citep{ehrlich_ocean_2000, blothe_manganese-cycling_2015}. The oxidation of Mn\textsuperscript{2+} is what would serve as a source of energy for the bacteria living in the nodules \citep{kuhn_composition_2017}. In addition, laboratory experiments have been conducted to try to produce manganese oxides through biological activity by bacteria and fungi. The mineral phases obtained in these experiments are similar to those detected in natural nodules, although the growth rates of these phases induced by bacterial activity are several orders of magnitude higher than those determined for the formation of natural nodules \citep{kuhn_composition_2017}. The study of the arrangement of microorganisms within the nodules has also allowed some authors to propose a biological origin \citep{wang_biogenic_2009, wang_marine_2009, wang_organized_2009}. These authors have found that the microorganisms are located only in the micronodules and not in the interstitial zones between them. They have therefore proposed that the formation of Fe-Mn micronodules is induced by bacterial activity. These micronodules continue to grow by  (non-biogenic) accretion and eventually assemble with other micronodules to form  centimetric nodules (Fig.~\ref{fig:nodules3}; \citet{wang_biogenic_2009, wang_marine_2009}).

The biological or abiogenic origin of these nodules is still under discussion. While deep-ocean polymetallic nodules do not display biological morphologies like some other objects described in this review, the debate around their biogenicity is whether or not microbes play a role in the metal oxidation processes and precipitation of the minerals that compose them. To resolve this question, further analyses of the nodules with modern high-resolution techniques are required \citep{hein_1311_2014}.

\subsubsection{Cave pearls}\label{sec:cavepearls}

\begin{figure}[tbp]
\begin{center}
\includegraphics[height=0.47\columnwidth]{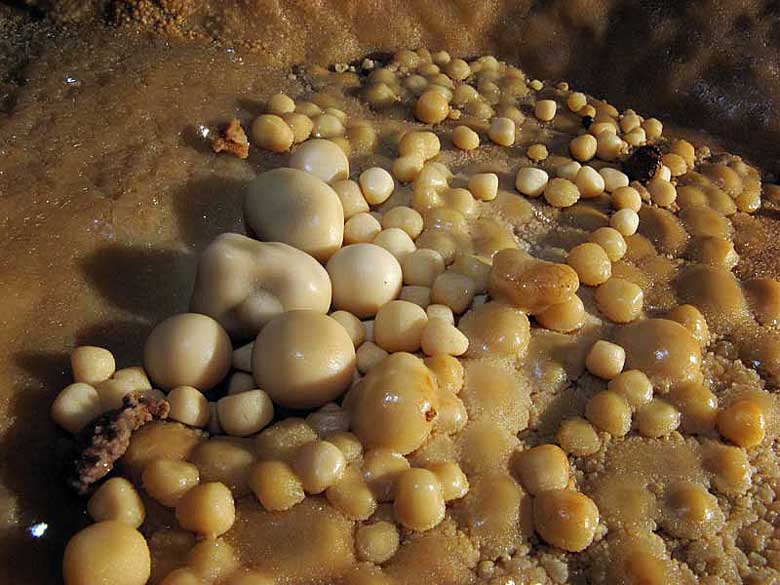}
\includegraphics[height=0.47\columnwidth]{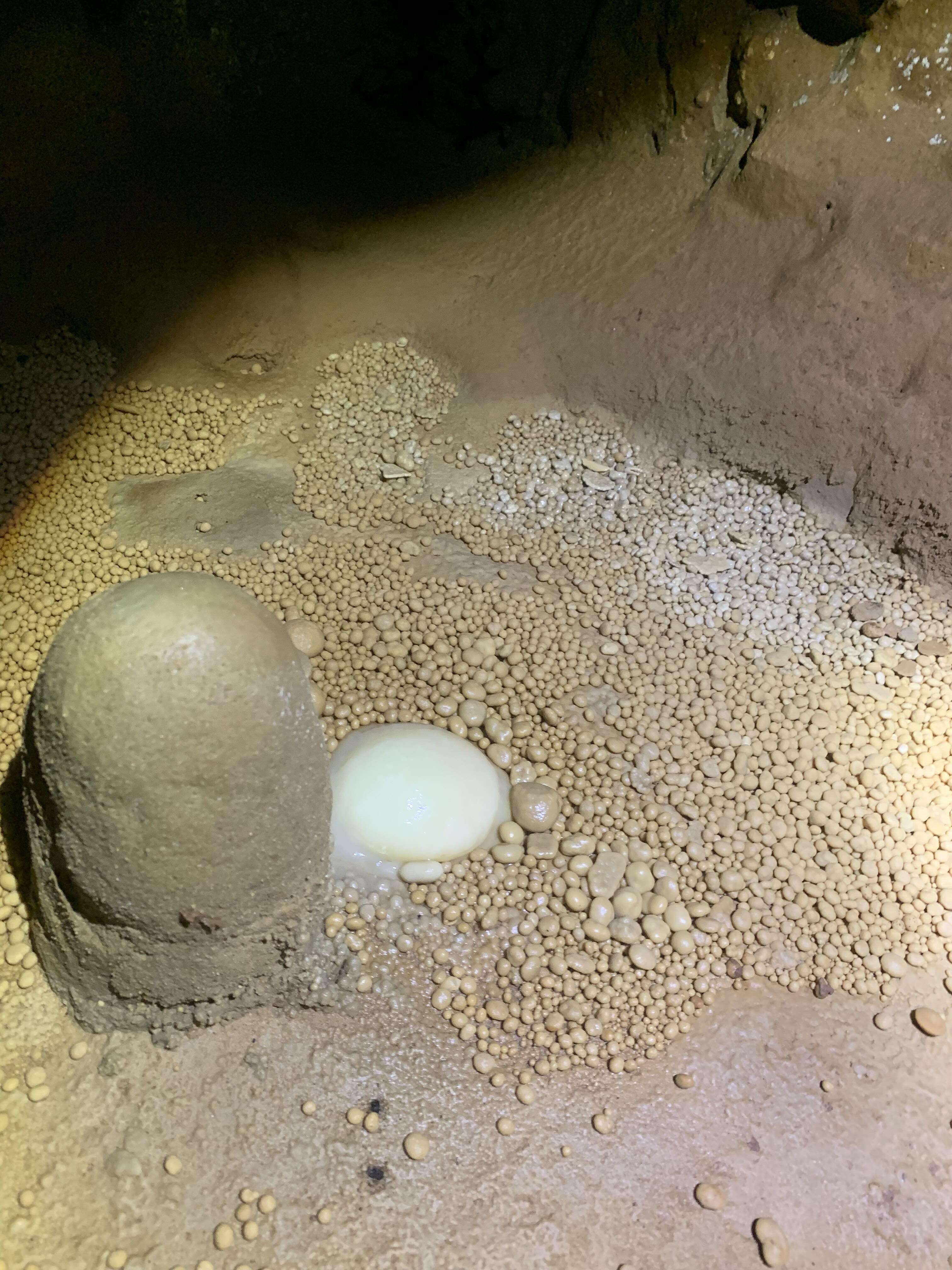}
\end{center}
\caption{Cave pearls in 
Diamond Caverns in Park City, Kentucky; Photo courtesy of Gary Berdeaux,  CC-BY-SA-4.0.
Cave pearls next to a stalagmite; 
Bcerrato1, 
CC-BY-SA-4.0.
}
\label{fig:cave_pearls}
\end{figure}

Cave pearls, Fig.~\ref{fig:cave_pearls}, are coated carbonate grains that grow in pools in caves where dripwater or small inflows keep the grains in intermittent motion. Classical descriptions emphasized their morphological similarity to ooids and pisoids and typically invoked purely physicochemical precipitation in agitated pools, with rolling and collisions selecting for nearly spherical shapes \citep{BakerFrostick1951,MackinCoombs1945,Thrailkill1963,HillForti1997}. In this abiogenic end‐member picture, the key control parameters are the degree of supersaturation, hydrodynamic energy, and the availability of free nuclei that are not cemented to the substrate.

More recent work has shown that many cave pearls may in fact be hybrid structures in which microbial activity and physicochemical self‐assembly are tightly coupled. Detailed petrography and microbiology studies document biofilms on pearl surfaces, calcified microbial filaments and cells, and fabrics that are best explained by microbially induced calcite precipitation rather than purely inorganic crystal growth \citep{Gradzinski2001,NorthupLavoie2001,Jones2010}. In some systems, bacteria within a pearl and its surface biofilm act as the main source of supersaturation, for example through chemolithoautotrophic consumption of $\text{CO}_2$, while the hydrodynamic environment still selects which grains are kept in motion and thus can grow as pearls \citep{Gradzinski2001}.

A good illustration of this ``integrated'' view is provided by the Grand Cayman pearls described by \citet{Jones2009}, who argued that cave pearls should be regarded as the combined product of abiogenic and biogenic processes. Similar conclusions emerge from Kanaan Cave in Lebanon, where concentric internal fabrics, stable isotopes and trace elements point to fluctuating growth conditions, with both fluid chemistry and microbial activity recorded in the lamination \citep{Nader2007}. In an underground limestone mine in Illinois and in mine pools studied more recently, cave pearls have been observed to grow and recrystallize on human time scales, with rapid textural overprinting and strong spatial variability in growth styles \citep{MelimSpilde2011,MelimSpilde2018,MelimSpilde2021}. These systems demonstrate that cave pearls are highly dynamic, far‐from‐equilibrium objects, in which self‐organized rolling and impact dynamics, chemical forcing, and microbial processes all contribute.

From the perspective of this review, cave pearls therefore occupy an intermediate position on the continuum between purely abiogenic self‐assembled patterns and strongly biologically controlled biominerals. Their geometry and internal fabrics encode both hydrodynamic constraints (agitation, pool morphology, residence times) and the presence or absence of microbial \emph{helpers} capable of sustaining local supersaturation and stabilizing specific crystal fabrics \citep{Jones2010,NorthupLavoie2001}. As such, they provide a useful natural experiment for comparing non‐biological and biologically influenced pattern formation within a single, well‐constrained system.

\subsubsection{Carbonate spherulites}\label{sec:spherulites}

\begin{figure}[tbp]
\begin{center}
\includegraphics[width=\columnwidth]{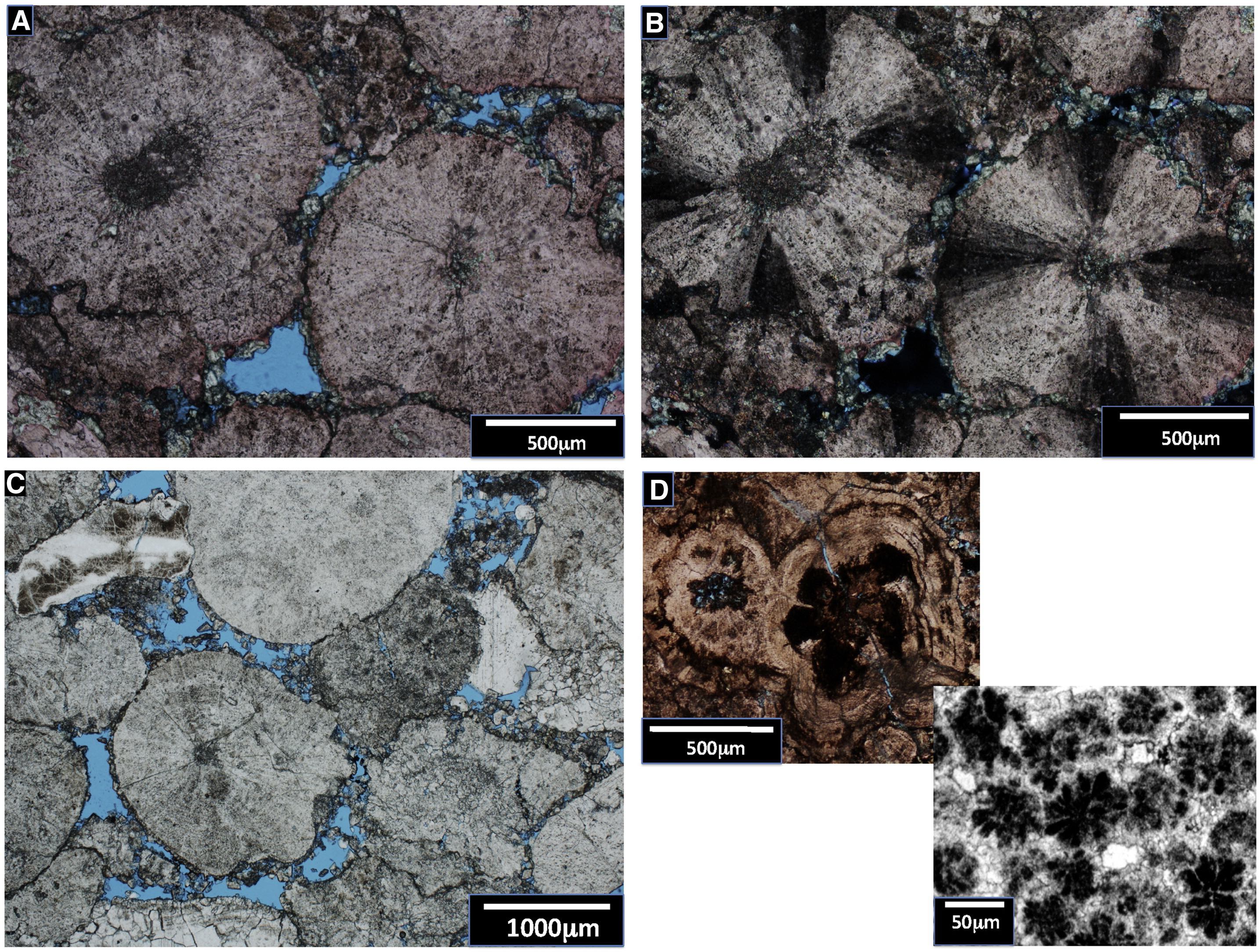}
\end{center}
\caption{Carbonate spherulites.
Photomicrographs of Pre-Salt spherulites (A–C and upper left of D) and Mammoth Hot Spring, Yellowstone National Park (lower right of D). (A) Views in plain light and (B) in crossed-nicols of the same field of view and show the very well-developed radical cortical structure around micritic and partly dissolved nuclei common in the Pre-Salt spherulites. Observe the paucity of sediment between the spherulites, the compaction, and associated dissolution due to pressure solution of these allochems. (C) The spherulites have a common micritic core as well as an abundance of open pore space between spherulites, spherulite-to-spherulite compaction, and a minor amount of diagenetic dolomite cement. (D) The Pre-Salt spherulite in the upper left centre displays a clover-leaf micritic core, similar to the clover-leaf accumulations of bacterial bodies comprising the peloidal layers in hot spring travertine deposits from Mammoth Hot Springs at Yellowstone displayed in the photomicrograph on the lower right \citep{chafetz2018origins}.
}
\label{fig:spherulite}
\end{figure}

Geological spherulites are small, rounded masses of radiating needle-like crystals (Fig.~\ref{fig:spherulite}). 
Spherulites are often found in vitreous igneous rocks, such as obsidian \citep{gardner2012compositional}, pitchstone \citep{kshirsagar2012spherulites}, and rhyolite \citep{hanson2020word}. They typically form through a process known as devitrification, in which glassy, non-crystalline substances restructure into crystalline forms. The formation of spherulites involves the rapid nucleation of high-temperature crystallization domains within glassy, silica-rich lavas.
This rapid nucleation results in the characteristic radiating, needle-like crystal structure. Such spherulites would seem to be entirely self assembled.
However, spherulites are also found in carbonate rocks \citep{kirkham2018thrombolites}. Here there is a debate over their abiotic or biological origin. There is evidence in favour of spherulite nuclei being created by microbial metabolism \citep{verrecchia1995spherulites}. Subsequent growth may be inorganic or could be controlled by microbial extracellular polymeric substances (EPS) which occur as relatively abundant inclusions within the spherulite crystals. This would suggest that carbonate spherulites may be at least partly microbially / biotically influenced \citep{kirkham2018thrombolites,chafetz2018origins}.

\begin{figure}[tbp]
\begin{center}
    \begin{subfigure}{0.47\textwidth}
    \includegraphics[width=\linewidth]{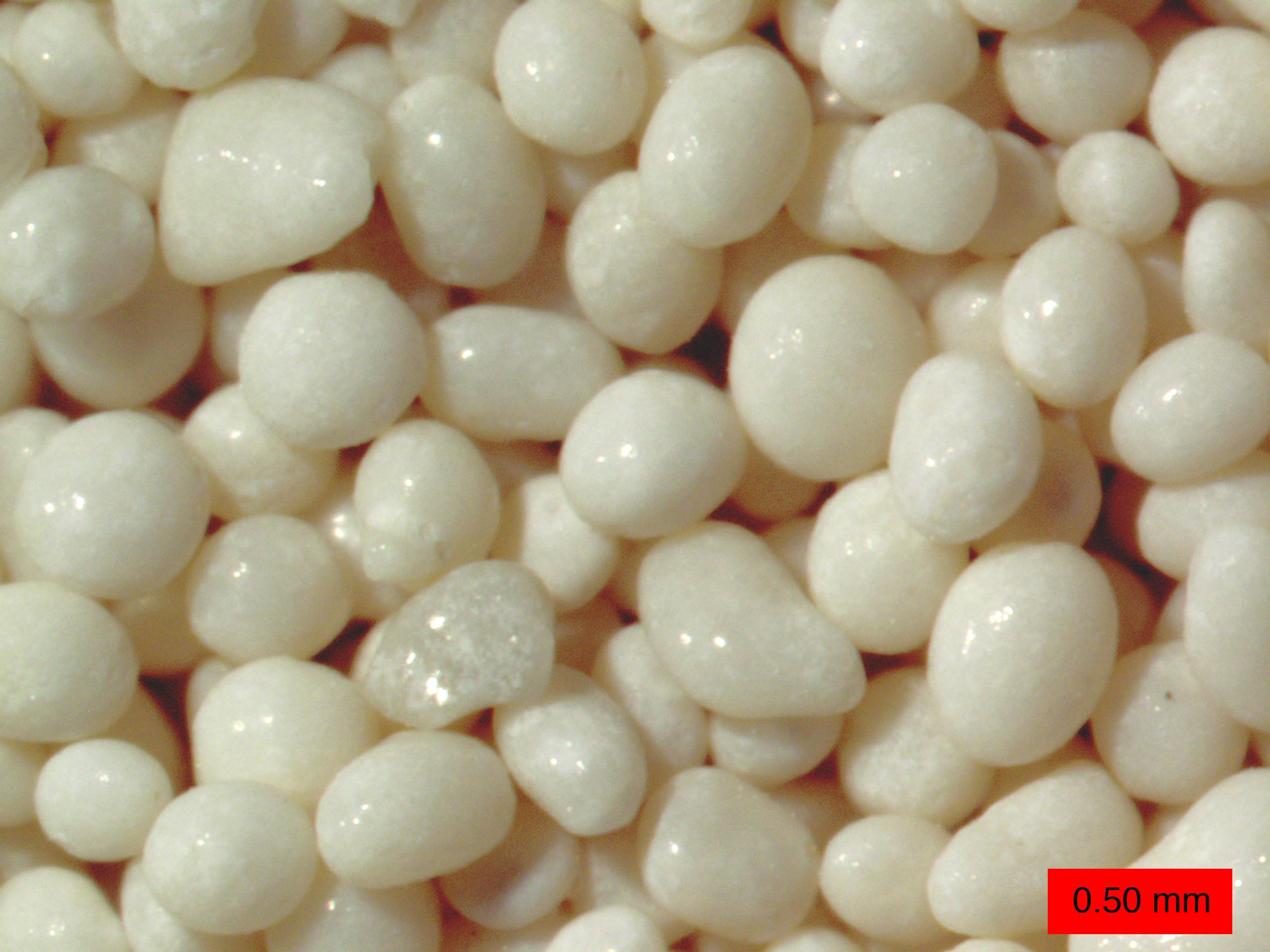}
    \caption{}
    \end{subfigure}
    \begin{subfigure}{0.51\textwidth}
    \includegraphics[width=\linewidth]{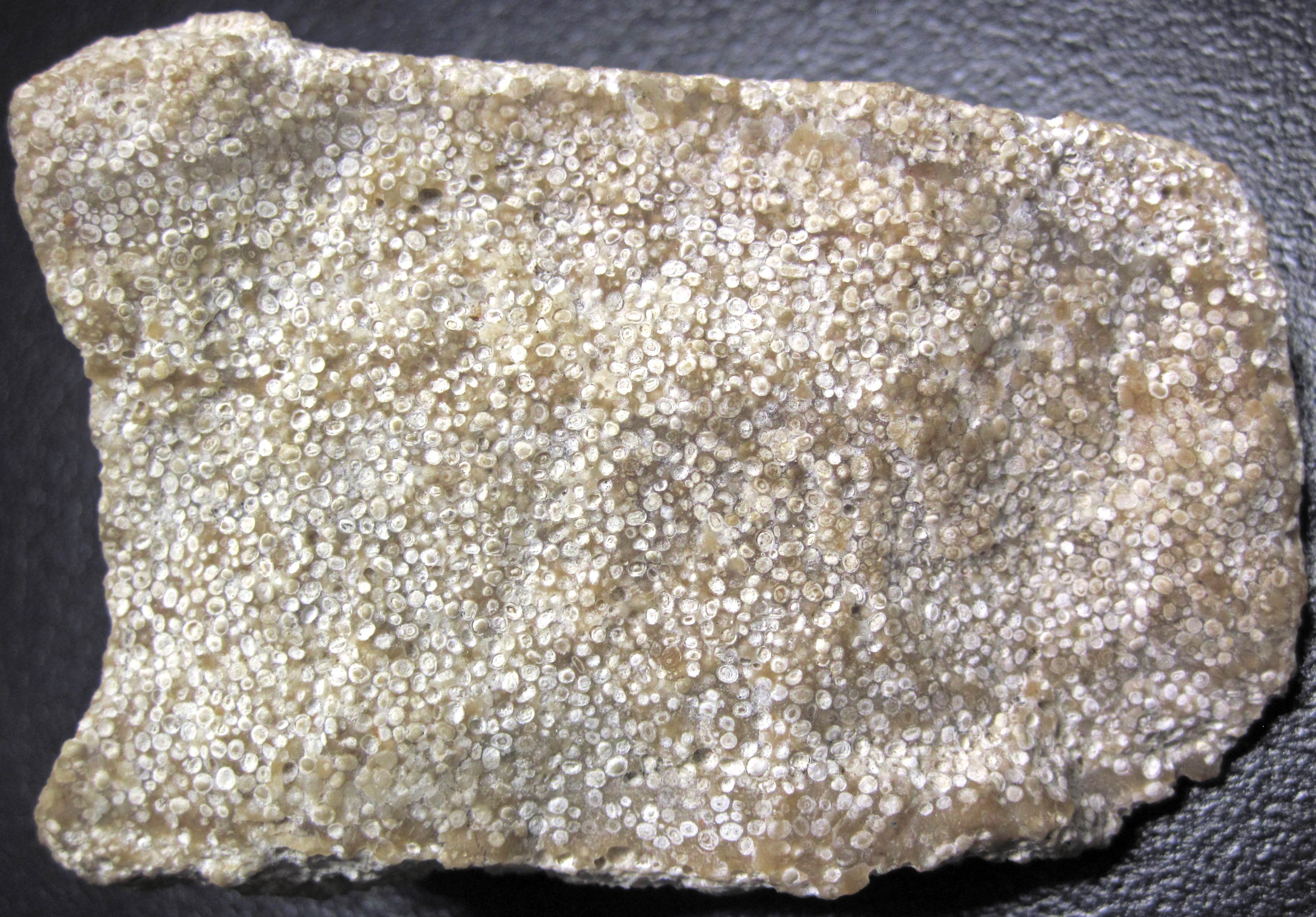}
    \caption{}
    \end{subfigure}
    \begin{subfigure}{0.99\textwidth}
    \includegraphics[width=\linewidth]{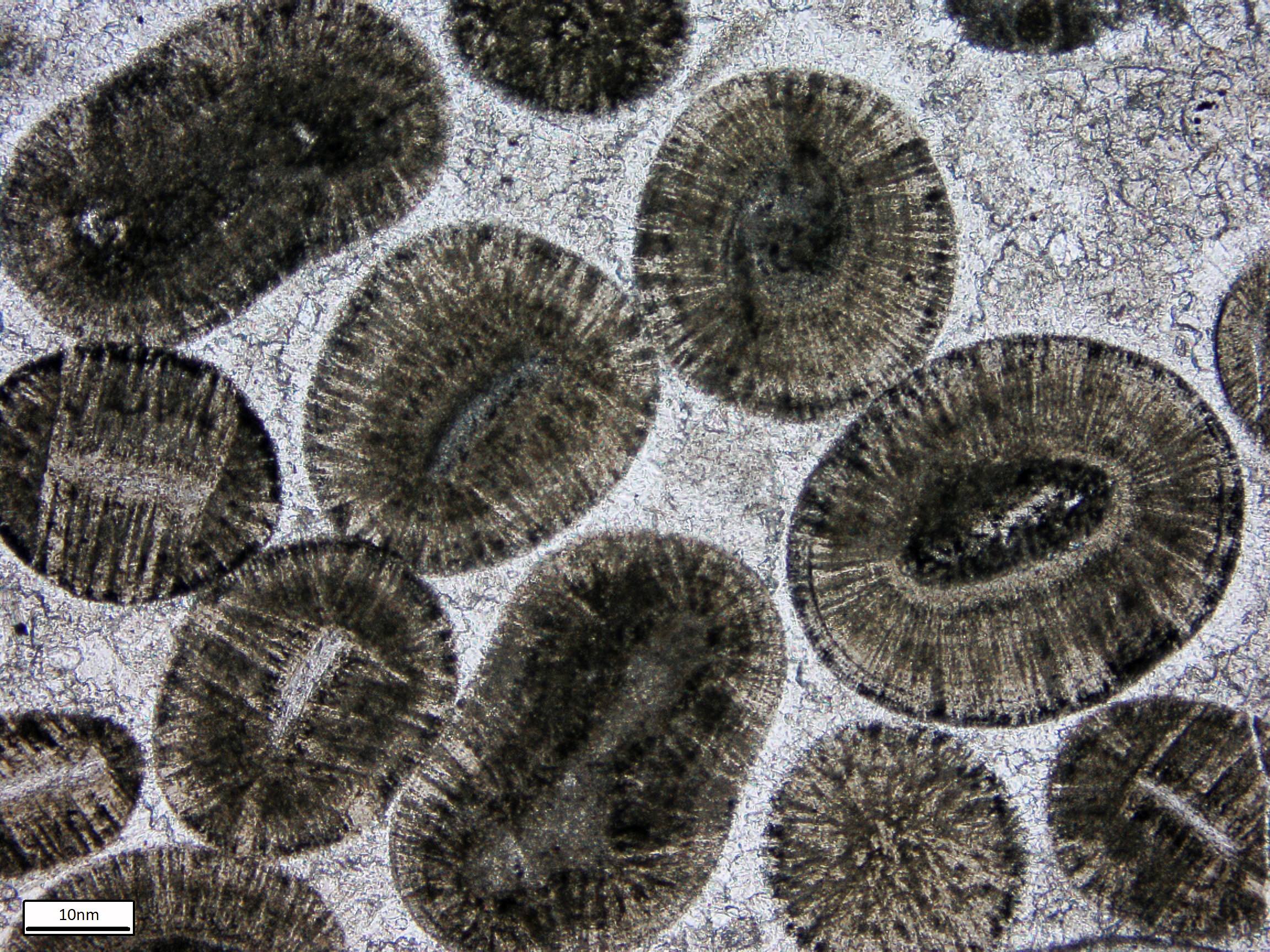}
    \caption{}
    \end{subfigure}
\end{center}
\caption{Ooids:
(a) Modern ooids from a beach on the Joulter Cays, The Bahamas (image:  Mark A. Wilson, public domain).
(b) Oolitic limestone from the Mississippian of Indiana, USA (image: James St. John, CC-BY-SA-4.0).
(c) Ooids in the Silurian Mifflintown Formation, PA (image:	Michael C. Rygel, CC-BY-SA-4.0).
}
\label{fig:ooid}
\end{figure}

\subsubsection{Ooids}\label{sec:ooids}

Ooids are small, spherical to ellipsoidal grains, typically of calcium carbonate, either calcite or aragonite, but sometimes of other minerals, which form in shallow marine environments (Fig.~\ref{fig:ooid}). They are characterized by concentric layers that build up around a nucleus, often a sand grain or a small fragment of a mollusc shell. When cemented together, they form oolite.
In ooids \citep{simone1980ooids} the question is not so much that there is pattern or structure that looks biological, but rather whether or not mineral precipitation results from microbial activity. 
Despite decades of research, their formation process is unresolved, with both abiotic \citep{lin2022holocene} and biological mechanisms \citep{Pei2024} being hypothesized.
In this sense, ooids represent an interesting counterexample to other systems we discuss here that are considered to `look' much more biological, but may in fact have an abiotic origin.

\subsubsection{Automicrite}

Automicrite is carbonate mud that precipitates \emph{in situ}, i.e., it forms essentially where it is found rather than being transported as detritus. Texturally it consists of microcrystalline calcite or aragonite and it acts as the mud‐grade matrix in many limestones. In contrast, \emph{allomicrite} denotes lime mud that has been transported before deposition~\citep{TuckerWright1990}. The distinction is important for our purposes, because automicrite is a key component of the \emph{mud-mound} carbonate factory, where fine carbonate is produced directly on the seafloor by a mixture of abiotic precipitation and microbially induced reactions~\citep{Schlager2003,Reijmer2021}.

Modern and ancient examples show that automicrite commonly forms at or just below the sediment–water interface, where small changes in pH and alkalinity, driven by organic-matter degradation and redox reactions, push pore waters into carbonate supersaturation~\citep{Munnecke2023}. In low-energy settings this leads to pervasive micrite coatings on grains, pore‐filling cements and microcrystalline mudrocks. In mud-mound and microbialite systems, automicrite is often intimately associated with microbial mats or biofilms: extracellular polymeric substances and decaying organic tissue locally modify carbonate chemistry and nucleation barriers, so that precipitation is biotically induced or biotically influenced rather than purely abiotic~\citep{Schlager2003,Reijmer2021}. The resulting fabrics include stromatolitic lamination and clotted or peloidal micrite (thrombolites), which at the outcrop scale appear as banded or nodular structures.

High-resolution petrography from submarine cave crusts provides a clear example of such microbially influenced automicrite. In Aegean marine caves, biogenic crusts built by sponges, corals, bryozoans and serpulids contain two micrite fractions: a detrital micrite derived from external mud and an autochthonous micrite  precipitated directly within microcavities of the framework~\citep{Guido2019}. The autochthonous micrite forms thin peloidal infills and cements associated with microbial biofilms and sulfate-reducing bacteria, which act as local carbonate pumps. Similar organomineralized automicrite has been described from other cryptic and low-energy marine environments and demonstrates that finely crystalline mud can originate from strongly microbially mediated processes while remaining texturally indistinguishable from purely abiotic seafloor cements.

On larger scales, automicrite participates in self-organized banding and layering through early diagenetic redistribution of carbonate mud. The dissolution of metastable aragonite and high-Mg calcite and the reprecipitation of low-Mg calcite within the shallow burial zone can concentrate micrite into discrete beds or couplets (limestone–marl alternations), hardgrounds and secondary carbonate mudrocks \citep{Munnecke2023}. These diagenetic feedbacks between mineral instability, microbial metabolism and permeability generate centimetre- to metre-scale patterns that mimic primary depositional banding. At the same time, recent work on nonskeletal aragonite mud production in tropical platforms suggests that a substantial fraction of carbonate mud can form by direct precipitation from seawater (whitings), with only weak biological influence,  emphasizing the continuum from abiotic to biotically induced automicrite~\citep{Geyman2022}.

Micritic limestones and mud-mounds therefore occupy an ambiguous middle ground. Their fine-grained fabrics are easy to recrystallize and overprint, and the same micrite-rich banding or nodular geometries can arise from (i) largely abiotic precipitation in supersaturated waters, (ii) organomineralization in microbial habitats, or (iii) early diagenetic reworking of originally skeletal sediments. Distinguishing between these endmembers requires combining microtextural criteria (wrinkled lamination, peloidal clots, calcified filaments, geopetal infills) with geochemical and stratigraphic context, rather than relying on the presence of micrite alone as a biosignature.

\subsection{Banded patterns}

\begin{figure}[tb]
\begin{center}
\includegraphics[height=0.295\columnwidth]{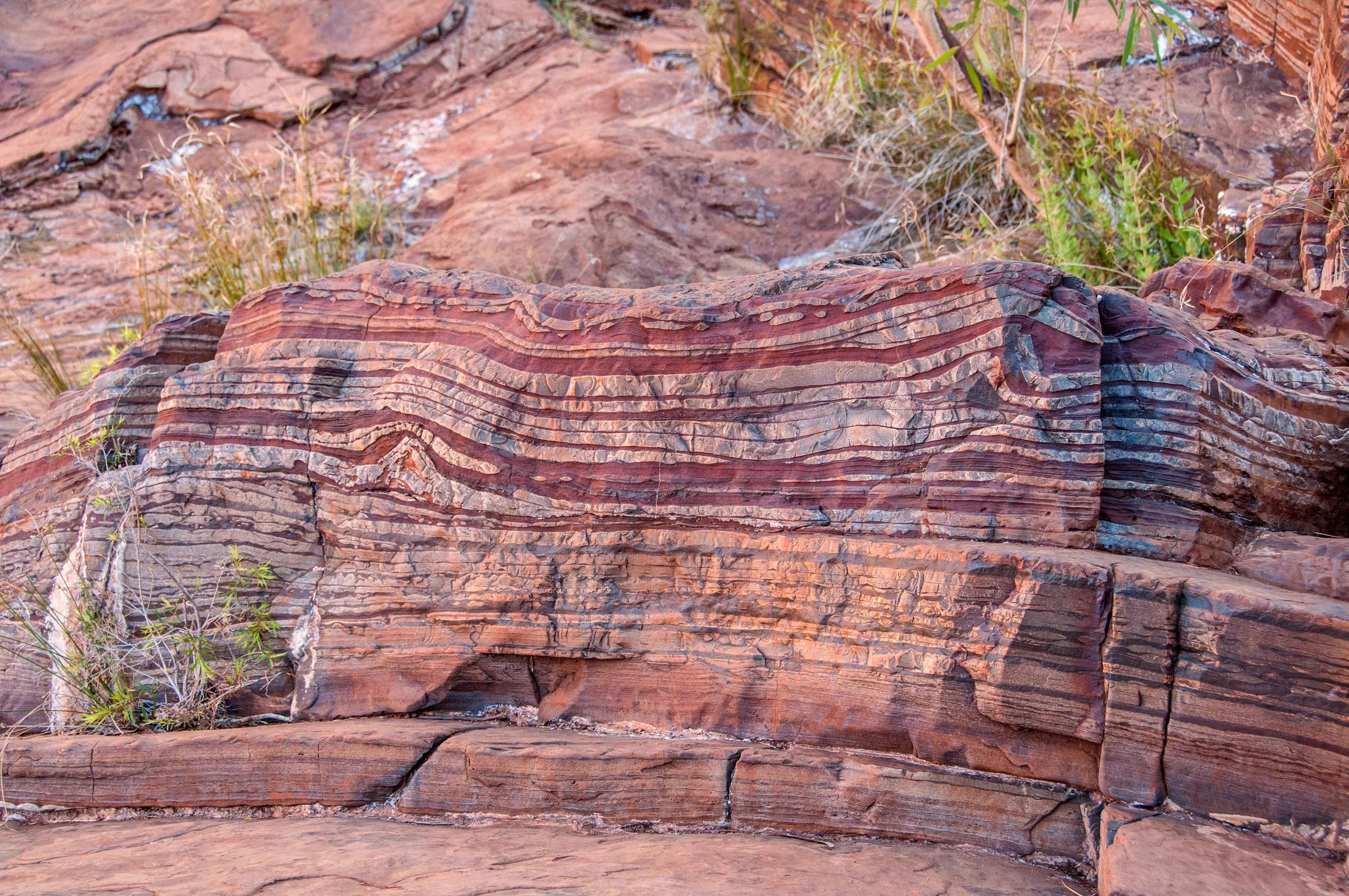}
\includegraphics[height=0.295\columnwidth]{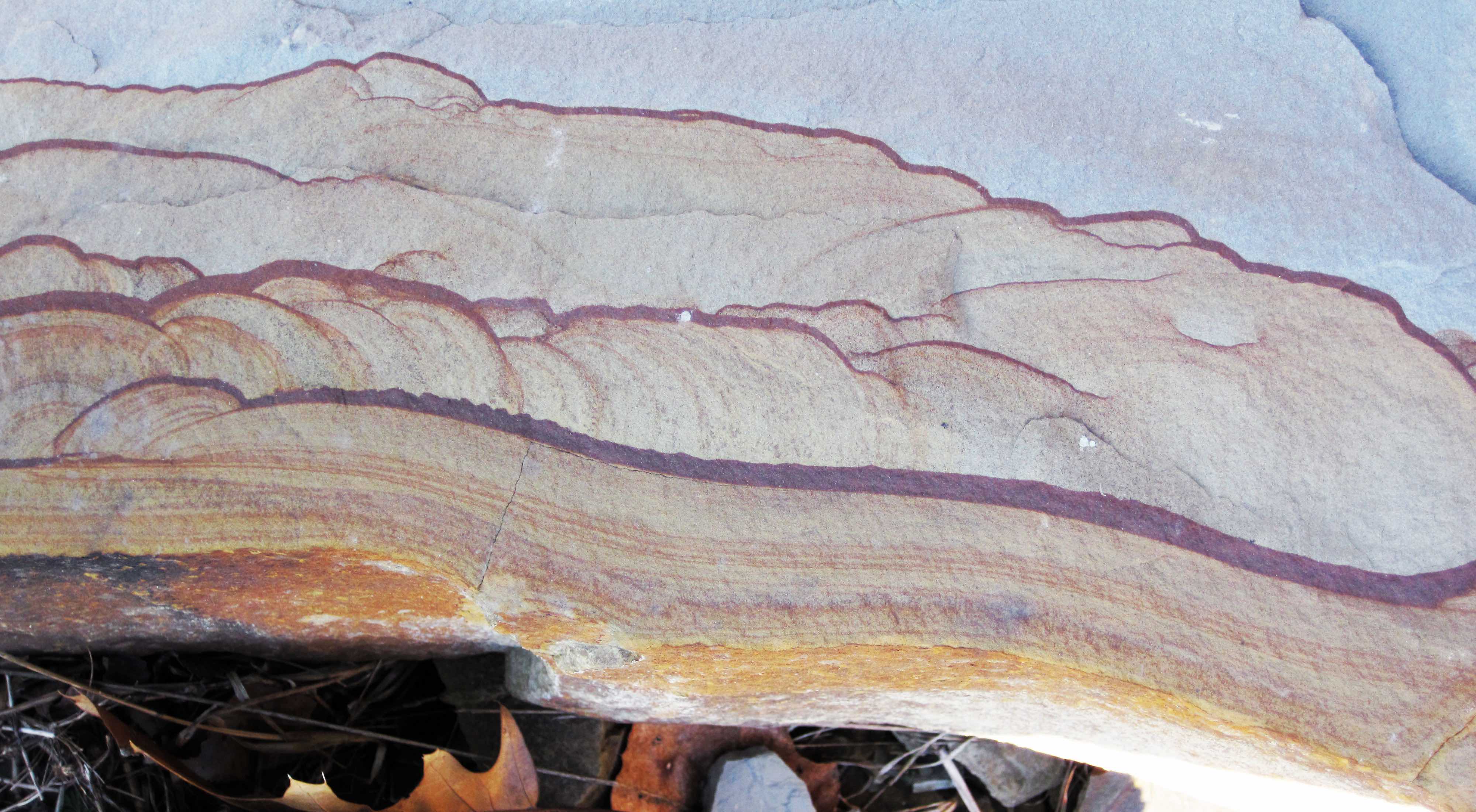}
\includegraphics[height=0.32\textwidth]{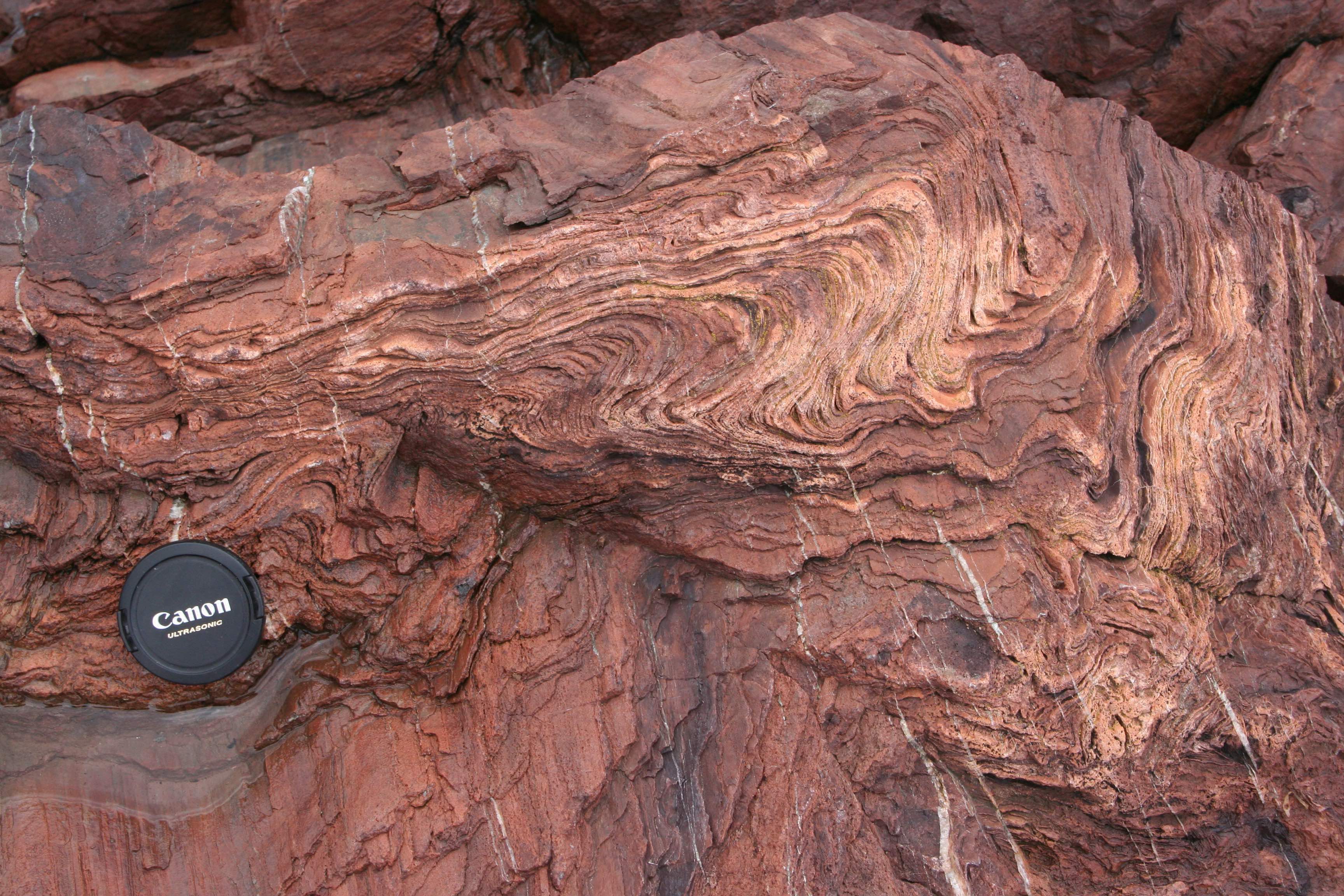}
\includegraphics[height=0.32\textwidth]{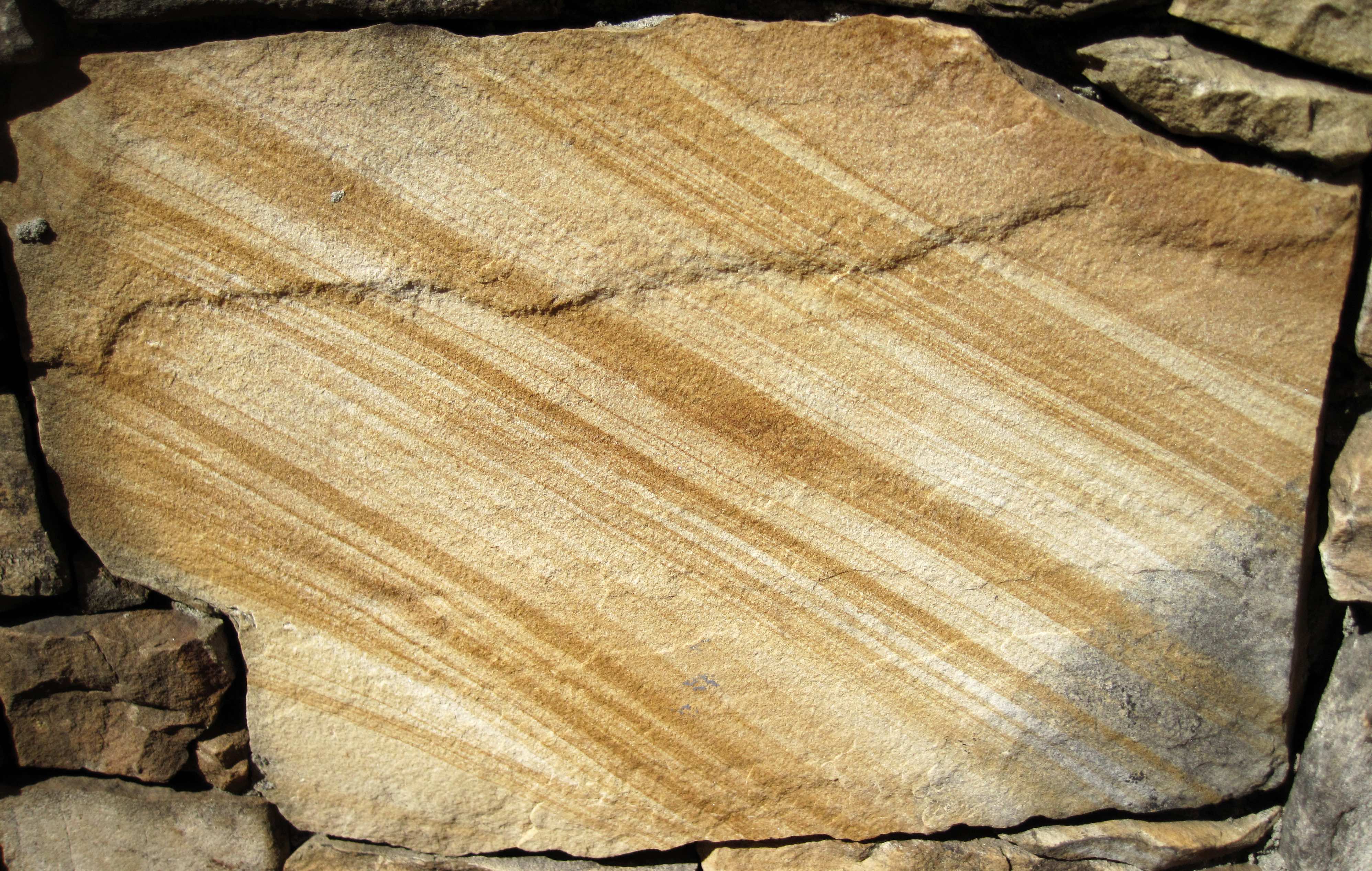}
\end{center}
\caption{Examples of banded structures in rocks. 
A) Banded Iron Formation at the Fortescue Falls, Western Australia (image: Graeme Churchard, CC BY 2.0).
B) Quartzose sandstone in the Mississippian of Ohio, USA (image: James St. John CC BY 2.0).
C) Flow banded rhyolite in the Dunn Point Formation (Ordovician) exposed near Arisaig, Nova Scotia (image: Michael C. Rygel, CC-BY-SA-3.0).
D) Sandstone from the Upper Paleozoic of Tennessee, USA (image: James St. John CC BY 2.0).
}
\label{fig:banding}
\end{figure}

Banded structures in rocks (Fig.~\ref{fig:banding}) are quasi-periodic layers of contrasting mineral composition, texture, porosity, or colour. They range from $\mu$m-scale oscillatory zoning in single crystals, through mm–cm bands in concretions, travertines and stromatolites, to metre-scale bands in classic banded iron formations (BIFs). A central question is whether such bands are intrinsic products of internally driven self-organization in reactive media; primarily extrinsic records of externally imposed environmental cycles; or some combination of both.

From the non-biological side, the theory of geochemical self-organization predicts that coupled reaction–transport systems can spontaneously form spatially periodic bands once feedbacks between dissolution/precipitation, porosity, and transport cross an instability threshold \citep{cartwright_self-organized}. This framework, developed in detail by Ortoleva and co-workers, unifies phenomena such as Liesegang bands, oscillatory crystal zoning and rhythmic dissolution--precipitation fronts in sedimentary basins \citep{R9}. Laboratory precipitation experiments further demonstrate that purely abiotic reaction--diffusion--advection systems can robustly generate banded and lamellar patterns in narrow geometries, even when external boundary conditions are steady \citep{Nakouzi2016PrecipitationPatterns}. On geological scales,  small-scale banding in BIFs is now widely interpreted as a manifestation of such internal geochemical dynamics: thermodynamic and kinetic models show that mixing of Fe–Si–rich hydrothermal fluids with seawater can undergo oscillatory precipitation of Fe- and Si-rich layers through nonlinear speciation feedbacks, without requiring an externally periodic driver \citep{Wang2009BIFInternalDynamics}. Similarly, self-organized iron-oxide cementation geometries and banding in red beds and sandstones---including nodules, rinds and scalloped fronts---can be reproduced by models of reactive infiltration and are observed in nature as diagnostic records of paleo-groundwater flow paths \citep{Wang2015IronOxideCementation}.

In contrast, many carbonate banded structures clearly encode extrinsic environmental forcing. Fluvial and lacustrine stromatolites, for example, exhibit alternations of micritic, peloidal and sparry laminae whose thickness and texture correlate with monitored seasonal changes in temperature, discharge and biotic productivity. Detailed work on recent fluvial stromatolites from the River Piedra (Spain) shows that micro-laminae couple systematically to hydrological and climatic variability at seasonal to sub-seasonal scales, allowing reconstruction of palaeo-flow regimes and climate from older stromatolitic successions \citep{Arenas2017StromLamination}. Travertine and tufa systems provide another canonical example: laminated travertines in hot-spring and riverine settings often display couplets that reflect daily or seasonal alternations between dense crystalline layers and porous, organic-rich layers, tied to day--night photosynthetic cycles, discharge fluctuations and temperature changes \citep{Okumura2012Travertine,Fouke2011Mammoth,Pentecost2005Travertine}. In these settings, band thickness and internal texture can be calibrated against directly measured environmental time series, making them high-resolution geologic \emph{tape recorders} of external forcing.

Rock varnish sits at an interesting intermediate point in this classification. This millimetre-thick Mn–Fe–rich coating on desert rocks is often composed of micron-scale micro-laminations (VMLs) that can be read as a high-resolution palaeo-environmental archive. Early work established that these micro-laminations are chemically distinct, reproducible over large regions, and potentially sensitive to changes in effective moisture or dust flux \citep{Dorn1984VarnishLaminations,Broecker2001RockVarnish}. More recent microbiological and geochemical studies have documented diverse microbial communities inhabiting varnish and demonstrated that Mn and Fe cycling in the coating is strongly microbially mediated \citep{Kuhlman2008}. As a result, the banding is best viewed as an emergent product of strong, nonlinear couplings between microbial metabolisms, redox chemistry, dust deposition and rare wetting events; i.e., a hybrid of intrinsic biogeochemical self-organization and extrinsic climatic pacing.

These examples illustrate that banded geological structures occupy a continuum between purely autogenic, internally generated patterns and purely allogenic, externally forced stratification. In some systems, e.g., centimetre-scale BIF banding, the spacing and regularity are dominated by internal feedbacks in reaction--transport space \citep{R9,Wang2009BIFInternalDynamics}. In others---e.g., certain stromatolites and travertines---the bands predominantly record periodicities in hydrology, temperature or light, but their expression is filtered and reshaped by microbial mats and precipitation kinetics \citep{Arenas2017StromLamination,Okumura2012Travertine}. Rock varnish exemplifies a case where intrinsically generated biogeochemical micro-laminations are paced and modulated by episodic environmental events \citep{Broecker2001RockVarnish,Kuhlman2008}. Recognizing where a given banded structure lies on this spectrum is essential before using it as a quantitative palaeo-environmental archive or as evidence for biological versus abiotic pattern formation.

\subsubsection{Stromatolites and stromatolite-like structures}\label{sec:stromatolites}

\begin{figure}[tbp]
\begin{center}
    \begin{subfigure}{0.47\textwidth}
    \includegraphics[width=\linewidth]{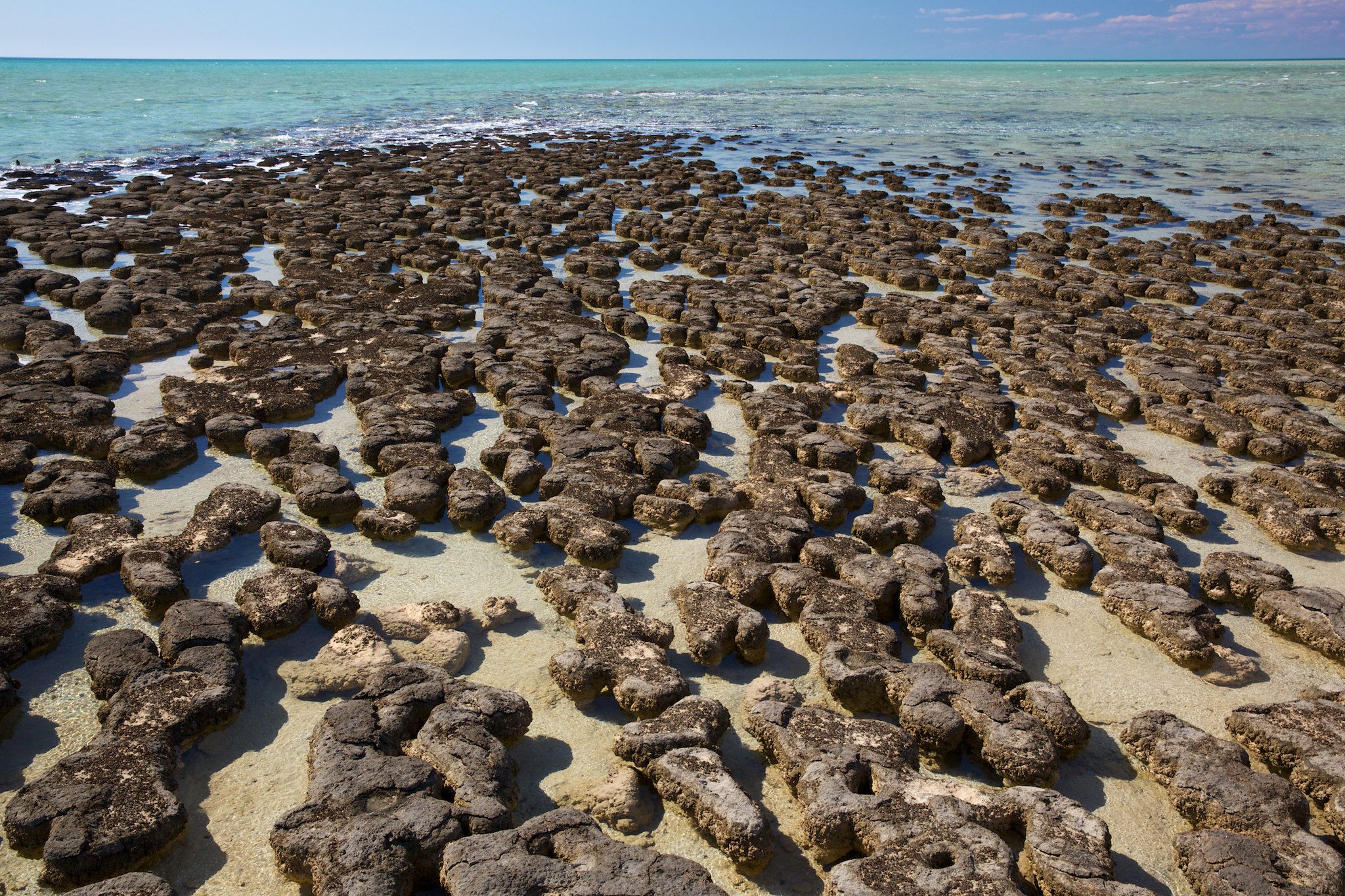}
    \caption{} 
    \end{subfigure}
    \begin{subfigure}{0.45\textwidth}
    \includegraphics[width=\linewidth]{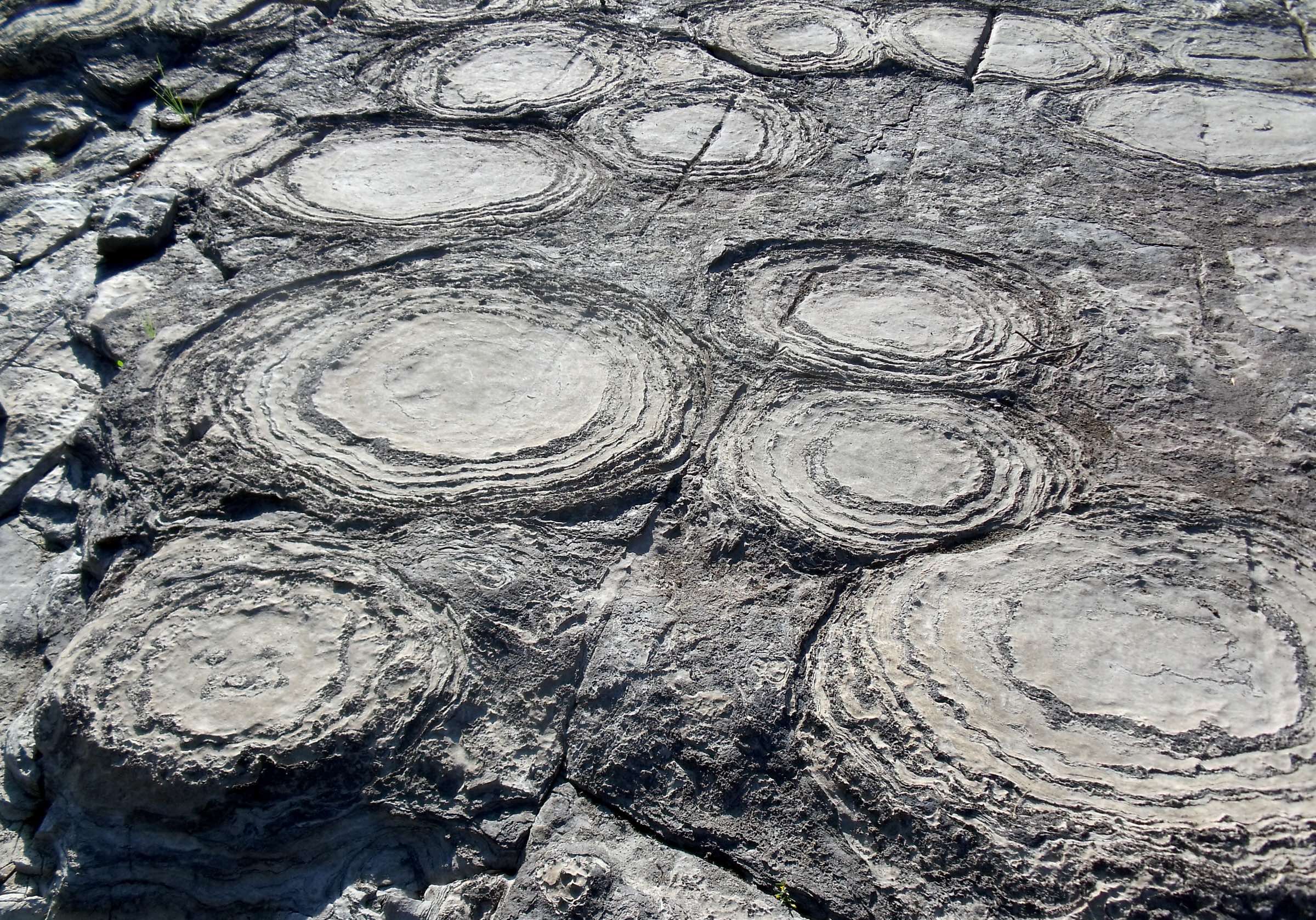}
    \caption{} 
    \end{subfigure}
    \begin{subfigure}{0.312\textwidth}
    \includegraphics[width=\linewidth]{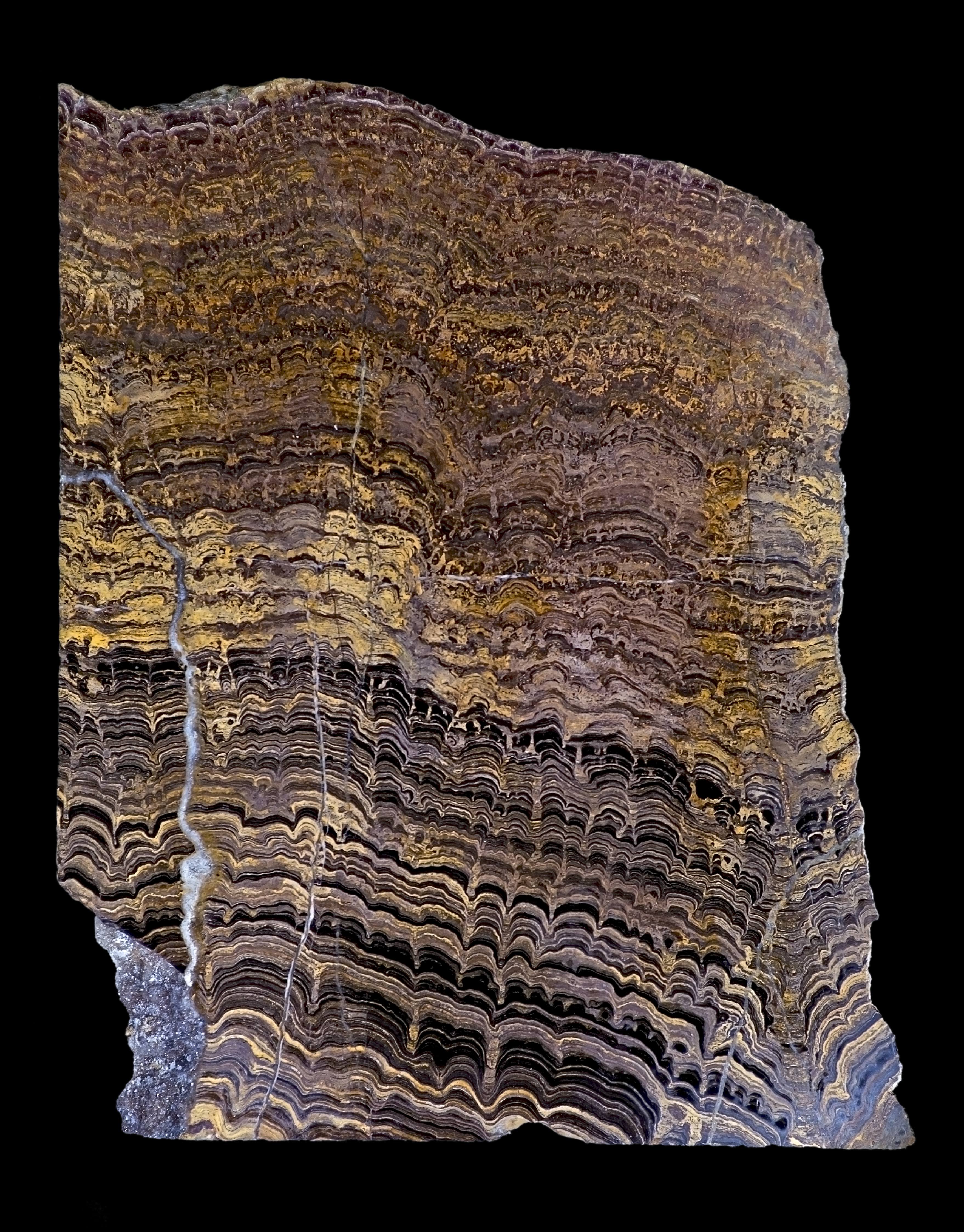}
    \caption{} 
    \end{subfigure}
    \begin{subfigure}{0.6\textwidth}
    \includegraphics[width=\linewidth]{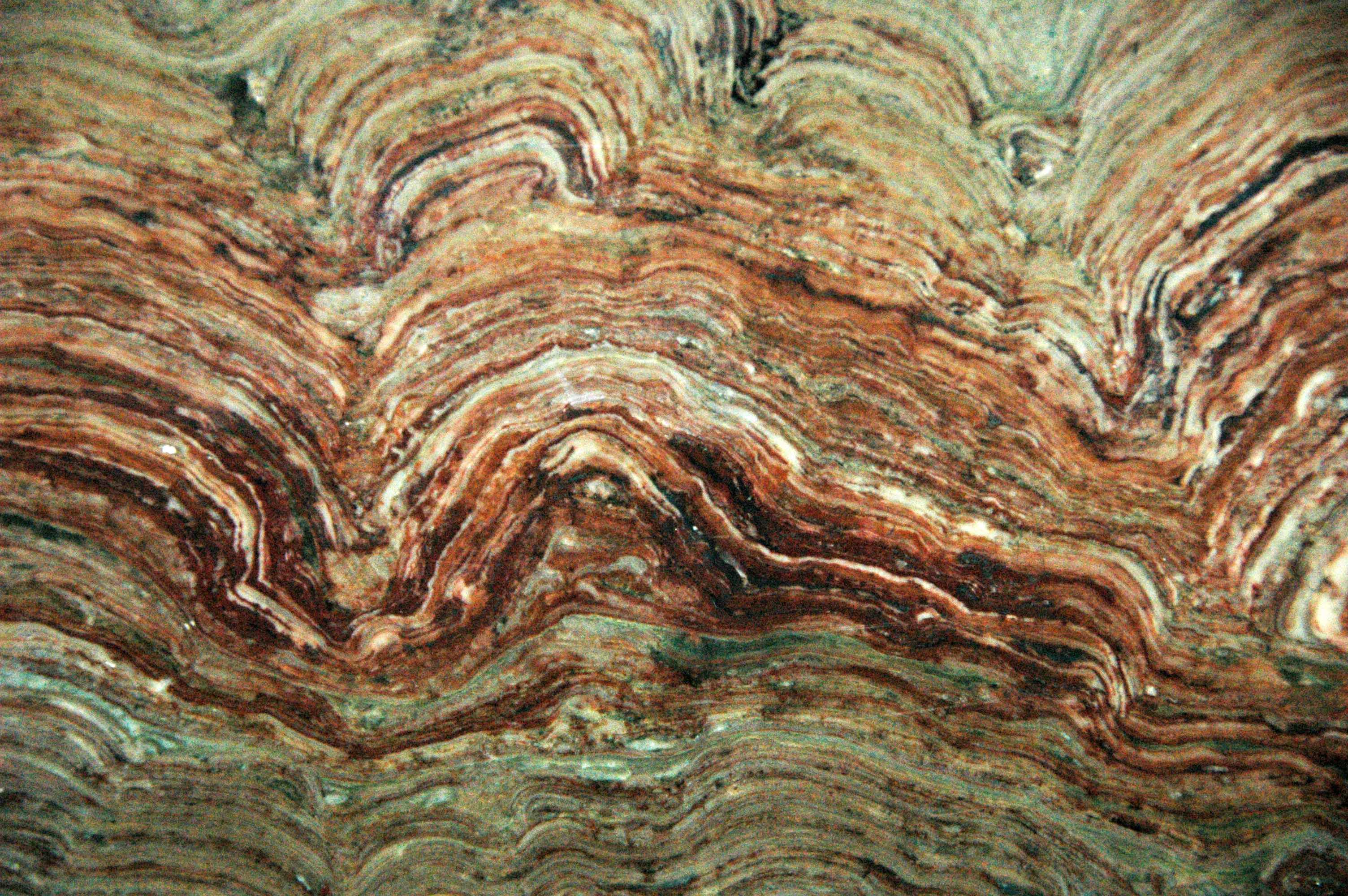}
    \caption{} 
    \end{subfigure}
\end{center}
\caption{
Stromatolites:
(a) modern, at Shark Bay, Hamelin, Western Australia (image: Julie Burgher, CC BY-NC 2.0);
(b) ancient, Ottawa, Canada (image:  Mike Beauregard, CC BY 2.0);
(c) Eastern Andes South of Cochabamba, Bolivia (image: Didier Descouens, CC-BY-SA-4.0);
(d) Glacier National Park, northwestern Montana, USA (image: James St. John, CC BY 2.0).
}
\label{fig:stromatolites}
\end{figure}

Geological structures that are debated as being abiotic or having a biological component are often smaller ones---microscopic biomorphs of various sorts, dendrites, vesicles, and so on---but stromatolites, Fig.~\ref{fig:stromatolites}, are large structures compared to those and both biological \citep{bosak2013meaning} and abiotic \citep{grotzinger1996abiotic} formation mechanisms have been proposed.
Microbialites are heterogenous organo-sedimentary deposits that form by the interplay of phylogenetically diverse benthic microorganisms with external physicochemical parameters, such as wind patterns, sediment supply, evaporation rate, water flow dynamics, pH, etc \citep{dupraz2009processes,chagas2016modern,suarez2019trapping}. As a consequence of this formation mechanism, microbialites record ecologic and environmental information and can host different types of biosignatures. 
The laminated microbialites that may project upwards as cones, domes, ridges and columns, are known as stromatolites, a term derived from the Greek words ``stromae'', for layer, and ``lithos'', for rock. Stromatolites have marked Earth's fossil record. They represent the oldest macroscopic, widely-accepted evidence of fossil life on Earth, found at Pilbara in Western Australia, dating back approximately 3.45 Ga, \citep{R1}. In addition, ever since the Archean, they have been continuously present in the rock record and were particularly widespread during the Proterozoic.

Stromatolites are found in different aqueous settings such as marine environments, freshwater habitats, hypersaline lakes, hot springs and crater lakes. While the grand majority of modern stromatolites are lacustrine, marine stromatolites are today forming at Hamelin Pool, Shark Bay, Australia, the Exuma archipelago, Bahamas and the Sheybara Island, Red Sea \citep{R2,R3, R4}. These laminated deposits are commonly carbonate-based (e.g., aragonite, dolomite, low-Mg calcite, high-Mg calcite, monohydrocalcite) although silicate minerals (e.g., stevensite, kerolite, amorphous Mg-silicate, clays and amorphous silica) and evaporitic minerals (e.g., gypsum, halite and mirabilite) are found as well \citep{R5}. With respect to their microbial diversity and structure, they host various taxa that perform different metabolic functions such as oxygenic and anoxygenic photosynthesis, heterotrophy, bacterial sulfate reduction, nitrification, etc \citep{R6}. Among them Cyanobacteria, oxygenic photosynthetic bacteria, are key residents of carbonate stromatolites capable of promoting Ca-/Mg-carbonate mineral precipitation as a result of  photosynthetic activity \citep{Dupraz_mats_2009}.

Over time, the term \emph{stromatolite} has been subject to various definitions. Some definitions focus on morphogenesis, such as lamination, regular spacing, and conical forms, while others highlight the role of biology in stromatolite formation (see \citet{R8}, and references therein). The need for these re-definitions arises from three main reasons. Firstly, the degree and quality of lamination vary within and among stromatolite deposits. Indeed, it is commonly observed that the lamination ranges from fine to crude, transitioning ultimately to thrombolytic or dendritic structures. Thus, some stromatolite deposits may not display the expected lamination. Secondly, abiotic processes alone can form laminated sediments and evaporites. Among others, self-organization processes related to crystallization, variable energy/sediment supply rates during deposition, diffusion-limited aggregation at high surface tension, ballistic deposition of colloids, diagenetic, tectonic and metamorphic processes, as well as wind activity and chemical weathering are some of the mechanisms known to form laminated deposits, domes and ridges \citep{R9,grotzinger1996abiotic,R11, MCLOUGHLIN2008Growth, R13,R14, R15, mcmahon2022false}. \citet{grotzinger1996abiotic} proposed a purely abiotic dynamical model for  surface growth for some Precambrian stromatolites, based on chemical precipitation, sediment fallout, diffusive rearrangement of suspended sediment and uncorrelated random noise. More recently, \citet{R17} showed that the formation of some stromatolite-like sinter deposits at the splash zones of the geysers of El Tatio, in Chile is controlled by wind  dynamics, thermally driven evaporation and rewetting cycles and to a lesser extent by possible biological activity. Thirdly, the exact role of  microbial communities in stromatolite formation is poorly understood. Although microorganisms can influence  microbialite formation by: i) trapping and binding sediment grains or salts leading to layered deposition, ii) serving as nucleation sites for mineral nucleation and growth, and iii) changing the chemical conditions of their (micro)environment (e.g., pH, ion concentrations), as a result of their metabolic activity (e.g., oxygenic and anoxygenic photosynthesis, sulfate reduction) thereby affecting mineral precipitation \citep{R18, R19, Dupraz_mats_2009}, the extent of  microbial vs.\ abiotic controls on stromatolite formation is still poorly understood.

\subsubsection{Banded iron formations}\label{sec:BIF}

 \begin{figure}[tbp]
\begin{center}
\includegraphics[width=\columnwidth]{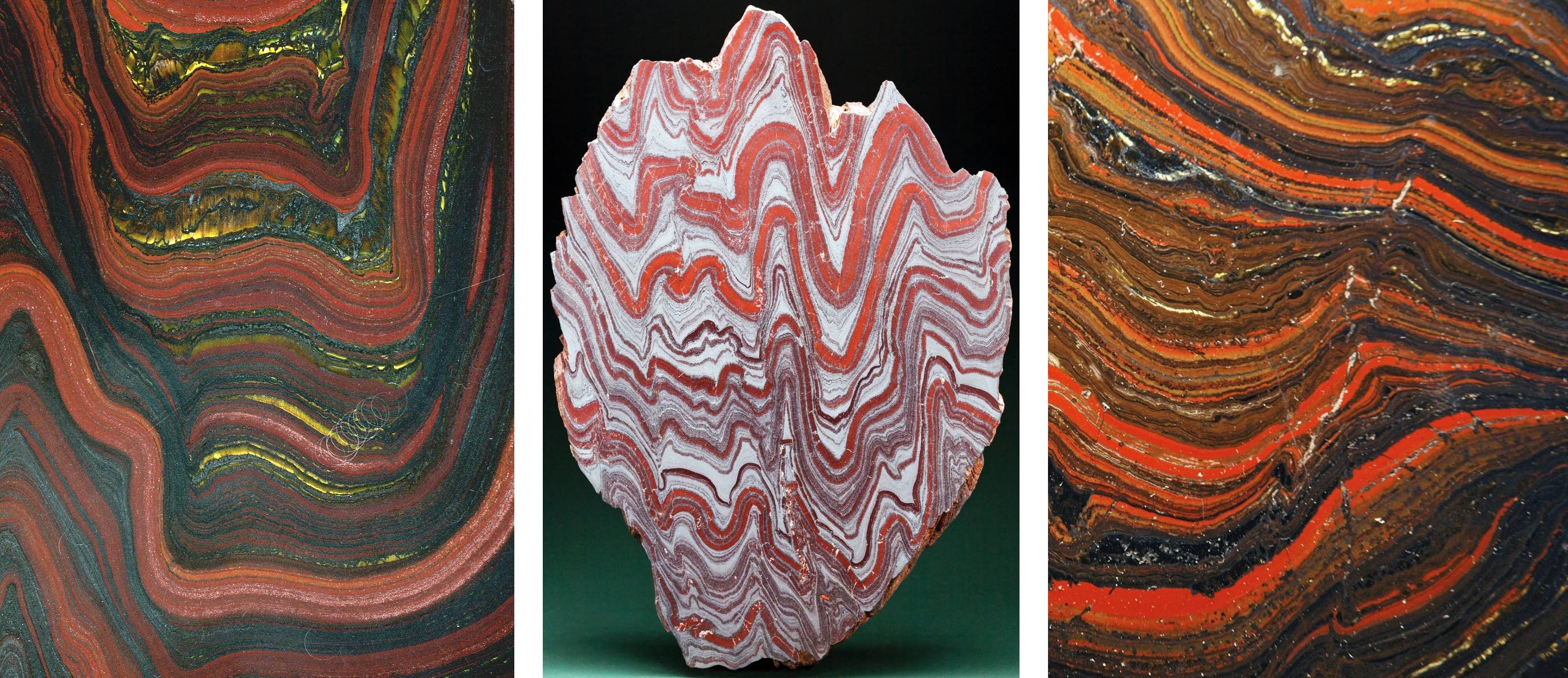}
\end{center}
\caption{Banded iron formations. Folded jaspilite banded iron formations, consisting of red chert (jasper) and grey/black hematite or magnetite bands. (A) and (C)  BIFs from Hamersley Range, Western Australia with  characteristic \emph{tiger-eye} jasper bands; photo by James St. John, distributed under CC license. (B) BIF from Minnesota, USA; public domain photograph provided by the Smithsonian Institution.}
\label{fig:banded_iron}
\end{figure}

Banded iron formations (BIF) are layered sedimentary rocks composed of alternating layers of iron-rich minerals, mainly hematite or magnetite, and silica, usually quartz, as illustrated in Fig.~\ref{fig:banded_iron}. They were formed between 3.8 and 1.8 billion years ago, during the Precambrian era, and are no longer forming today; thus they are an extinct rock species. BIFs are important because they record early Earth's oxygenation events, particularly the Great Oxygenation Event.

Because of their age, most BIFs have been around long enough to have been subjected to one or more orogenic (mountain-building) events. As such, most BIFs are folded and/or metamorphosed to varying degrees, as illustrated in Fig.~\ref{fig:banded_iron}.  BIFs are known from around the world, and are a major source of iron ore, crucial to modern industry. Some of the most famous and extensive BIF deposits are found in the vicinity of North America’s Lake Superior Basin, in western Australia, in northern China and in Brazil. 

The formation of BIFs likely involves both biological and nonbiological mechanisms. In ancient oceans, iron deposition was enabled by ferrous iron-rich seawater, facilitated by a low-oxygen atmosphere, minimal sulfate concentrations, and significant hydrothermal iron input. While it has been shown that the primary minerals forming some Archean iron formations were Fe(II)-rich clays \citep{Johnson2018,Rasmussen2021}, with Fe(III) minerals forming as later, metamorphic replacements,  research on BIF deposition mechanisms has widely focused on the oxidative process converting dissolved Fe(II) to solid iron (oxyhydr)oxides.

The classic model of iron deposition \citep{Cloud1973} proposes that ferric iron, \(\text{Fe(OH)}_3\), precipitated at the boundary between oxygenated shallow waters and upwelling, iron-rich reduced waters
\[
2\text{Fe}^{2+} + 0.5\text{O}_2 + 5\text{H}_2\text{O} \rightarrow 2\text{Fe(OH)}_3 + 4\text{H}^+ .
\]
In this model, oxygen is thought to have been produced by planktonic oxygenic photosynthesizers. Cyanobacteria are widely considered the primary source of this oxygen in Archean oceans, as they perform oxygenic photosynthesis and no evidence for eukaryotic fossils exists before approximately 1.9 billion years ago. Although some studies suggest oxygenic photosynthesis emerged in the Neoarchean era, direct fossil evidence for cyanobacteria from this period is still lacking.

Alternatively, anoxygenic photosynthesis, which does not produce oxygen, might have dominated in the more reducing conditions typical of the Archean era, oxidizing Fe(II) without atmospheric oxygen. The significance of this pathway has been recognized for over a century \citep{Harder1919}, but  interest in this process has grown considerably over the past years with a surge in research on microbial iron oxidation, making this model increasingly prevalent.

There are three main pathways of metabolic microbial iron oxidation: micro-aerophilic, anoxygenic photosynthetic, and nitrate-dependent. Micro-aerophilic Fe(II) oxidation, which plays a key role in modern iron springs and hydrothermal vent systems, involves bacteria such as \emph{Gallionella ferruginea}, \emph{Leptothrix ochracea}, and \emph{Mariprofundus ferrooxydans}. These bacteria utilize ferrous iron as an electron donor and oxygen as the electron acceptor, taking in carbon dioxide, which is then reduced to organic carbon via chemoautotrophy
\[
6\text{Fe}^{2+} + 0.5\text{O}_2 + \text{CO}_2 + 16\text{H}_2\text{O} \rightarrow \text{CH}_2\text{O} + 6\text{Fe(OH)}_3 + 12\text{H}^+.
\]
Microaerophilic Fe(II) oxidizers are widespread in marine systems, including iron-rich hydrothermal vents like those on Loihi Seamount \citep{Emerson2002,McAllister2011}. Under low oxygen conditions, these microbial Fe(II) oxidizers can dominate the iron cycle, as their oxidation rates can be over 50 times faster than abiotic rates~\citep{Sogaard2000}. They are also found at the chemocline in ferruginous lakes, such as Pavin Lake in France, where they contribute to ferric iron-rich sediment deposition \citep{Lehours2007}.

Anoxygenic photosynthetic oxidation, also known as photoferrotrophy, is another metabolic Fe(II) oxidation pathway thought to have been common on early Earth and potentially linked to BIF deposition \citep{Hartman1984}. This process uses Fe(II) as a reductant for carbon dioxide fixation
\[
4\text{Fe}^{2+} + 11\text{H}_2\text{O} + \text{CO}_2 \rightarrow \text{CH}_2\text{O} + 4\text{Fe(OH)}_3 + 8\text{H}^+ .
\]
The presence of organisms capable of photoferrotrophy was predicted before they were first cultured in the early 1990s \citep{Widdel1993}, and since then, diverse strains of anoxygenic Fe(II)-oxidizing phototrophs have been identified, including purple sulfur, purple nonsulfur, and green sulfur bacteria. 

Finally, the third pathway for metabolic microbial iron oxidation involves nitrate-dependent Fe(II) oxidation, which can be microbially mediated \citep{edwards2003, straub1996}. In this pathway, Fe(II) is oxidized through metabolic coupling with nitrate reduction
\[
    10 \, \text{Fe}^{2+} + 2 \, \text{NO}_{3}^{-} + 24 \, \text{H}_2\text{O} \rightarrow 10 \, \text{Fe(OH)}_3 + \text{N}_2 + 18 \, \text{H}^+.
\]
Nevertheless, laboratory experiments suggest that pure cultures of nitrate-reducing iron-oxidizers are challenging to cultivate, indicating a possible need for a microbial consortium to complete this process \citep{Blothe2009}.

In general, there is substantial evidence  that metabolic Fe(II) oxidation contributed significantly to iron formation (IF) deposition. Modelling suggests that even small populations of microaerophilic or photosynthetic iron oxidizers could facilitate IF deposition~\cite{Chan2016}. Moreover, photosynthetic Fe(II) oxidation under mild ocean mixing and circulation conditions could nearly exhaust upwelling ferrous iron \citep{kappler2005b}. Modern instances, such as Lake Matano’s iron-rich sediment accumulation and IF-like deposits at the Loihi hydrothermal field, illustrate the role of metabolic iron oxidation in iron-rich aquatic systems today \citep{Crowe2008, Emerson2002}. This implies a similar process likely occurred in ancient iron-rich waters.

Although microbial fossils are common in Proterozoic and younger iron oxide-rich rocks, they are absent in Archean BIFs. Structures from microaerophilic Fe(II) oxidizers appear only after the Paleoproterozoic, possibly due to increased oxygen levels that promoted oxide-coated formations \citep{planavsky2009}. The absence of such fossils in the Archean could result from the reduced atmospheric oxygen, which limited oxide structure formation without excluding microbial activity \citep{hallbeck1990}. Geochemical evidence, such as rare earth element signatures, supports the idea that a discrete redoxcline likely formed only after 2.4 Ga, suggesting anoxic or microaerophilic oxidizers’ involvement in early BIF deposition \citep{edwards2003, straub1996}.

Nonbiological, abiogenic mechanisms could also drive BIF formation, allowing iron to precipitate even in prebiotic conditions. \citet{Cairns-Smith1978} proposed that ferrous iron might have been oxidized by UV photons, which could reach Earth's surface prior to atmospheric oxygen and ozone formation. This photochemical reaction, occurring in acidic waters with UV wavelengths (200–300 nm), can be expressed as
\[
2\text{Fe}^{2+} + 2\text{H}^+ + h\nu \rightarrow 2\text{Fe}^{3+} + \text{H}_2.
\]
This model was then expanded \citep{braterman1983, anbar1992} to neutral pH conditions, where Fe(OH)$^+$ species can be oxidized by UV wavelengths (300--450~nm), producing ferric oxyhydroxide precipitates. Estimated oxidation rates from this process range from 0.5 to 2.3 $\times$ 10$^{13}$ mol Fe(II) annually, which could potentially explain iron deposition.

Alternatively, \citet{Foustoukos2008} proposed that volcanic-hosted BIFs associated with volcanic massive sulfide (VMS) deposits may result from subsurface phase separation, forming oxidized, brine-enriched solutions upon eruption. Though not fully verified, this hypothesis may explain the link between Archean jaspers and Cu-rich VMS deposits in deep-water environments.

The rhythmic structure of BIF consists of alternating iron-rich and silica-rich bands at a mesoscopic scale (mm to cm). Silica-rich mesobands primarily contain a dense mosaic of microcrystalline quartz, with some fine-grained iron carbonates, silicates, or oxides typically present. Despite their composition, silica-rich bands maintain a significant iron content (5--25\%).

The origin of BIF bands is debated. In his original paper, \citet{Cloud1973} suggested that banding resulted from fluctuations in cyanobacterial populations due to self-poisoning by oxygen, as cyanobacteria had not yet evolved enzymes such as superoxide dismutase to survive in an oxygenated environment, making them vulnerable to free-radical damage from oxygen. This led to a coupled dynamical system prone to intrinsic oscillations: a larger bacterial population increased oxygen levels, which subsequently reduced bacterial populations, and so forth.

Another concept proposed by \citet{trendall1970} suggests that the bands are secondary features resulting from the dissolution and reprecipitation of silica during the burial and compaction of strata, accompanied by associated dehydration. Finally, \citet{Morris1993} proposed that the bands relate to pulsed outputs of mid-ocean ridge (MOR) or hotspot hydrothermal activity. During periods of low activity, the water in BIF depositional environments had relatively low iron content, leading to silica-rich precipitation; conversely, during periods of higher MOR or hotspot activity, iron-rich bands were deposited.

\subsubsection{Desert or rock varnish}\label{sec:varnish}

\begin{figure}[tbp]
\begin{center}
\includegraphics[width=\columnwidth]{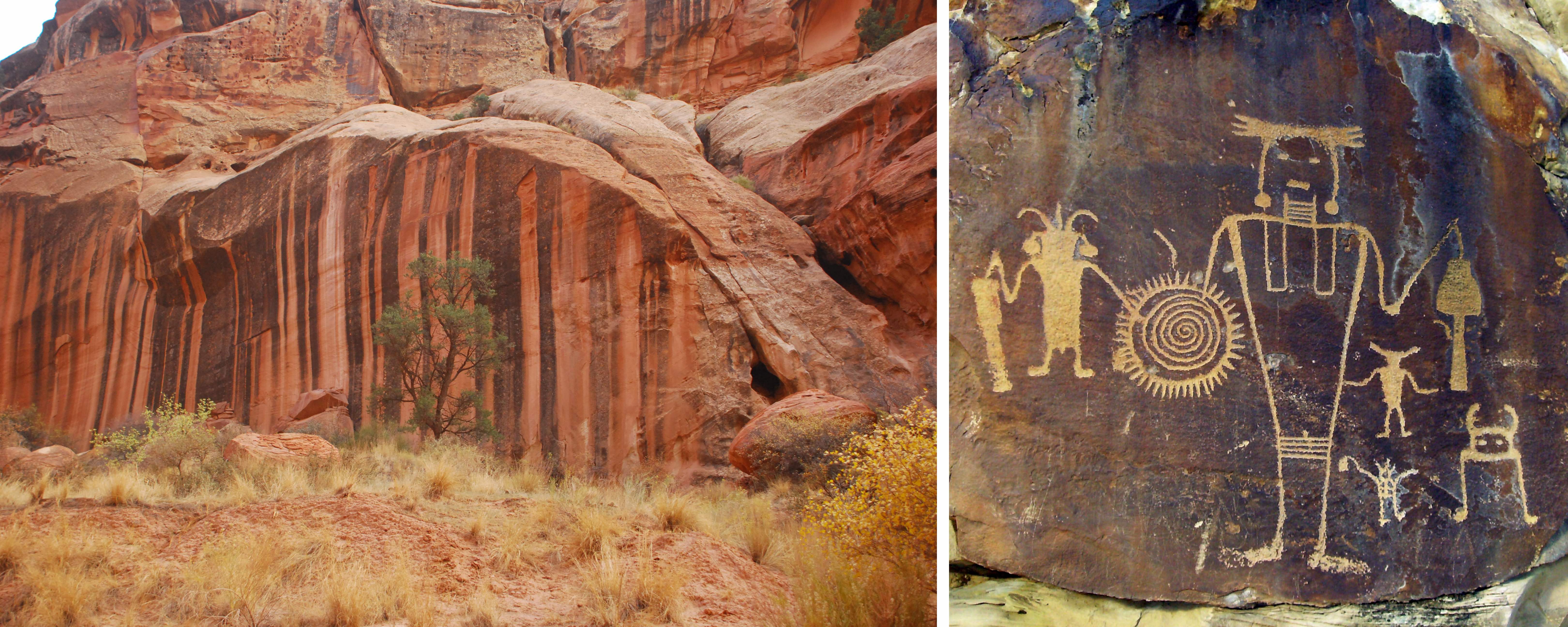}
\end{center}
\caption{(A) Bleeding rock, covered with desert varnish, in Capitol Reef National Park, USA (B) Utah Petroglyphs in rock varnish, Dinosaur National Monument, USA. The photo in (A) is by Rosa (lorea2006) on Flickr, distributed under CC BY-ND 2.0 license. The photo in (B) is by Jay Gannett, distributed under CC BY-SA 2.0 license.}
\label{fig:varnish}
\end{figure}

Rock varnish, also known as desert varnish, Fig.~\ref{fig:varnish}, is a thin, dark coating that forms on exposed rock surfaces, predominantly in arid and semi-arid environments, with thicknesses ranging from 5 to 600 \textmu m. Accumulating at a rate of a few micrometres to tens of micrometres per thousand years, it is one of the slowest-forming sedimentary deposits. Composed of up to 70\% clay minerals, about 30\% manganese and iron oxides, and trace elements including rare earth elements (REEs), rock varnish is notably enriched in manganese;  often 50–100 times higher than surrounding soils, dust, or underlying rock~\citep{Potter1977,Potter1979}. Its metallic sheen, attributed to Mn and Fe oxyhydroxides, makes it a striking feature, often used as a canvas for petroglyphs (Fig.~\ref{fig:varnish}B)  and a proxy for paleoclimatic studies. Despite extensive research, and despite the fact
 that this field began in the 19th century  with Humboldt and Darwin
\citep{dorn2012revisiting},
the formation of rock varnish remains a geochemical enigma, with several hypotheses proposed to explain its development, particularly the extreme Mn enrichment. 

The abiotic hypothesis posits that rock varnish forms through purely physico-chemical processes, primarily the leaching and precipitation of Mn and trace elements from atmospheric dust during wetting events like rain or fog. Small changes in pH (around 5.7–8) and redox conditions (Eh $\sim$ 0.8) mobilize $\text{Mn}^{\text{2+}}$, which oxidizes to insoluble $\text{Mn}^{\text{4+}}$ and precipitates as Mn oxyhydroxides, cementing clay minerals to the rock surface. Dust accumulates in micro-basins on rocks, where leaching and adsorption of Mn, Ba, Pb, and REEs onto clays occur~\citep{Engel1958,Perry1978,Thiagarajan2004}. Studies, such as that by \citet{Goldsmith2014}, show high enrichment of Mn, Co, Ni, Pb, and REEs in varnish compared to local dust, consistent with scavenging by Fe-Mn oxyhydroxides in aqueous environments. In the Negev Desert, Mn, Ba, and Pb are enriched $\sim$ 100× relative to settling dust, with Mn mobilized at pH $\sim$ 8. On the other hand, \citet{Lu2019} and \citet{Xu2019} suggest photo-catalysis, where solar radiation drives Mn oxidation in dust, as a possible mechanism of rock varnish formation. However, these abiotic models struggle to explain  extreme Mn enrichment without biological catalysts, as Mn oxidation is kinetically slow in typical desert conditions, and it does not fully account for organic compounds or microbial signatures in varnish.

The silica gelation hypothesis, proposed by \citet{Perry2006}, suggests an abiotic process where silica dissolves from minerals, transitions from a sol to a gel state on rock surfaces, and hardens to trap Mn, Fe, and organic material. High silica content and amorphous silica phases (opal-A) in some varnishes support this idea, as does the presence of organic compounds and water \citep{Perry2006}. However, critics, including \citet{Dorn2007}, argue that it fails to explain extreme Mn enrichment, the dominance of clay minerals, or high Fe concentrations in varnish on Fe-poor rocks, and it does not account for varnish in colder environments with limited heat and sunlight.

In contrast, the biotic hypotheses argue that microorganisms, such as bacteria and fungi, actively oxidize $\text{Mn}^{\text{2+}}$ to $\text{Mn}^{\text{4+}}$, concentrating Mn and Fe to form varnish. Cyanobacteria, Actinobacteria, Proteobacteria, and microcolonial fungi like {\it Dematiaceous hyphomycetes} are commonly found in varnish and may use Mn oxidation metabolically, possibly as an antioxidant strategy to survive UV radiation and desiccation~\citep{Dorn1981,Taylor1983}. \citet{Lingappa2021} propose that Cyanobacteria sequester Mn as a catalytic antioxidant, leaving Mn-rich residues upon death that oxidize into varnish cements over millennia. Metagenomic studies reveal distinct microbial communities in varnish, with higher abundances of Actinobacteria, Proteobacteria, and Cyanobacteria compared to surrounding soils. Cultured bacteria, such as {\it Geodermatophilus} and {\it Bacillus}, oxidize Mn in laboratory conditions, as noted by \citet{Hungate1987}. Amino acids (e.g., d-alanine, glutamic acid) and DNA in varnish, reported by \citet{Perry2003,Kuhlman2008}  further suggest microbial activity. However, low levels of Mn-oxidizing enzymes in some varnish samples and the slow growth rate of varnish (microns per millennia) challenge the extent of active microbial involvement.

Finally, the  polygenetic hypothesis integrates biotic and abiotic processes, proposing a sequence where dust deposition provides clay minerals, microorganisms concentrate Mn and Fe, microbial decay mobilizes these elements at the nanoscale, and physicochemical processes cement oxides into clays. \citet{dorn2024rock} supports this model, noting that high-resolution electron microscopy shows Mn and Fe concentrated on bacterial sheaths, later mobilized into clay minerals \citep{Krinsley2017}. Studies in the Sonoran Desert indicate {\it Proteobacteria} and {\it Actinobacteria} mobilize Mn/Fe, with radiocarbon dating suggesting recent microbial activity alongside older varnish layers~\citep{MartinezPabello2021}. This model explains both organic compounds from microbes and inorganic enrichments from dust leaching, making it quite versatile. However, the exact balance of biotic and abiotic contributions remains unclear, and variability in varnish composition across environments complicates universal application.

The formation of rock varnish remains unresolved due to its slow growth, extreme Mn enrichment, and complex composition. The abiotic hypothesis struggles with Mn oxidation kinetics, while the biotic hypothesis is limited by sparse evidence of active Mn oxidation in situ. Even though the polygenetic model, integrating microbial initiation and abiotic cementation, is currently favoured for its ability to reconcile organic and inorganic signatures~\citep{chaddha2024biotic,dorn2024rock}, further research is needed to quantify microbial roles, understand nanoscale processes, and reconcile formation across diverse environments. Rock varnish’s applications as a paleoclimate proxy, dating tool (e.g., cation-ratio dating, radiocarbon dating), and Martian analogue underscore its significance, though dating methods remain controversial due to environmental variability. Advances in metagenomics, microscopy, and trace element analysis will likely refine our understanding of this multifaceted biogeochemical coating, clarifying how microbial, chemical, and environmental processes may each contribute to rock varnish formation.

\subsubsection{Tufa and travertine structures}\label{sec:travertines}

\begin{figure}[tbp]
\begin{center}
    \begin{subfigure}{0.49\textwidth}
    \includegraphics[width=\linewidth]{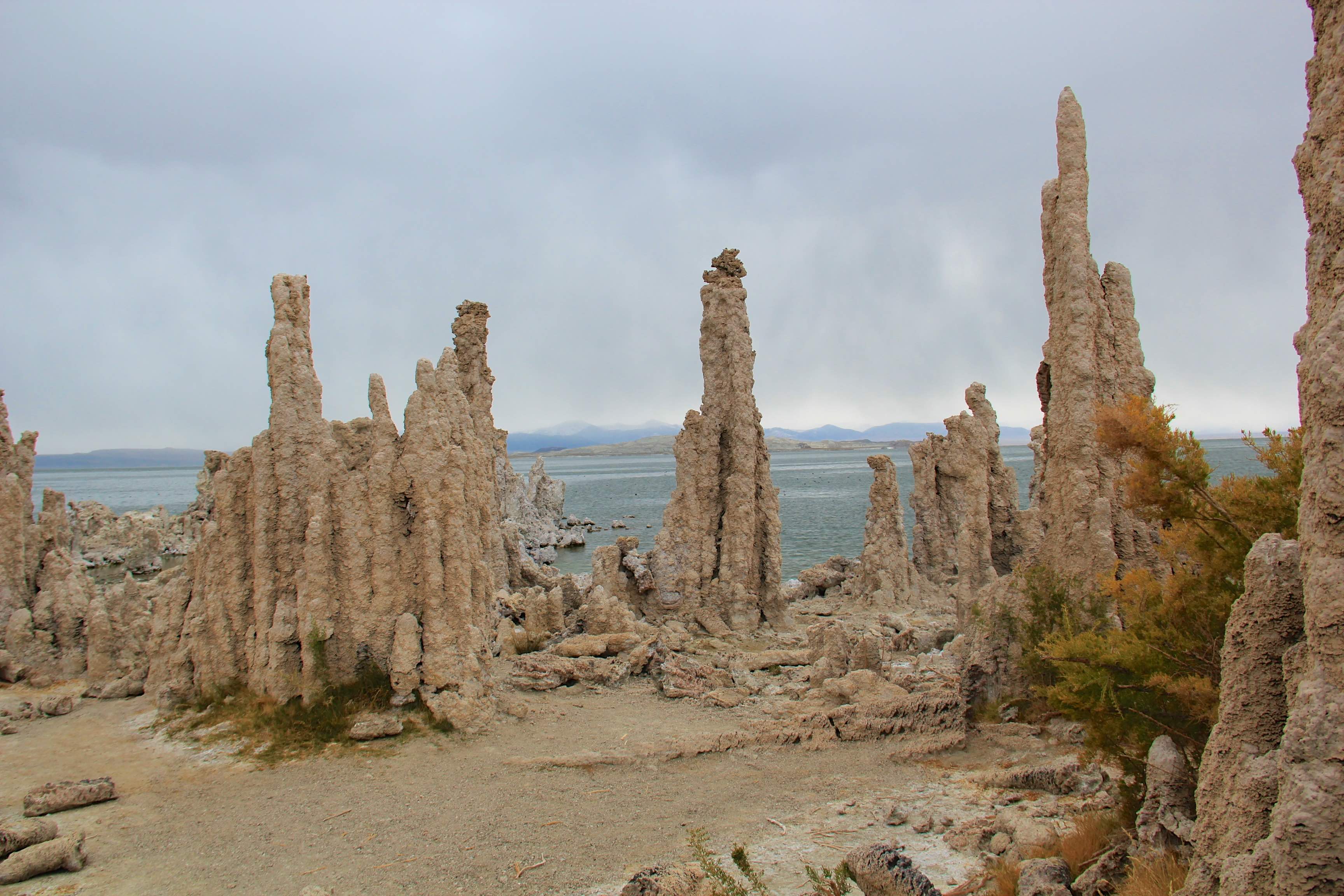}
    \caption{}
    \end{subfigure}
    \begin{subfigure}{0.49\textwidth}
    \includegraphics[width=\linewidth]{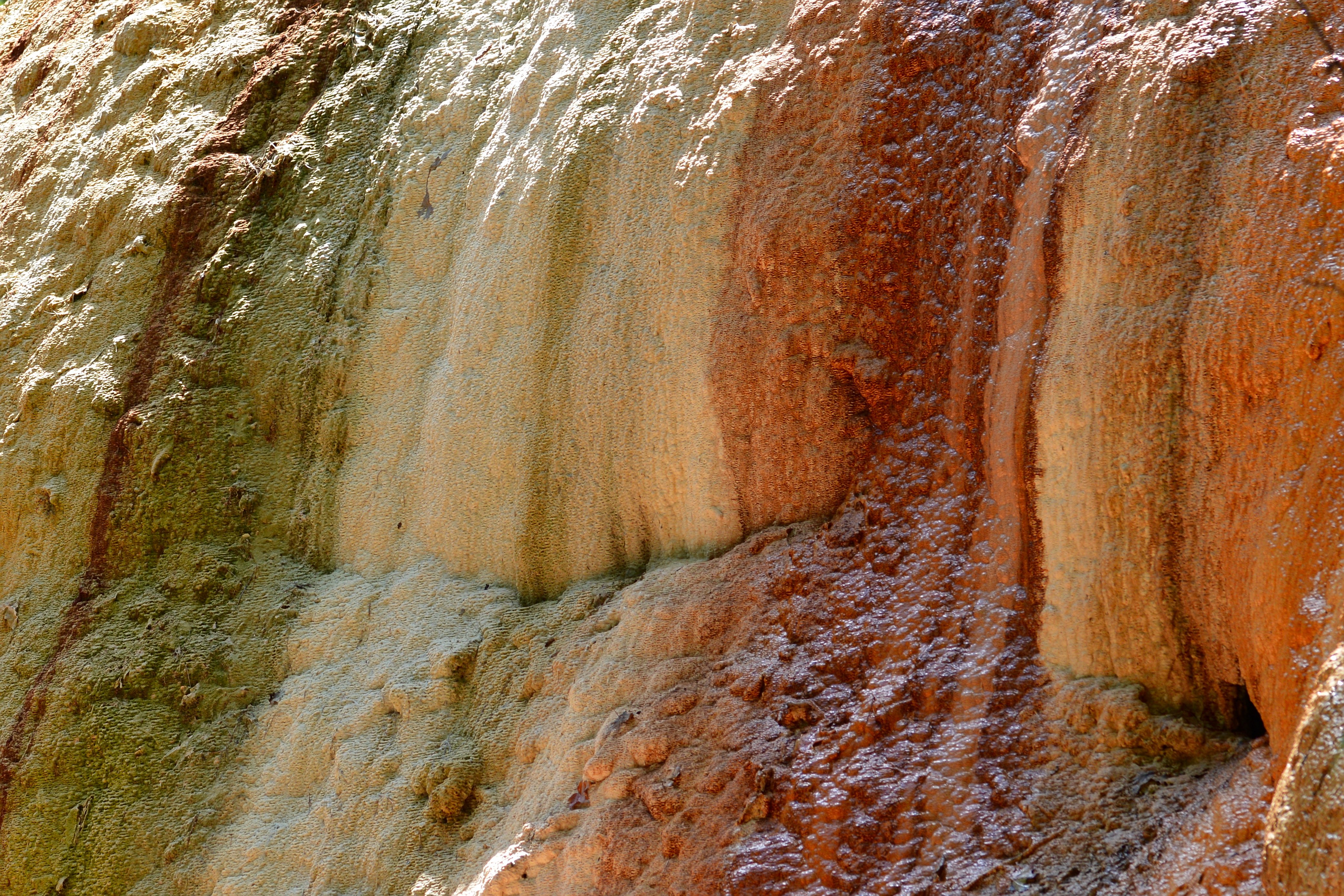}
    \caption{}
    \end{subfigure}
\end{center}
\caption{
Tufa and travertine. 
(a) Mono Lake, California (image: David P. Fulmer, CC-BY-2.0)
(b) Orenda mineral spring tufa deposits, Saratoga Springs, New York (image: Ryan Hodnett, CC-BY-SA-4.0)
}
\label{fig:tufa_travertine}
\end{figure}

Tufas \citep{pedley2009freshwater} and travertines \citep{zhijun2019biological} are freshwater carbonate deposits whose formation is debated between abiotic and biological mechanisms. The terms are often used interchangeably, but some use the term \emph{tufa} to describe all ambient-temperature freshwater carbonate deposits, and reserve \emph{travertine} for hydrothermal freshwater carbonate deposits \citep{pedley2009tufas}. 
Abiotic processes of CO$_2$ degassing, temperature changes and evaporation alone lead to the supersaturation of calcium carbonate and its subsequent precipitation. On the other hand, biological processes such as the activity of photosynthetic microorganisms and bacteria can significantly enhance and influence carbonate deposition by altering the local chemical environment and providing nucleation sites, affecting for instance CaCO\textsubscript{3} microtextures and polymorphism. 

Microbial influences on CaCO\textsubscript{3} mineralization have  been documented for many travertine systems  \citep{Shiraishi_2008, Perri_2012, Okumura_Textures_2013, Kano_Travertine_2019, Della_Porta_Travertine_2021}, producing recognizable sedimentary fabrics and textures \citep{GUO_1992, Kano_Travertine_2019}. 
For instance, small-scale variations of CaCO\textsubscript{3} polymorphism in travertines have been interpreted as resulting from localized physicochemical changes occurring in micro-environments at the surface of the growing travertines, most likely under the influence of microorganisms and their EPS. \citet{Peng_2013} described the co-precipitation of calcite, aragonite and amorphous calcium carbonate within distances of a few microns in hot-spring deposits, and proposed that microbial biofilms growing on the carbonates were forming microdomains, within which specific physicochemical conditions developed as a result of microbial activity, influencing CaCO\textsubscript{3} precipitation and polymorphism. Biofilms contain abundant EPS, which influence CaCO\textsubscript{3} precipitation principally due the presence of negatively charged, Ca\textsuperscript{2+}-binding organic functional groups \citep{Dupraz_mats_2009}, and have been shown exert a control on CaCO\textsubscript{3} polymorphism in laboratory experiments \citep{Tourney_2008, Tourney_2009}. Microlaminations formed by alternating layers of aragonite and calcite in hot spring travertines in Japan have been interpreted as resulting from diurnal cycles affecting microbial EPS \citep{Okumura_Processes_2013}: during the day, EPS built by photosynthetic organisms would bind Ca\textsuperscript{2+}, reducing supersaturation and promoting calcite formation, while at night, decomposition of EPS by heterotrophs would release Ca\textsuperscript{2+} and lead to aragonite precipitation. Similarly, CaCO\textsubscript{3} polymorphism in Italian hot-spring travertine deposits has been interpreted to reflect microbial diurnal control \citep{GUO_1992}, but there aragonite was thought to grow during the day.

\subsubsection{Geyserite structures}\label{sec:geyserites}

\begin{figure}[tbp]
\begin{center}
\includegraphics[width=0.66\columnwidth]{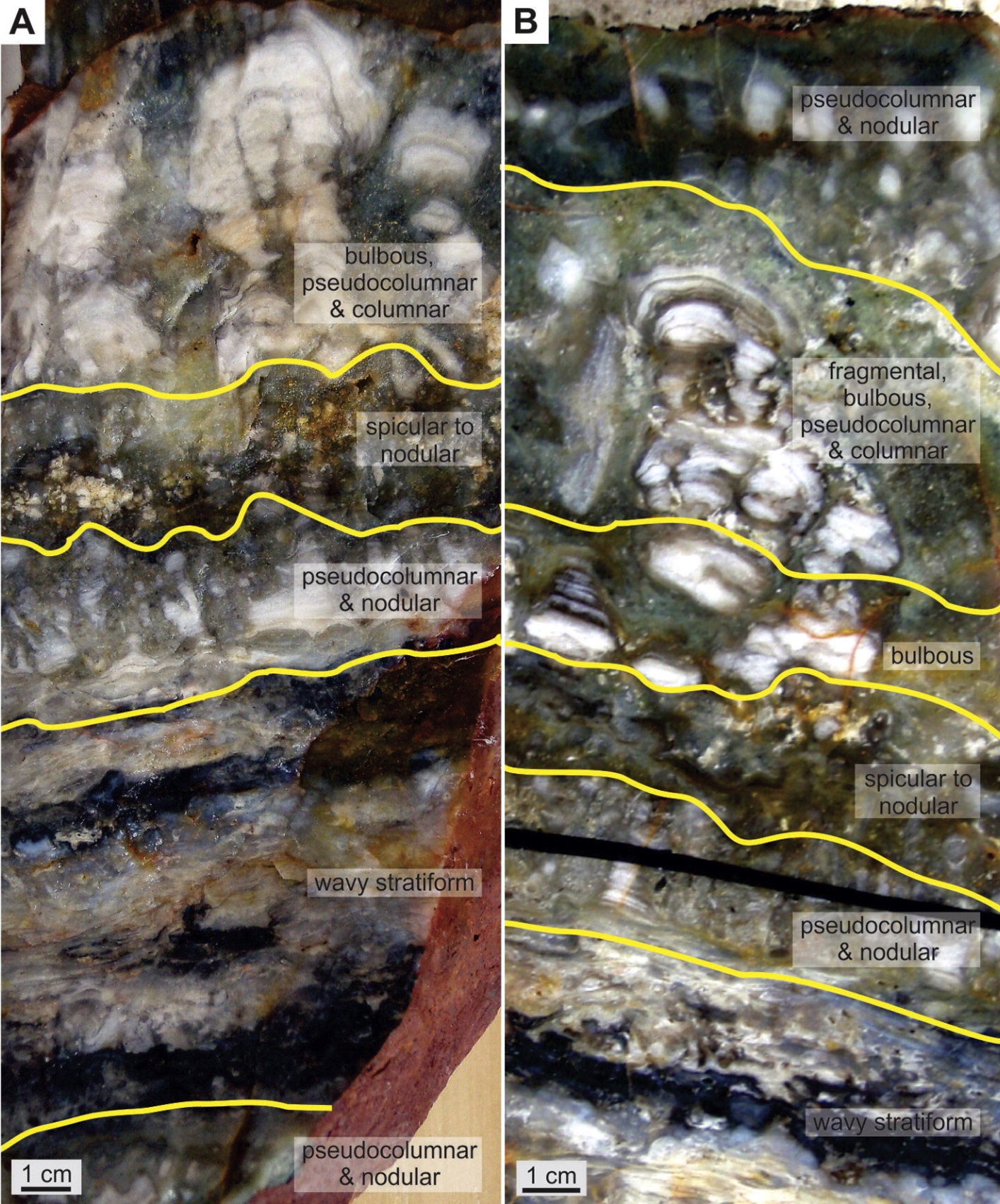}
\end{center}
\caption{
Jurassic geyserites from Deseado Massif volcanic province, Argentina (taken from \citet{CAMPBELL201544}). Both A and B are polished rock sections, showing some of the different textures (labelled) that can be observed in geyserites.
}\label{fig:geyserite}
\end{figure}

Geyserites are rocks formed in the vicinity of geysers, so these rocks are rare both nowadays and in the geological record. They are composed of almost pure silica, in particular, amorphous opal (opal-A) with varying degrees of hydration, which can recrystallize during diagenetic processes to quartz or, less commonly, chalcedony \citep{Jones2004Water, CAMPBELL201544}. The criteria for identifying these rocks in the geological record were outlined by \citet{Walter1972hot, Walter1976Chapter}. In outcrops, geyserites can be identified by spicular, bulbous, (pseudo)columnar, beaded, and/or stratiform textures occurring within botryoidal (centimetric) or banded (metric) deposits (Fig.~\ref{fig:geyserite}). Although diagenetic recrystallization processes may mask or obliterate the smaller textures, at the microscale, they may exhibit very fine laminations with occasional cross-stratification, bridges, and cornices.
These rocks are a clear indicator of high-temperature geothermal activity in terrestrial volcanic environments \citep{CAMPBELL201544}. Although it was once thought that microbial activity was not present in such geological environments, more recent studies have identified bacteria and archaea in water temperatures above 75\textdegree{}C \citep{Blank2002Microbial}. However, the role of biological activity in the formation of geyserites is still unclear \citep{CAMPBELL201544}. Numerous studies can be found supporting the abiogenic formation of these rocks \citep{Walter1976Chapter, MCLOUGHLIN2008Growth, Braunstein2001Relationship, Lowe2003Microstructure}, while others show that the presence of microorganisms acts as sites for heterogeneous silica nucleation \citep{Brian1997Biogenicity, handley2005abiotic, Handley2008Silicifying}, thus facilitating their formation.

\subsubsection{Kinneyia-type wrinkle structures}

\begin{figure}[tbp]
\begin{center}
\includegraphics[width=1\columnwidth]{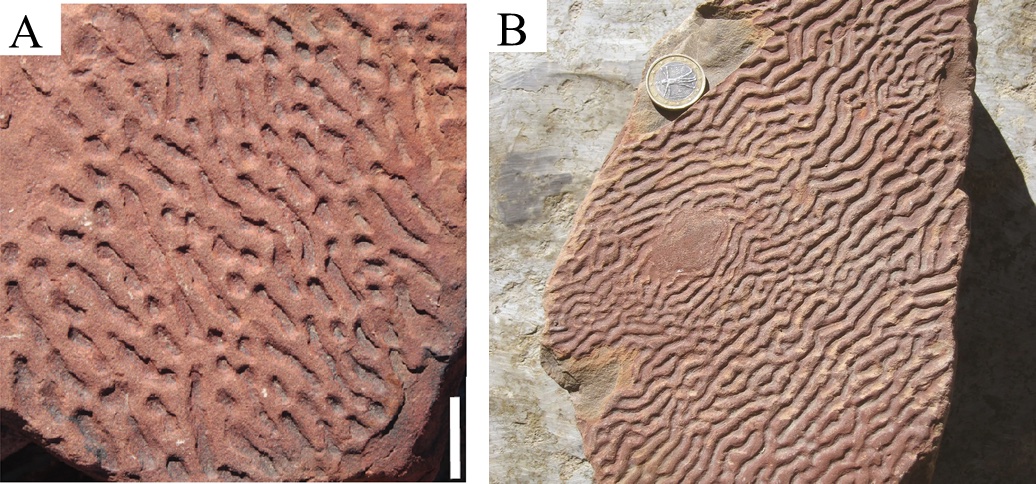}
\end{center}
\caption{Proterozoic Kinneyia structure from farm Arimas (A) \cite{turk2022paleontology} and farm Neuras (B), Namibia \cite{HerminghausEtAl2016KinneyiaFlow}
}\label{fig:Kinn}
\end{figure}

Kinneyia (Fig.~\ref{fig:Kinn}) was originally erected by \citep{Walcott1914Kinneyia} as a genus of fossil algae based on Precambrian material from the Belt Supergroup, but is now widely regarded as a microbially induced sedimentary structure (MISS) rather than a discrete body fossil. In modern usage, ``Kinneyia'' denotes a class of wrinkle structures  \citep{PoradaBouougri2007WrinkleReview,PoradaEtAl2008KinneyiaPalaios}: ripple-like, millimetre-scale ridges with flat or gently rounded crests and steep flanks, commonly arranged in honeycomb or reticulate patterns. They occur on the upper surfaces of fine-grained sandstone and siltstone  beds, typically in shallow subtidal to intertidal siliciclastic settings, where they are closely associated with other MISS and with evidence for cohesive microbial mats stabilizing the sediment surface \citep{PoradaBouougri2007WrinkleReview,DaviesEtAl2016MISSReview}.

Ichnotaxonomically, Kinneyia is usually treated as a mat-related trace fossil that records deformation of a microbial mat–sediment pair, rather than the remains of the mat itself \citep{PoradaEtAl2008KinneyiaPalaios,Stimson2017WrinkleIchnotax}. In the German literature the term \emph{Runzelmarken} is often used for similar wrinkle structures, and is now generally taken to include Kinneyia-type morphologies \citep{PoradaBouougri2007WrinkleReview,Peterffy2016JurassicWrinkle}. These structures thus sit squarely within the MISS concept: they require a cohesive, viscoelastic microbial mat for their formation, but their geometry reflects deformation by physical flows and sediment loading rather than purely biological growth features \citep{DaviesEtAl2016MISSReview}.

Several formation mechanisms have been proposed. Earlier models emphasized gas doming and degassing beneath a degraded mat, mat shrinkage, or postburial soft-sediment deformation \citep{PoradaBouougri2007WrinkleReview,PoradaEtAl2008KinneyiaPalaios}. More recent work, combining flume experiments and linear stability analysis, has shown that the characteristic Kinneyia wavelength and crest geometry can be reproduced by a shear-induced Kelvin–Helmholtz-type instability in a viscoelastic film subject to overlying flow \citep{ThomasEtAl2013KinneyiaShear,HerminghausEtAl2016KinneyiaFlow}. In these models the microbial mat is treated as a soft viscoelastic layer whose thickness sets the dominant wavelength, while the overlying water flow supplies the shear needed to amplify initial perturbations. Preferential trapping of sand or silt in the troughs then enhances topography and improves preservation \citep{ThomasEtAl2013KinneyiaShear}. 

From the standpoint of this review, Kinneyia-type structures are a clear example of strong, nonlinear couplings between microbial mats and hydrodynamics: the presence of a cohesive biofilm is essential, but the detailed pattern is selected by a physical flow instability in a soft-matter layer, rather than by any genetically encoded biological template. They therefore occupy an intermediate position in our spectrum, alongside other MISS where biological stabilization and physical self-organization are inseparable.

\subsubsection{Moonmilk}\label{sec:moonmilk}

\begin{figure}[tbp]
\begin{center}
\includegraphics[height=0.33\columnwidth]{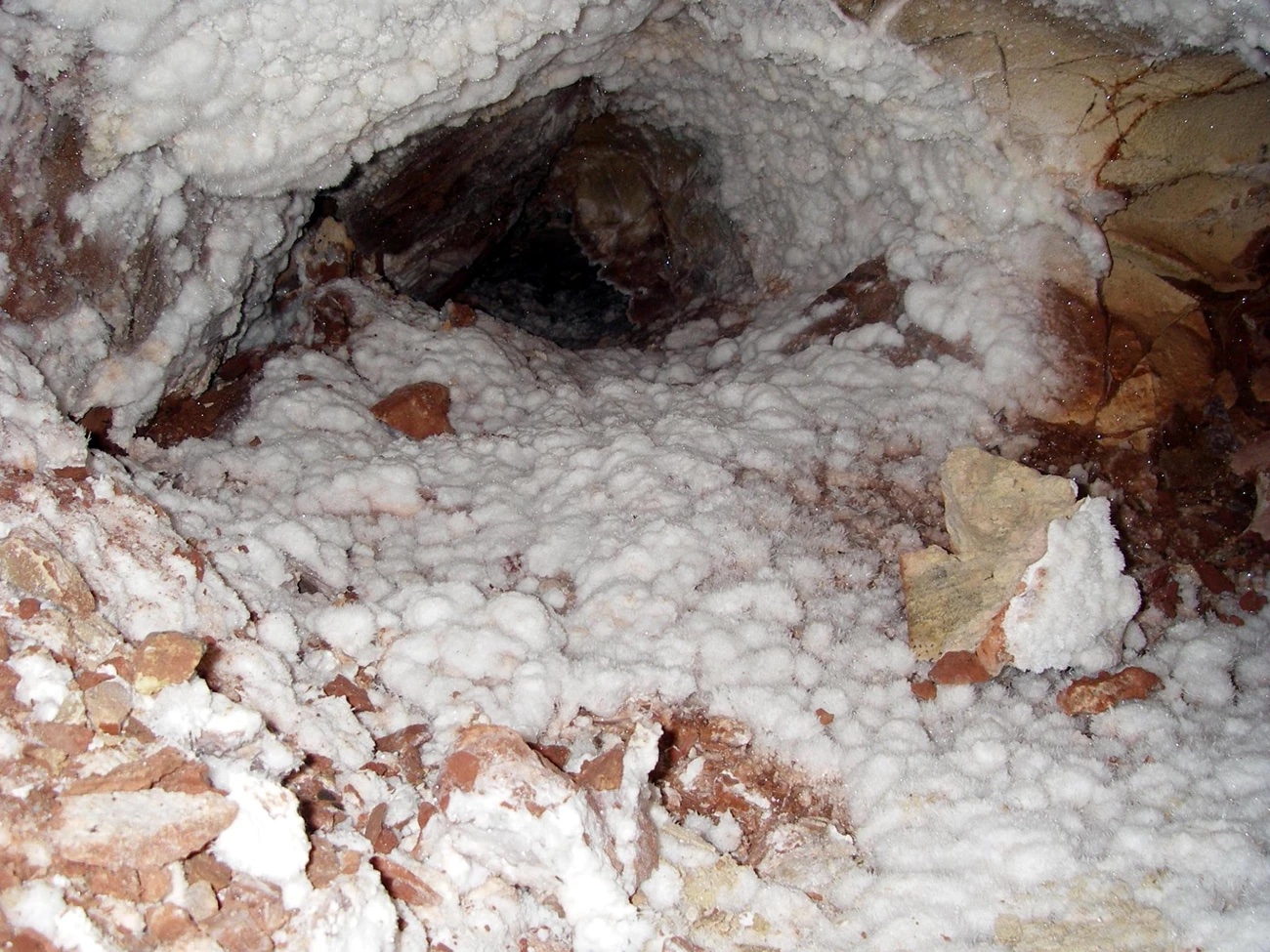}
\includegraphics[height=0.33\columnwidth]{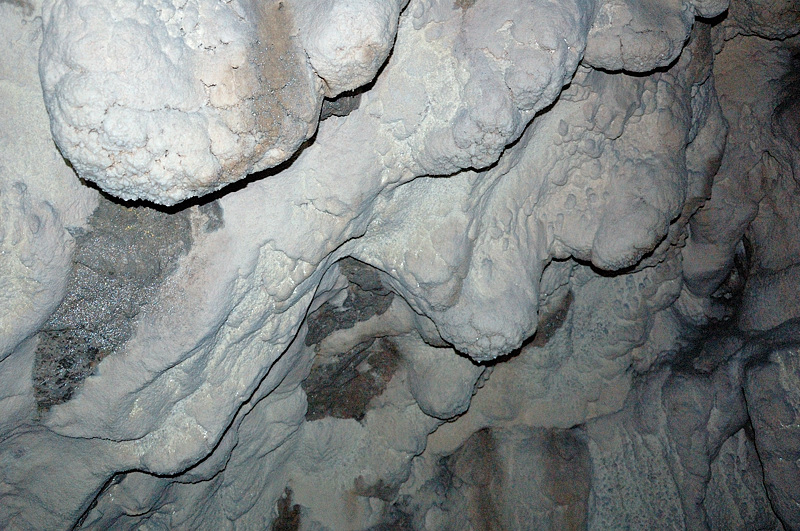}
\end{center}
\caption{Moonmilk: moonmilk deposits in the Wind Cave National Park, South Dakota.
NPS photo, public domain.
Moonmilk in the Bergmilchkammer, Austria. Doronenko, CC BY-SA 3.0.
}
\label{fig:moonmilk}
\end{figure}

Moonmilk, Fig.~\ref{fig:moonmilk}, refers to a cave mineral formation. It is a generic term used to refer to a secondary mineral deposit of varied composition, but characterized by a very specific texture. Moonmilk is a white, soft, porous, plastic and wet speleothem formed of very small fibrous crystals \citep{borsatoCalciteMoonmilkCrystal2000, canaverasOriginFiberCalcite2006, millerOriginAbundantMoonmilk2018}. It is mainly composed of carbonates, calcite, hydromagnesite, aragonite, vaterite, and huntite \citep{borsatoCalciteMoonmilkCrystal2000}. However, moonmilk composed mainly of gypsum or gibbsite has also been described \citep{gazquezPrecipitacionMoonmilkProceso2012}.

The discussion about the possible biotic or abiotic origin of moonmilk is a long-standing debate. Theories about the possible biotic origin of moonmilk can be divided into two broad categories: those that argue that the origin of this structure is due to the indirect role of bacteria, and those that argue for the direct role of bacteria. The former postulate that bacteria only play a role in dissolving the host rock, which releases calcium and carbonate ions into the cave water, promoting the subsequent precipitation of moonmilk. However, this hypothesis has been losing ground in favour of the idea of a direct role of bacteria in the precipitation of calcium carbonate
\citep{cirigliano2018calcite, ronca2023biogenic}. Theories of biological origin began from the observation of bacteria, algae, and fungi in moonmilk deposits. However, no direct evidence has yet been found of carbonate crystals being precipitated by bacteria in moonmilk \citep{ronca2023biogenic}.

On the other hand, theories proposing the abiogenic origin of moonmilk also acknowledge the presence of bacteria, but note that they are not ubiquitous and, in many cases, are not even detectable, as in moonmilk composed of magnesium carbonates \citep{northup1997microorganisms}. The abiogenic origin of moonmilk proposes that this structure may have been formed by precipitation from an atmosphere saturated with water vapour or by capillary flows saturated with carbonates \citep{borsatoCalciteMoonmilkCrystal2000}. Both crystallization pathways would allow the formation of fibrous crystals.

This open debate even allows for the possibility that both formation mechanisms could have been proposed for moonmilk formed in the same place, as in a spring-water tunnel in Porto (Portugal)  \citep{millerOriginAbundantMoonmilk2018}.

\subsection{Spirals and helices}\label{sec:spirals}

\begin{figure}[tbp]
\includegraphics[width=\textwidth]{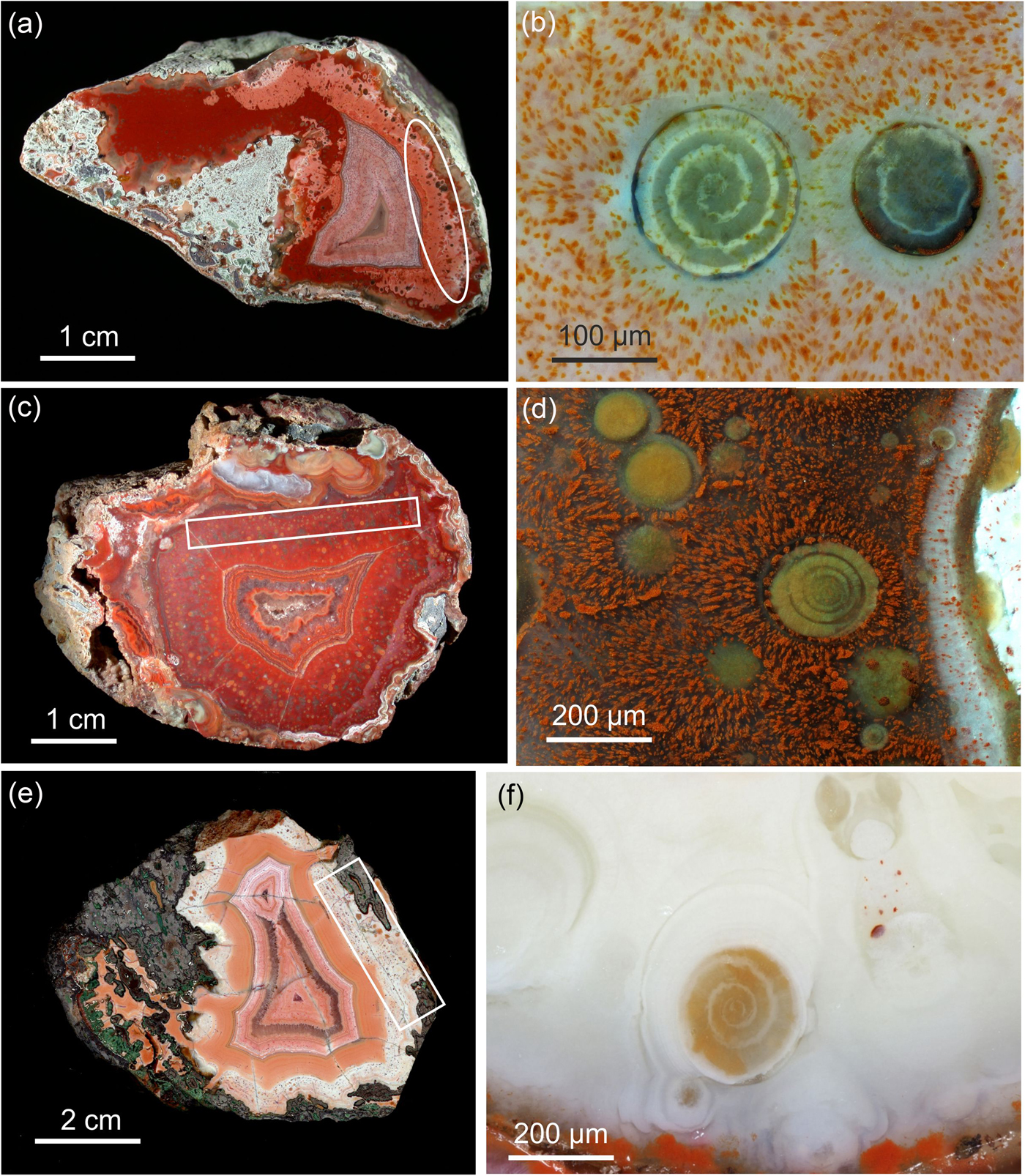}
\caption{Polished agate halfs and related micrographs of magnified areas with spiral growth (see marked boxes in a, c, e) in samples from different locations of the Saar-Nahe region of Germany \citep{gotze2019micro}.}
\label{fig:spiral_agate}
\end{figure}
 
Spiral and helical patterns are common in biology. We can think of climbing plants winding helically as they climb, the intestines of vertebrates, or indeed the structure of DNA itself. Some bacteria  adopt helical morphologies (e.g., cyanobacteria of the genus \textit{Spirulina}). Others produce extracellular helical mineralized structures called twisted stalks. These helices, formed by microaerophilic Fe-oxidizing bacteria, consist of polysaccharides and iron oxide minerals, and can be preserved as fossils in the geological record \citep{Chan2010Lithotrophic, Picard2015Experimental}. 
Thus it is not surprising that spirals, in two dimensions, and helices, in three, when encountered in a geological setting, are often taken to be fossils. Indeed, this is frequently the case. Gastropod molluscs with helical shells, and cephalopods such as ammonites and the modern-day Nautilus with spiral shells all leave extensive fossil records \citep{cartwright2024arms}.
Likewise, spiral foraminiferal tests are extremely numerous \citep{berger1969planktonic,thomsen2008coccolith}.
Spiral coprolites \citep{williams1972origin,mcallister1985reevaluation} are another instance.

Yet it is not necessarily  the case that geological spirals and helices must be biological in origin, as they can also be produced by abiotic physico-chemical processes.
Silica-carbonate biomorphs (Sec.~\ref{sec:biomorphs}) and carbon-sulfur biomorphs (Sec.~\ref{sec:carbon-sulfur}) can form spirals and helices, as can physical and chemical pattern-forming processes such as Liesegang patterns and the Belousov--Zhabotinsky reaction \citep{cartwright_self-organized}. Helictites in caves, despite their name, are not generally helical, but they sometimes are \citep{davis2019helictites}.
Micrometric helices are found in a geological setting in agate and chalcedony \citep{gotze2019micro}; Fig.~\ref{fig:spiral_agate}. Their mechanism of formation is not clear. They seem to be 3D helices, rather than 2D spirals. They are apparently abiotic, composed of silica plus traces of iron. The quartz crystallites are arranged rather like the carbonate nanorods in silica-carbonate biomorphs described in Sec.~\ref{sec:biomorphs}. Might acidic conditions have leached out the carbonates and left silica?  Like in silica-carbonate biomorphs, the initial formation of nanorod crystallites may be part of the mechanism here. If these are not silica-carbonate biomorphs, we are left with the question of how they form.

\subsubsection{Coprolite-like siderite masses}\label{sec:coprolite}

\begin{figure}[tbp]
\begin{center}
    \begin{subfigure}{0.9\textwidth}
    \includegraphics[width=\linewidth]{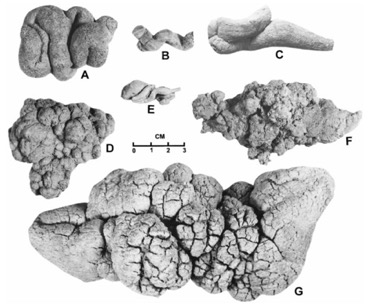}
    \caption{} 
    \end{subfigure}
    \begin{subfigure}{0.9\textwidth}
    \includegraphics[width=\linewidth]{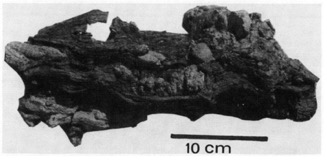}
    \caption{} 
    \end{subfigure}
\end{center}
\caption{
Coprolite-like siderite masses. 
(a) Different samples taken from \citet{mustoe2001}.
(b) Coprolite-like siderite masses enclosed in a carbonized wood fragment \citep{spencer1993}.
}\label{fig:coprolite}
\end{figure}

Coprolites are fossils from the feces of animals (invertebrates, mammals, dinosaurs, fish, etc) \citep{hantzschel1968coprolites,hunt2012vertebrate,knaust2020invertebrate}. These fossils have very characteristic morphologies that generally makes it difficult to confuse them with other types of rocks. One particular instance is the helical structure of so-called \emph{spiral coprolites} \citep{williams1972origin,mcallister1985reevaluation}.
However, at the beginning of the 20th century, rocks with morphologies similar to those of coprolites were found in the state of Washington, USA (Fig.~\ref{fig:coprolite}a), which have been the subject of controversial debates about their origin \citep{spencer1993,yancey2013}. The two main theories about their origin are: i) that they may be true fossils, and ii) that they may be pseudofossils formed by the deformation of plastic sediment.

Researchers supporting the first theory argue that these remains may be fossil feces, i.e., coprolite \citep{roberts1958,amstutz1958}, or fecal matter preserved in the gut, i.e., cololith \citep{seilacher2001}. Researchers supporting the abiotic origin of these samples suggest that their formation is due to episodes of fluidization of organic-rich clay sediments \citep{roberts1958,spencer1993,mustoe2001}. This fluidization may be due to a high sedimentation rate, coseismic events or methanogenic processes. The morphologies of these samples would be produced by the extrusion of the fluid material in hollow logs (Fig.~\ref{fig:coprolite}b) or in the overlying sediment.
What most authors agree on is that the original material (either biological or argillaceous) has been subsequently transformed into siderite \citep{postma1982}
\begin{center}
4FeOOH + CH$_2$O + 3H$_2$CO$_3$ $\to$ 4FeCO$_3$ + 6H$_2$O
\end{center}
Siderite masses with coprolite-like forms have been found in other localities beyond the USA \citep{seilacher2001,brachaniec2022}, including Canada, Madagascar, China and Poland, and although in those cases they have always been catalogued as coprolite/cololite, some researchers have left the door open to abiotic processes \citep{brachaniec2022}. According to some authors, in order to have clear evidence that these remains correspond to biological activity, evidence of food should be found in them---e.g., seeds, tissues or skeletal remains---or they should be related to fossil remains \citep{spencer1993,yancey2013}. Exceptionally preserved microbial cells in some phosphatic coprolites can also serve as an argument for biogenicity, but their identification requires the use of nano-scale analytical methods \citep{Cosmidis2013Nanometer‐scale}.

\section{Chemical biomimetism: Self-organized abiotic chemical systems that produce life-like forms}
\label{sec:self-organized}

There are  completely abiotic self-organized chemical systems which may be found in geology and form biomimetic structures that could be mistaken as being of  biological origin. In this section, we give an overview of  instances of these.
We must point out that there is necessarily a difference in time-scale between laboratory experiments and many geological patterns. This is a problem for reproducing geological patterns in the lab that may form over a time-scale far longer than a student grant; far longer indeed than a human lifetime. 
It is also not known whether all these chemical pattern-forming systems given below appear in geological settings, or not. The first laboratory system we discuss, chemical gardens, has geological counterparts in  hydrothermal vents, and may be found in some agates. 
The second system, silica-carbonate biomorphs, may perhaps also be seen in agates,
as we mentioned in Sec.~\ref{sec:spirals}.
The other systems we discuss are not today known to appear in a geological setting. They are proposed as self-organizing chemical systems that are found to produce self-assembled structures in laboratory experiments, and which are hypothesized as being plausible to occur under geological conditions.

\subsection{Chemical gardens}\label{sec:chemgardens}

\begin{figure}[tb]
\includegraphics[width=\textwidth]{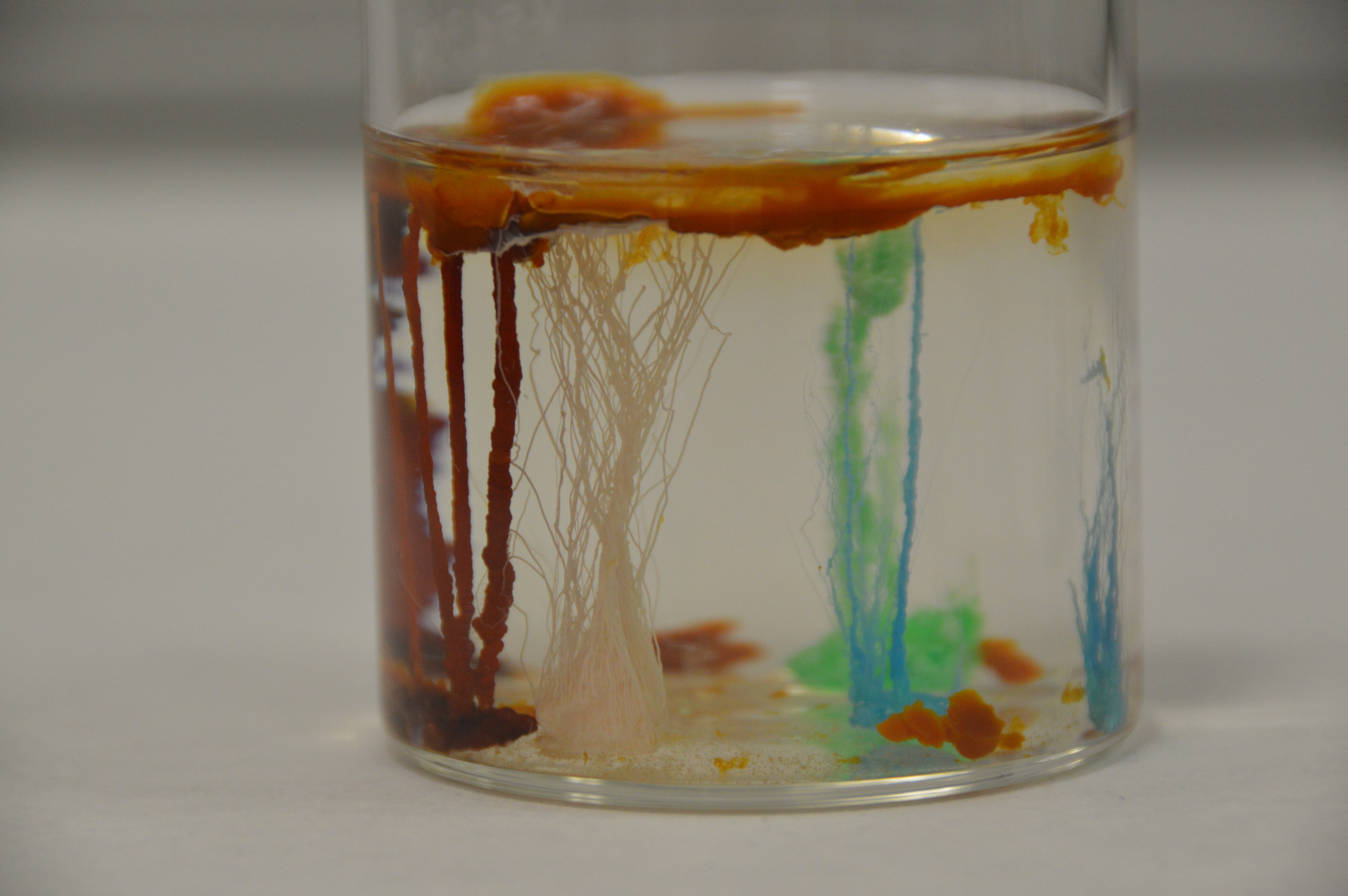}
\caption{A chemical garden (image: Julyan Cartwright)}
\label{fig:chem_garden}
\end{figure}

Chemical gardens are abiotic biomimetic structures forming tubular solids that may be twisted into multiple forms imitating plants (Fig.~\ref{fig:chem_garden}). These inorganic hollow tubes have been studied for several centuries. Their formation often starts with a solid seed crystal of a metal salt placed into a solution of silicate, phosphate, carbonate or a one of a variety of anions whose metal salts are relatively insoluble. The liquid phase begins to dissolve the surface of the seed crystal, and chemical reaction forms the insoluble salt product, which precipitates around the seed. The solid product forms a sheet covering the seed, but possessing a degree of porosity, creating a semipermeable membrane. The external solvent enters inside this membrane by osmosis. The entrance of external solvent provokes an increase of volume, stressing the membrane until rupturing it. Then, the internal solution exits under pressure and buoyancy forces. When the internal components of this solution contact with the external solution further precipitation occurs forming the walls of the tubes. These forces of osmosis and buoyancy plus precipitation change constantly provoking an alteration of growth rates, direction and morphology, yielding structures that mimic biological gardens. This phenomenon is a self-organizing non-equilibrium process of chemical reaction operating under advective fluid  dynamics.

This process has fascinated many researchers over the centuries beginning with \citet{glauber1646} until now. The phenomenon was from its beginnings in alchemy and proto-chemistry related with then-current ideas of the processes of life by researchers such as \citet{boyle} and \citet{newton}. Ideas linking chemical gardens and life persisted through the 19th century with work of \citet{traube1867} on artificial cells, to reach a peak with the research at the end of the 19th and beginning of the 20th centuries, including by \citet{herrera1903} with \emph{plasmogenesis}  and by \citet{leduc1911} with \emph{synthetic biology}.  

With  developments in physics, chemistry and instrumental techniques, concepts have been introduced such as osmosis, buoyancy, fluid dynamics, explaining the formation mechanisms of this phenomenon \citep{barge2015chemical}. In the last years, the interest in these systems has increased as a clear example of self-organized patterns and also for educational purposes.
Many studies on this phenomenon have been performed at the laboratory scale \citep{pimentelChemobrionicsDatabaseCategorisation2023}. However, these structures can be found also in nature as geological samples, though the time scale is clearly different. The same components and driving forces occur in nature, such as, semipermeable membranes, osmotic pressure, concentration gradients, pH gradients, buoyancy, pressure, fluid dynamics, and precipitation.  Some examples of geological chemical gardens are, e.g., hydrothermal vents in the sea-floor \citep{russell1989vitro,russell2019prospecting}, brinicles in sea-water \citep{brinicle}, corrosion processes \citep{brau2018filament}, inorganic vesicles \citep{holler2023hybrid}, and embedding patterns between two mineral phases, such as possible chemical gardens in agates \citep{knoll2022petrified}.

\subsection{Silica-carbonate biomorphs}\label{sec:biomorphs}

\begin{figure}[tb]
\centering
\includegraphics[width=\textwidth]{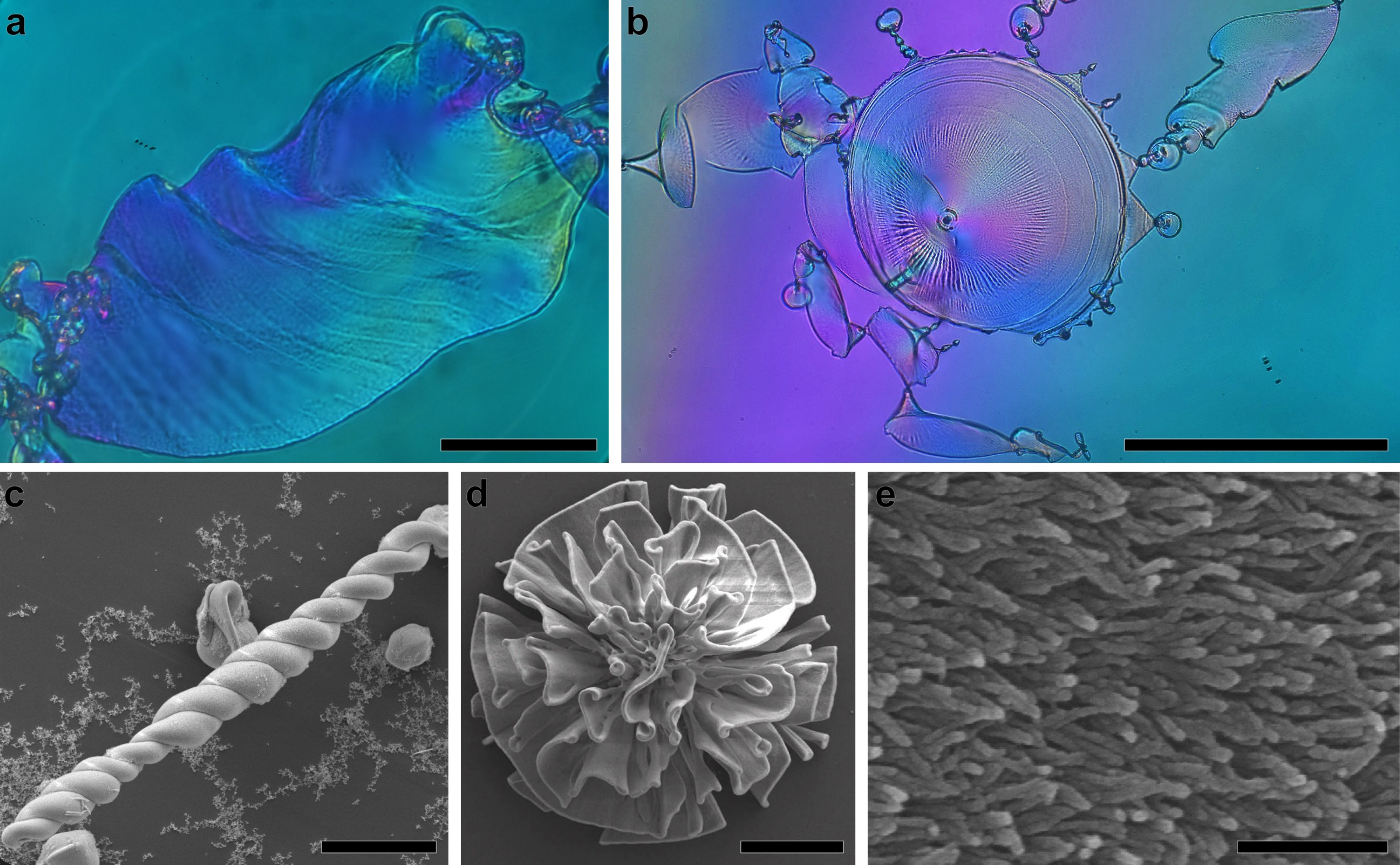}
\caption{Silica-carbonate biomorphs. Polarized light mircorgraphs of a thin sheet (a) and interconnected funnels (b). SEM images of a double helix (c) and  coral-like silica-carbonate biomorphs (d). Each microstructure is composed of nanorods (e) coaligned in the direction of the growing structure. Scale bars are a) 100~$\mu$m, b) 500~$\mu$m, c,d) 20~$\mu$m, and e) 300~nm. (images: Pamela Knoll).
}
\label{fig:SCbiomorph}
\end{figure}

Silica-carbonate biomorphs are crystalline microstructures with smoothly-curved, biomimetic shapes. They are composed of thousands of metal carbonate nanorods which self-organize into a plethora of life-like shapes such as sheets, helices, funnels, and corals (Fig.~\ref{fig:SCbiomorph}). First synthesized from the diffusion of a metal-salt solution (calcium-, barium-, and strontium-based salts) into a silica gel medium, silica-carbonate biomorphs will also form in solution from an initially homogeneous mixture of sodium metasilicate and the above mentioned metal salts. In both methods, CO$_2$ from the air is incorporated into the alkaline gel (pH $\sim$10.5) or solution (pH 10--12) and reacts with the metal ions to crystallize aragonite-type metal carbonates, although more recently biomorphs composed of monohydrocalcite were formed at elevated temperatures. During the first phase of growth a globule forms as growth along the c-axis of an initially pseudo-hexagonal crystal rod is interrupted by polymeric silica along growing face. This induces successive fractal branching into a spherical shape from which all biomorphs nucleate. From the globule the biomorphs emerge with the attachment of spherical nanoparticles (nanodots) to the structure at the active growth front, a narrow zone of a lower pH from the bulk solution. These nanodots coalesce to form the characteristic nanorods (Fig.~\ref{fig:SCbiomorph}e) with a transition zone between dots to rods from the active edge to older regions of the microstructures. While the mechanism behind the microscale architectures has not been fully elucidated, simulations using nonlinear reaction-diffusion models have successfully produced the shapes seen experimentally. For example, flat sheets have edges that are well described by logarithmic spirals forming their leaf-like shape. The same experimental shapes are simulated using a two-variable set of reaction-diffusion equations for subexcitable media. Furthermore, the simulations were extended into three-dimensional space reproducing a range of coral-like architectures using a three reaction step model.

\subsection{Carbon-sulfur biomorphs}\label{sec:carbon-sulfur}

\begin{figure}[tb]
\includegraphics[width=\textwidth]{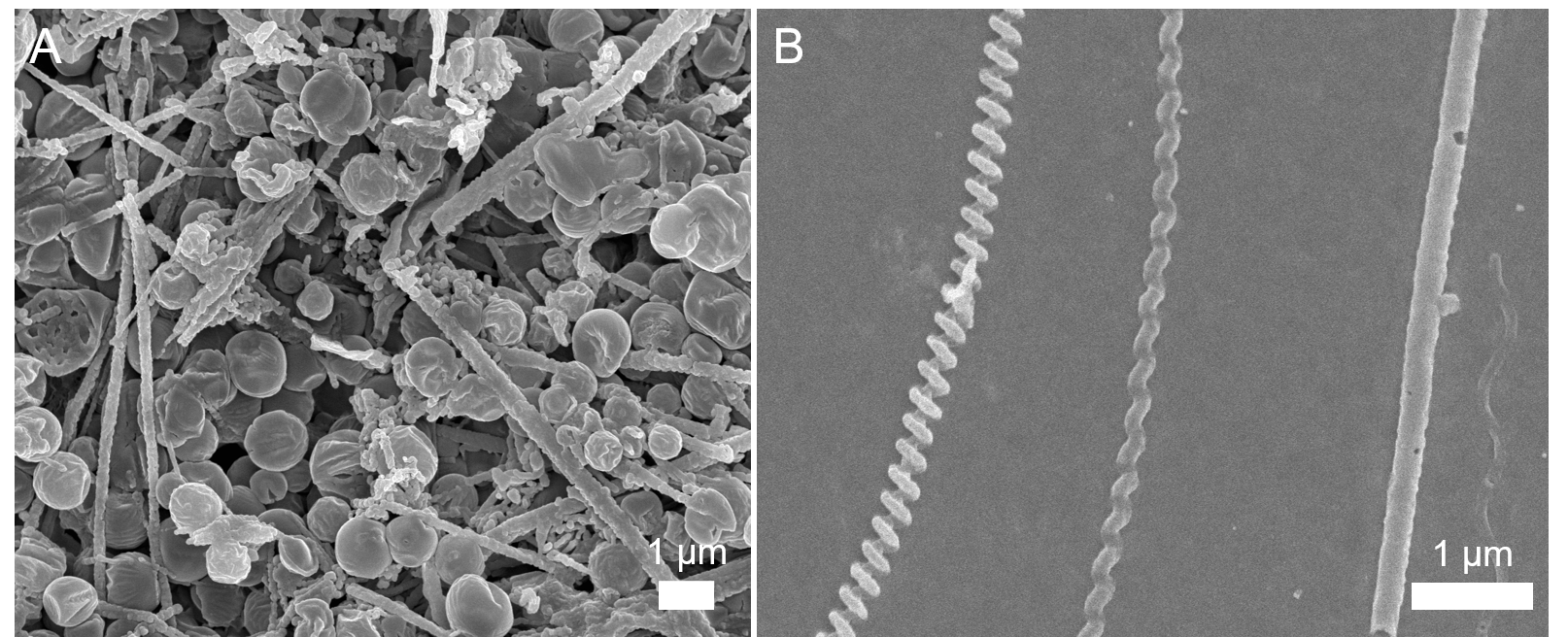}
\caption{Carbon-sulfur biomorphs (SEM images). (A) Spherical and filamentous biomorphs. Note the wrinkly aspect of the spheres, due to deformations of the organic envelopes in the vacuum of the SEM chamber. (B) Twisted and rectilinear filamentous biomorphs. (images: Chrissie Nims (A) and Julie Cosmidis (B))}
\label{fig:CSbiomorph}
\end{figure}

Carbon-sulfur biomorphs are microscopic objects with a core--shell structure, composed of elemental sulfur minerals (the core) encapsulated within an organic envelope (the shell); Fig.~\ref{fig:CSbiomorph}. They spontaneously form following the slow oxidation of sulfide in solutions that contain dissolved organics \citep{cosmidis2016self}. Although their formation mechanism has not yet been fully elucidated, it is thought to involve a 3-step process: (1) the oxidation of sulfide to elemental sulfur, which upon further reaction with sulfide can form polysulfides, (2) the reaction of sulfide and polysulfides with the dissolved organics, resulting in the formation of large macromolecules, including amphiphiles, through sulfurization, and (3) the self-assembly of long-chain amphiphiles at the hydrophobic surface of elemental sulfur minerals, resulting in the formation of the organic shell \citep{cosmidis2019formation}. 
These microstructures may exhibit a wide range of morphologies including spheres, flexible filaments, rigid filaments branching at 45 and 90 degree angles, `corkscrews', `test tube brushes', or `beaded necklaces' \citep{cosmidis2016self,cosmidis2019formation}. Carbon-sulfur biomorphs have been formed in the laboratory from a range of organic compounds ranging from complex mixtures (e.g., yeast extract, peptone) to simple molecules (e.g., glucose), including prebiotic organics (e.g., glycine). They are typically obtained under geochemical conditions (pH, temperature, salinity) relevant to natural aqueous environments, so it is expected that they could be forming in the environment whenever sulfide and dissolved organics are present in solution \citep{cosmidis2019formation}. Carbon-sulfur particles with a core-shell structure have  been observed in microbial cultures of sulfur-oxidizers as well as natural biofilms forming in a sulfidic cave, and it has been suggested that the self-assembly process at the origin of carbon-sulfur biomorphs may be used by bacteria to stabilize elemental sulfur minerals (an energy resource) in their extracellular medium \citep{cosmidis2022microbes,cron2019elemental,cron2021organic}.
It is not clear whether, after formation, carbon-sulfur biomorphs may persist in the geological record. Elemental sulfur is unstable under diagenetic conditions, but recent experimental fossilization experiments suggest that the organic shells of the biomorphs can be preserved through silicification under certain conditions, showing better morphological and chemical preservation potential than microbial cells, which can commonly be found fossilized in ancient cherts \citep{nims2021organic}.

\subsection{Organic biomorphs}\label{sec:organic}

\begin{figure}[tbp]
\begin{center}
\includegraphics[width=\columnwidth]{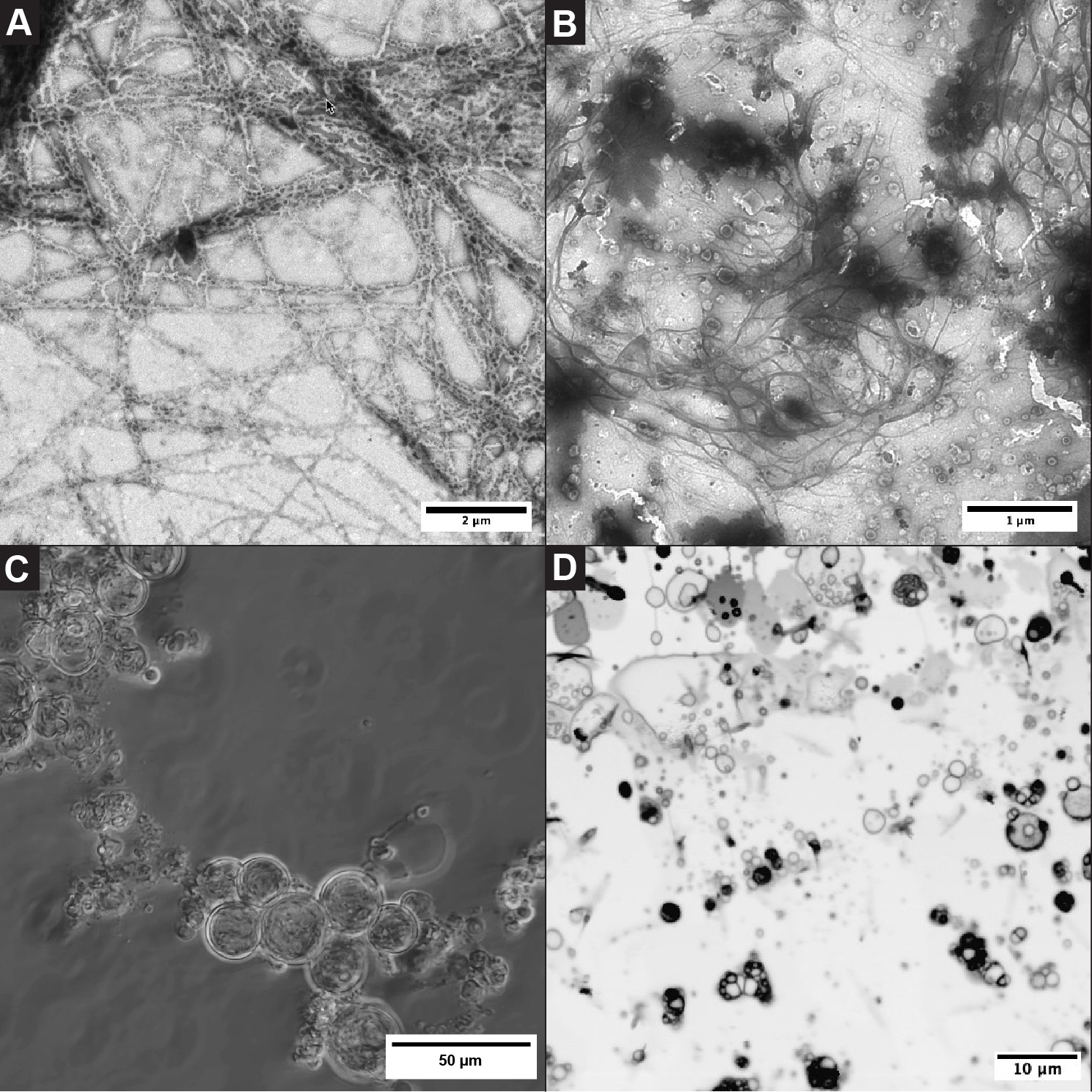}
\end{center}
\caption{
Organic biomorphs. Transmission electron microscopy (A, B), phase contrast microscopy (C), and confocal microscopy (D) micrographs of organic biomorphs formed from single chain amphiphilic molecules. A, filamentous chains, B, individual protocell membranes and associated dendritic structures, C, protocell membrane clusters, D, individual and clustered protocell membranes (images: Sean Jordan).  
}\label{fig:org_bioporph}
\end{figure}

Organic biomorphs are structures  that resemble biological morphologies comprised of abiotic organic molecules. Most data on organic biomorphs come inadvertently from research in prebiotic chemistry and the origin-of-life field, featuring organic molecules likely present on the early Earth. Many of the products of prebiotic chemistry experiments could potentially satisfy currently employed biogenicity criteria, perhaps most notably those of investigations into prebiotic cell membrane and protocell formation \citep{mcmahon2022false, mcmahon2022fundamental}. The list of prebiotically plausible organics that would have been available for protocell formation is ever-growing due to results from both laboratory experiments and analyses of real-world samples \citep{mccollom2013miller}. Prebiotic synthesis experiments have shown the formation of carboxylic acids, amino acids, sugars, and nucleotides \citep{mccollom2013miller}. For example, Fischer-Tropsch-type (FTT) syntheses under hydrothermal conditions produce numerous fatty acids, alcohols and alkanes containing 6 to 40 carbon atoms \citep{mccollom1999lipid}, all of which are suitable membrane-forming components. Carboxylic acids and amino acids of abiotic origin have been detected in rock samples from Earth (e.g., \citet{mccollom2007abiotic,menez2018abiotic}), while numerous organic molecules including both aliphatic and polyaromatic hydrocarbons (PAHs), hydroxy acids, polyols and amino acids have been detected in meteorites (e.g., \citet{elsila2021amino,furukawa2019extraterrestrial,martins2007indigenous,martins2008extraterrestrial,sephton2002organic}). Protocell membranes have been formed directly from organics contained within meteorite samples \citep{deamer1985boundary}. The meteorite flux to the Earth was much higher during the Hadean than it is today, and likely spiked between ca.\ 4.1 and 3.8 Ga during the Late Heavy Bombardment  \citep{hopkins2015protracted}. The vast amount of organics delivered to the early Earth during this time represents a significant source of organic molecules. Coupled to organics formed \emph{in situ} on the early Earth, it is probable that a wide range of compounds would have been available for membrane formation.  During tests investigating the assembly and stability of protocell membranes, the formation of a variety of unique structural assemblages has been observed \citep{jordan2019promotion,jordan2019isoprenoids,jordan2024prebiotic}. Morphologies include small (10--20 $\mu$m) to large ($>$ 100 $\mu$m) clusters of protocells, filaments composed of protocells linked together in a linear assembly, combined mineral/protocell structures, and protocell clusters linked by possible organic wires (Fig.~\ref{fig:org_bioporph}). The appearance of these structures seems to be affected by environmental conditions; however, this has not been systematically investigated. Many of these organic biomorphs are reminiscent of living organisms and ancient microfossils. Further work is needed to determine the preservation potential of organic biomorphs and to characterize their morphological and chemical signatures following diagenesis.

\section{Abiotic or biological?}\label{sec:abiotic-biological}

We have highlighted in this review that morphology is often a poor indicator of biogenicity, as physical and chemical self-organization can result in morphologies that are common to biological and abiotic systems. Lab-grown chemical gardens and biomorphs are now well-known examples of this fact. 
Fortunately, morphology is rarely used alone to assert the biogenicity of a structure or object found in the geological record. The well-publicized debates about the origin of the putative microfossils in the Apex Chert Archean rocks \citep{schopf_microfossils_1993, Wacey_Chert_2016} or the Martian meteorite ALH84001 \citep{McKay1996-ALH84001, Bradley1997-NoNano}, involved discussions around chemical (e.g., organic, inorganic, isotopic) and mineralogical signals \citep{MYu2000-AH84001, Golden2001-ALH, Martel2012-biomimetic}, even if in the public eye the interpretation of segmented filaments looking  in electron micrographs like microbes was the main part of the story. Modern protocols for life detection in ancient rocks or extraterrestrial objects, for instance, the \emph{Ladder of Life Detection} \citet{Neveu2018-LifeDectection}, rightly stress  that rigorously detecting microbial life requires combining different sorts of evidence with increasing levels of strength, including textures and fabrics, isotopic fractionations indicating metabolism, the presence of complex organics with properties such as large enantiomeric excess, or evidence of growth and Darwinian evolution. However, it has been pointed out that for many biogenicity criteria in use today, abiotic structures have either been shown to satisfy them or data are currently unavailable \cite{mcmahon2022fundamental}. Among all the potential biosignatures, morphological \emph{biofabrics} are considered the least reliable, although, unfortunately, among the easiest to detect \citep{Neveu2018-LifeDectection}.

\subsection{Types of biomineralization: controlled/induced/influenced}\label{sec:biomin}

In this review, we have described geological formations forming recognizable patterns and for which the involvement of life has been debated. In these cases, morphology often formed part of the discussion, but often it revolved around the role of microbial activity in the formation of the minerals composing the geological structures in question. Microbes are known to cause the precipitation of minerals through different mechanisms, a process generally called microbial biomineralization \citep{cosmidis2022microbes}. In some cases, the biomineralization is said to be \emph{biologically controlled}, meaning that nucleation and growth of the minerals are tightly directed by the cell’s molecular machinery, encoded in the genome, and the resulting minerals have well defined, recognizable properties such as narrow ranges of sizes and shapes, near perfect crystal structures, or distinct elemental and isotopic compositions \citep{Amor2020-Magnetotactic, Mathon2024-biosign}. When they can be recognized in rocks, such features of controlled biominerals are in principle as solid evidence of life as the Eukaryotic biominerals commonly recognized as fossils, such as shells and skeletons; the challenge remaining that they are much smaller and harder to detect, and possibly more prone to diagenetic transformations. The biogenicity of minerals is more difficult to establish when the mineralization mechanism is said to be \emph{biologically induced} or \emph{biologically influenced}. In the former case, mineral precipitation is caused by chemical changes in the vicinity of the cells, induced by their metabolic activity; e.g., a pH increase. The latter process is a more passive one, in which the cells and their extracellular polymeric materials create charged surfaces on which heterogeneous mineral nucleation can happen. These two mechanisms generate minerals that may lack distinct features, meaning that their mineralogical properties such as morphology, structures, or composition, may not differ greatly from those of minerals formed abiotically in similar geochemical conditions.

\subsection{Fossils, trace fossils, and fossil structures}

The foregoing discussion in some ways mirrors the distinction between fossils \textit{sensus stricto}, which are preserved remains of organisms in rocks, and \emph{trace fossils}, a term that has been employed since the 1950s and that can be defined as ``a sedimentary structure resulting from the activity of an animal moving on or in the sediment at the time of its accumulation'' \citep{simpson1956trace}. 
Trace fossils are 
also known as ichnofossils.
Examples of such sedimentary structures  are tracks, trails, burrows, borings, and fecal pellets, resulting from the physical action of the organisms on the local sediments, or waste products of their metabolic activity. Like actual fossils, controlled biominerals are preserved \emph{body parts} or microorganisms and relatively easily recognized, while biologically induced and influenced minerals are, like trace fossils, more indirect evidence of biological activity, and more difficult to identify. 
Things become more complicated when we consider single-celled organisms that modify their environment. If a stromatolite is biogenic in origin, should it be termed a fossil or a trace fossil? Some call stromatolites a third term,  \emph{fossil structures}; they seldom preserve any actual fossilized cells.

\subsection{Microbe-mediated structures}

It can  be argued that, due to the ubiquity of microbial life in Earth-surface environments, microbes by their mere presence and chemistry-altering metabolic activity influence---at least to some extent---all sedimentary and low-grade diagenetic processes. Indeed, while the existence of life is difficult to establish on other celestial bodies, its absence from a particular environment on Earth is almost equally difficult to demonstrate  \citep{Belilla2022-Dallol}. Microbial life has evolved the capacity to thrive in virtually all available ecological niches on our planet \citep{Merino2019-Extremes}, making truly abiotic low-temperature ($\lesssim 120$°C) systems almost inexistent. It is for this reason that we need to resort to experiments under controlled sterile conditions in the laboratory to study purely abiotic chemical self-organization processes. It may also be for this reason that both biological and abiotic mechanisms have been invoked for the geological structures that we have described: physico-chemical and biological processes are always occurring hand-in-hand in nature on this planet, and their relative influences are hard to disentangle. The comparative  importance of biological versus physico-chemical processes in pattern formation may also vary through the deposition and diagenesis history of a geological formation; for instance, mineral nucleation and initial growth may be biologically influenced or induced, but sedimentary accumulation of the minerals, followed by reworking and diagenesis may subsequently  be  driven by abiotic processes, partially overprinting primary biological signals, so that the observable macro-scale pattern  may be due to abiotic processes, even if the minerals that compose it are biogenic. In these cases, micro- to nano-scale analyses may be necessary to establish biogenicity.

\subsection{Mineral evolution}

Although it had previously been assumed that the diversity and distribution of minerals had not changed throughout the Earth's history, it has been known since the second half of the 20th century that this is not the case. \citet{zhabin1981mineralevolution} introduced the idea of mineralogical changes throughout geological history. \citet{yushkin1982evolutionary} observed that the number of minerals in the Earth had increased since its formation. However, it was not until the beginning of the 21st century that this theory became more widely accepted. \citet{Hazen2008Mineral} proposed that in addition to the appearance of new minerals, physicochemical changes in the Earth since its formation had led to the appearance of new groups of minerals. Roughly speaking, this evolution can be divided into three stages. The first, following planetary accretion, saw the formation of sulphides, phosphides, iron-nickel alloys, oxides and silicates. An estimated 250 different minerals were formed during this first stage. The second stage relates to the emergence of plate tectonics, which led to the formation of granites and metamorphic rocks. During this stage, a large number of sulphides and silicates were created, resulting in the formation of about 1\,250 new minerals. The third and final stage is linked to the appearance of photosynthesis and the increase of oxygen in the atmosphere. This changed the atmosphere of the planet from a reducing  to an oxidizing one, increasing the number of minerals in the oxide group, and allowing the appearance of new groups of minerals such as carbonates, sulphates and phosphates. Many of these are directly related to life. This last stage increased the number of minerals on Earth to the more than 6\,000 known today. This theory of mineral evolution allows us to deduce that planets or moons with different evolutionary histories to Earth's will have different mineral compositions that vary depending on their proximity to the star they orbit, their size, whether or not they have plate tectonics, and whether or not they support life \citep{hazenMINERALECOLOGYCHANCE2015, hazenEvolutionarySystemMineralogy2019}.

\subsection{The need for a better understanding of abiotic pattern-formation mechanisms in geology}

In addition to the need for a better understanding of biological mineral- and pattern-formation mechanisms, we argue for a greater understanding of the non-biological processes that can lead to complex pattern formation in geological systems; this is reviewed in a companion publication \citep{cartwright_self-organized}. Indeed, most strategies for life detection in an early Earth or extraterrestrial context include ruling out potential false positives; see for instance NASA's \emph{Ladder of Life Detection} \citep{Neveu2018-LifeDectection} and the \emph{Confidence of Life Detection} scale \citep{Green2021Call}. The latter, for example,  includes as a critical step in assessing the validity of a putative biosignature, that ``all known nonbiological sources of [the assessed] signal [have been] shown to be implausible in that environment''.  This concept is also at the centre of recently proposed definitions for biosignatures, such as that proposed by \citet{Gillen2023Call}, stating that a biosignature is ``any phenomenon for which biological processes are a known possible explanation and whose potential abiotic causes have been reasonably explored and ruled out''. This naturally leads to the problem of \emph{unconceived alternatives} \citep{Vickers2023Confidence}---a version of \emph{unknown unknowns}---what if we are not aware of all the possible abiotic processes likely to occur in the environment of interest,  potentially leading to the formation of biosignature mimics, e.g., biomorphs?. One  example of the paucity of research in this area is the fact that most biomorphs are described by morphology alone, but very little is understood about their chemical signature and whether or not that is distinguishable from true biosignatures. Researchers have thus called for a better understanding of the geologic \emph{abiotic baseline} above which biological signals can be unambiguously identified \citep{McMahon2024Astrobiology}. Establishing this abiotic baseline is  difficult in the case of early Earth environments, and  more so on other celestial bodies where crucial geochemical parameters are only loosely constrained.

\subsection{When abiotic and biological pattern-formation processes work hand in hand}

As we have discussed, the abiotic or biological distinction is not absolutely clear-cut. Some formations are indeed induced by biology but their subsequent formation process is abiogenic.
How should we characterize an instance in which the raw materials had a biogenic origin, but their formation into a given structure is entirely controlled by physical and chemical processes alone?
Many minerals on Earth exist only in part or wholly through biological processes. If such minerals subsequently form part of structures without any further biological intervention, how should we classify the resulting structure?

Examples in which both abiotic and biological mechanisms may have intervened include, e.g.,
banded iron formations (abiotic mechanism: precipitation of iron from ancient oceans through chemical processes, possibly related to oxygen levels;
biological mechanism: influence of early microbial life, such as photosynthesizing cyanobacteria, in oxidizing iron and facilitating its deposition); 
micritic limestones
(abiotic mechanism: direct precipitation of fine-grained calcium carbonate from seawater;
biological mechanism: microbial mediation and the breakdown of calcareous algae and other organisms contributing to micrite formation);
phosphorites
(abiotic mechanism: precipitation of phosphate minerals from seawater due to chemical changes in the water column;
biological mechanism: accumulation of organic matter and microbial activity concentrating phosphate and facilitating its precipitation); and
chert
(abiotic mechanism: silica precipitation from supersaturated waters, often associated with hydrothermal activity;
biological mechanism: silica deposition influenced by the dissolution of siliceous skeletons of microorganisms like diatoms and radiolarians).
We should distinguish these instances from some of the other  cases of biological or abiotic pattern formation we have discussed earlier. In those cases, a given pattern is either abiotic or biological, whereas in these, the existence of the mineral in nature (i.e., its atomic scale make-up) may be owing to biological activity, but not its structure above the molecular scale, what we are calling its pattern.

\subsection{The origin of biogenic structure and the origin of life}

One  aspect of the relationship between abiotic and biological pattern formation is that the former may very well have begot the latter. That is to say, life may have come into being in incubators that were formed through abiotic pattern formation processes. In this way, some of the geological patterns we have discussed here may have influenced the evolution of life as initially abiotic structures that have been taken over by life to become biological structures. Early cells may have begun to biomineralize, for instance,  to manipulate the mineral membranes that formed their compartments in order to
control their metabolism.

The hypothesis that clays are involved in the origin of life is now many decades old. A little more recent is the hypothesis that hydrothermal vents at the ocean floors---which are geological instances of the self-organized chemical system called a chemical garden---may have been the environment in which life emerged. Bringing these two ideas together is 
 the emergence of fougerite, green rust,  in hydrothermal vents as a possible catalyst for the origin of life. Although fougerite is not a clay, it is a  layered double hydroxide in which the interlaminar space is negatively charged rather than positively charged as in clays.

Another example of this is the spheroidal shape of many prokaryotic cells. Supramolecular bilayer structures known as lipid vesicles are suggested to be likely candidates for the first protocell membranes \citep{lai2020protocells}. These are often spheroidal but can also produce filamentous and rod-like morphologies under certain conditions \citep{jordan2024prebiotic}, all shapes resulting from spontaneous self-assembly of amphiphilic organic molecules which is driven by thermodynamics. The subsequent spheroidal (cocci), rod (bacilli), or filamentous shapes of prokaryotic cells then may well be a sort of \emph{evolutionary baggage}---what \citet{10.1098/rspb.1979.0086} called a \emph{spandrel}---not initially selected for the advantage over another form, but simply conserved throughout the transition from chemistry to biology due to their inevitability under certain environmental conditions.

\subsection{Isotopic biosignatures}

Isotopic biosignatures are specific ratios of isotopes; isotopic fingerprints that indicate the presence of life because living organisms preferentially use certain isotopes in their metabolic processes \citep{Eroglu2023,desmarais2025stable}. These signatures are found in the ratios of stable isotopes such as  carbon (\({}^{13}\)C/\({}^{12}\)C), nitrogen (\({}^{15}\)N/\({}^{14}\)N),  sulfur (\({}^{34}\)S/\({}^{32}\)S), and iron (\({}^{56}\)Fe/\({}^{54}\)Fe), and can be used to study past and present life by comparing isotopic patterns from biological sources to those from abiotic processes. 

It could be argued that isotopic signatures are one area where the abiotic baseline is particularly poorly characterized. There is a substantial lack of isotopic data relating to both inorganic and organic biomorphs, likely due in combination to the speciality of the techniques involved and the fact that this field of research is still in its infancy.

\subsection{The potential for artificial intelligence / machine learning techniques}

Among interesting new developments in the field of mineral biosignature identification is the use of machine learning methods to classify patterns, in this case, isotopic and chemical patterns in minerals  \citep{Figueroa2024-MachineLearning}, according to their biogenic versus abiogenic origin. In the future, classification algorithms may be trained to identify patterns with a biological origin based on visual information, using images of geological structures. An obvious limitation here is that such algorithms would need to be trained on sufficiently large, labelled datasets, which require that humans agree on the biogenicity of the geological patterns in question in the first place. Another limitation is that such machine learning models may have low interpretability, meaning that humans may not be able to understand how the models make their classification decisions, limiting their generalizability and robustness.

As we finished writing this text, a first step in this direction has just appeared \citep{wong2025organic}. In this paper, the authors used a pyrolysis–gas chromatography–mass spectrometer to chemically analyse a large number of rock samples containing fossil remains and completely abiotic material. They trained a machine leaning algorithm (Random Forest) on a dataset of 272 samples. The dataset was divided into 9 binary categories and further split into cross-validated training set (using 75~\% of the data in the set) and testing set (25~\%), achieving more than 90~\% accuracy for each individual models. Using this method, samples originated at least 2.52 billion years ago were recognized by the model as samples originated from photosynthetic organisms. Meanwhile, signatures recognized by the model as compatible with life were identified in tested rock samples dating back 3.33 billion years. The results are promising, though the very high dimensionality of the dataset (8,189 features after reduction) compared to the number of samples for each binary tests suggests that some degree of overfitting is still possible, even with cross-validation. If the generalizability of the models is confirmed, this method may help resolve some of the scientific controversies presented in this review.

\subsection{Bio-, or non-bio- ?}

The idea throughout this work has been to look at structures that may, or may not, have a biological influence.
Take the instance of stromatolites: while a few years ago there was a great deal of discussion about possible abiotic formation mechanisms, today stromatolites are usually considered as an example of microbe-mediated mineral precipitation. This can be done passively, when bacteria provide nucleation sites for the minerals via their extracellular polymeric substances, EPS, that have Ca$^{2+}$ binding affinity, or, they induce mineral supersaturation as a result of their metabolic processes. However, abiotic processes can easily form laminated structures, morphologically similar to the biological ones. In this case we speak of stromatolite-like structures. These can be also colonized by bacteria but these bacteria do not participate in the microbialite formation. In this fashion, stromatolites are another case of whether to define a structure by its morphology or by its mechanism of formation. In that respect stromatolites are like many other examples in geology where this debate is played out.

There is a synergetic effect between chemical and biological processes. In modern microbialites we can look at their microbes and their genes and we see that they perform certain metabolic functions that promote mineral precipitation and stromatolite accretion. And this is something that can be confirmed experimentally using microbes and undersaturated solutions. A problem arises with ancient microbialites for which DNA data are not available and undisputable microbial traces are almost never found, while diagenesis, paramorphosis, metamorphosis etc, alter the initial characteristics of the rock. Using isotopes, results are not always clear so the debate biological vs.\ abiotic remains open. 

Thus it may very well be the case that on Earth today all stromatolites are 
microbe-mediated, but in an abiotic environment one might be able to form stromatolites---or ``stromatolite-like'' structures---through abiotic self-organization. That question is important for the early Earth prior to the emergence of life, and it is important if the day comes that a planetary rover finds something that looks like a stromatolite on Mars, or some other place in the Solar System. Logic indicates that the two aspects are not logically incompatible: today, on our planet full of microbes, all stromatolites can be microbe mediated; but in the absence of biological organisms, stromatolites or stromatolite-like structures may form abiotically. And the same is true for many of the systems we have considered here.

 One can quite easily conceive that on the early Earth, when microbes were just beginning to colonize environments, one happened to arrive at an abiotic stromatolite-like structure. It found the surface a good place to be, as the abiotic growth mechanism of the structure brought it nutrients as it sat on the mineral growth front. It, in turn, evolved to mineralize those nutrients and made the now microbe-mediated stromatolite accrete faster than the old abiotic ones. In this way, terrestrial stromatolites became microbe-mediated mineral structures.

\subsection{Conclusion}\label{sec:conclusions}

The problem of distinguishing geological patterns that have abiotic pattern formation mechanisms from other geological patterns that have biological origins remains open.
This question is perhaps of greatest importance for studying the origin of life on Earth and for the search for life on other celestial bodies.   
It is clearly problematic if people should call structures on Mars  evidence for life on the basis of morphology alone. But surely it is also a problem that there might be biogenic processes producing some geological formation, and people would not see them as evidence of life if they do not look particularly  biogenic.

\section*{Acknowledgements}

This review paper originated at the workshop ``Geochemobrionics: Self-Organization in Geological Systems'' organized by Sean McMahon at the University of Edinburgh in 2022. The workshop was funded by the European COST (Cooperation in Science and Technology) programme, action number CA17120, Chemobrionics.
JC acknowledges support from the  European Union's Horizon 2020 ERC Starting grant program (BioFacts, grant agreement 101076666) 
JHEC and CISD acknowledge support from the Spanish Ministerio de Ciencia, Innovaci\'on y Universidades through grant PID2024-160443NB-I00.
FJH acknowledges Grants PID2020-114355GB-I00 and PID2023-146856NB-I00, funded by MICIU/AEI/10.13039/501100011033 and ERDF "a way of making Europe", EU.
EK acknowledges funding from the European Union's Horizon 2020 research and innovation programme under the Marie Sklodowska-Curie grant agreement No 101031812 `NanoBioS'. PS acknowledges the support of the National Science Centre (Poland) under research Grant 2022/47/B/ST3/03395. 

\bibliographystyle{elsarticle-harv} 
\bibliography{geochemobrionics}

\end{document}